\documentclass[%
aps,
onecolumn,
12pt,
tightenlines,
nofootinbib,
superscriptaddress,
floatfix,
prd
]{revtex4-2}

\usepackage{siunitx}
\usepackage{physics}
\usepackage{braket}
\usepackage{amsmath}	
\usepackage{amsthm}		
\usepackage{amssymb}	

\usepackage{mathrsfs, mathtools}

\usepackage{letltxmacro} 

\usepackage{color}
\usepackage{bm}

\usepackage{tgpagella}

\usepackage{graphicx}

\usepackage{fancyhdr}

\usepackage{lastpage}

\usepackage{titlesec}

\usepackage{rotating}
\usepackage{setspace}

\usepackage[printonlyused]{acronym}
\usepackage{aas_macros}
\usepackage{xspace}

\usepackage{datetime}

\usepackage{lettrine}

\usepackage{anyfontsize}

\usepackage[colorlinks=true
,urlcolor=DARKBLUE
,anchorcolor=DARKBLUE
,citecolor=DARKBLUE
,filecolor=DARKBLUE
,linkcolor=DARKBLUE
,menucolor=DARKBLUE
]{hyperref}

\titleformat{\section}
       {\centering\normalfont\fontsize{14}{17}\bfseries}{\thesection}{1em}{}

\numberwithin{equation}{section}

\makeatletter 
    \renewcommand{\thefigure}{{\bf\@arabic\c@figure}}
\makeatother

\usepackage{tcolorbox}

\definecolor{grey}{rgb}{0.4,0.4,0.4}
\definecolor{dullmagenta}{rgb}{0.4,0,0.4}
\definecolor{darkblue}{rgb}{0,0,0.4}
\definecolor{midblue}{rgb}{0,0,0.7}
\definecolor{midred}{rgb}{0.5,0,0}
\definecolor{orange}{rgb}{1,0.5,0}
\definecolor{lightbrown}{rgb}{0.75,0.5,0.25}
\definecolor{tan}{cmyk}{0.14,0.42,0.56,0}
\definecolor{djunglegreen}{cmyk}{0.99,0,0.52,0}
\definecolor{lightgreen}{rgb}{0,1,0}
\definecolor{olivegreen}{cmyk}{0.64,0,0.95,0.40}
\definecolor{midgreen}{rgb}{0.0,0.675,0.0}
\definecolor{darkgreen}{rgb}{0,0.5,0}
\definecolor{pink}{rgb}{1,0.078,0.57}

\definecolor{MONZA}{HTML}{CF000F}
\definecolor{DARKBLUE}{HTML}{00008b}
\definecolor{DARKMAGENTA}{HTML}{8b008b}

\newcommand{\vs}{\vspace}

\makeatletter
\let\oldr@@t\r@@t
\def\r@@t#1#2{%
\setbox0=\hbox{$\oldr@@t#1{#2\,}$}\dimen0=\ht0
\advance\dimen0-0.2\ht0
\setbox2=\hbox{\vrule height\ht0 depth -\dimen0}%
{\box0\lower0.4pt\box2}}
\LetLtxMacro{\oldsqrt}{\sqrt}
\renewcommand*{\sqrt}[2][\ ]{\oldsqrt[#1]{#2}}
\makeatother

\newcommand{\la}{\ensuremath{\leftarrow}}

\newcommand{\be}{\begin{equation}}
\newcommand{\ee}{\end{equation}}
\newcommand{\ba}{\begin{eqnarray}}
\newcommand{\ea}{\end{eqnarray}}

\smallskipamount
\settimeformat{ampmtime}

\def\ga{\mathrel{\raise.3ex\hbox{$>$\kern-.75em\lower1ex\hbox{$\sim$}}}}
\def\la{\mathrel{\raise.3ex\hbox{$<$\kern-.75em\lower1ex\hbox{$\sim$}}}}

\newcommand{\Msun}{\ensuremath{\,\rm{M}_{\odot}}\xspace}
\newcommand{\Ms}{\Msun}
\newcommand{\msun}{\Msun}

\newcommand{\Zs}{{\rm Z}_\odot}

\newcommand{\pc}{{\rm pc}}

\acrodef{BH}{black hole}
\acrodefplural{BH}[BHs]{black holes}

\acrodef{BBH}{Binary black hole}
\acrodefplural{BBH}[BBHs]{binary black holes}

\acrodef{IMBH}{intermediate-mass black hole}
\acrodefplural{IMBH}[IMBHs]{intermediate-mass black holes}

\acrodef{SMBH}{supermassive black hole}
\acrodefplural{SMBH}[SMBHs]{supermassive black holes}

\acrodef{PBH}{primordial black hole}
\acrodefplural{PBH}[PBHs]{primordial black holes}

\acrodef{BHXB}{black hole X-ray binary}
\acrodefplural{BHXB}{black hole X-ray binaries}

\acrodef{GC}{globular cluster}
\acrodefplural{GC}[GCs]{globular clusters}

\acrodef{NC}{nuclear cluster}
\acrodefplural{NC}[NCs]{nuclear clusters}

\acrodef{OC}{open cluster}
\acrodefplural{OC}[OCs]{open clusters}

\acrodef{YMC}{young massive cluster}
\acrodefplural{YMC}[YMCs]{young massive clusters}

\acrodef{MLT}{mixing-length theory}
\acrodef{HR}{Hertzsprung-Russell}
\acrodef{MS}{main sequence}
\acrodef{ZAMS}{zero-age MS}
\acrodef{CHeB}{core helium burning}
\acrodef{RGB}{red giant branch}
\acrodef{RG}{red giant}
\acrodef{RSG}{red super-giant}
\acrodef{HB}{horizontal branch}
\acrodef{AGB}{asymptotic giant branch}
\acrodef{TP-AGB}{thermally-pulsating asymptotic giant branch}
\acrodef{LBV}{luminous blue variable}
\acrodef{WR}{Wolf-Rayet}

\acrodef{WD}{white dwarf}
\acrodefplural{WDs}{white dwarfs}
\acrodef{NS}{neutron star}
\acrodefplural{NSs}{neutron stars}
\acrodef{CO}{compact object}
\acrodefplural{COs}{compact objects}

\acrodef{CC}{core collapse}
\acrodef{ECSN}{electron-capture supernova}
\acrodef{PI}{pair instability}
\acrodef{PISN}{pair instability supernova}
\acrodef{PPISN}{pulsational pair instability supernova}
\acrodef{compas}{\texttt{COMPAS}}
\acrodef{mobse}{\texttt{MOBSE}}
\acrodef{cBH}{Black hole}
\acrodefplural{cBH}[BHs]{Black holes}

\acrodef{sBH}{stellar-mass black hole}
\acrodefplural{sBH}[sBHs]{Stellar-mass black holes}

\acrodef{IMBH}{intermediate-mass black hole}
\acrodefplural{IMBH}[IMBHs]{intermediate-mass black holes}

\acrodef{SMBH}{supermassive black hole}
\acrodefplural{SMBH}[SMBHs]{Supermassive black holes}

\acrodef{IMRI}{intermediate-mass ratio inspiral}
\acrodefplural{IMRI}[IMRIs]{intermediate-mass ratio inspirals}

\acrodef{BBH}{binary black hole}
\acrodefplural{BBH}{binary black holes}

\acrodef{GC}{globular cluster}
\acrodefplural{GC}[GCs]{Globular clusters}

\acrodef{NSC}{nuclear star cluster}
\acrodefplural{NSC}[NSCs]{Nuclear star clusters}

\acrodef{OC}{open cluster}
\acrodefplural{OC}[OCs]{open clusters}

\acrodef{YMC}{young massive cluster}
\acrodefplural{YMC}[YMCs]{Young massive clusters}

\acrodef{AGN}[AGN]{active galactic nucleus}
\acrodefplural{AGN}[AGNs]{active galactic nuclei}

\acrodef{IMF}{initial mass function}

\acrodef{GW}{gravitational wave}
\acrodefplural{GW}{gravitational waves}
\acrodef{LVK}{LIGO-Virgo-KAGRA collaboration}
\acrodef{LISA}{Laser Interferometer Space Antenna}

\acrodef{VMS}{very massive star}
\acrodefplural{VMSs}[VMS]{very massive stars}

\acrodef{ZAMS}{zero-age main sequence}
\acrodef{MS}{main sequence}
\acrodef{WD}{white dwarf}
\acrodefplural{WDs}{white dwarfs}
\acrodef{NS}{neutron star}
\acrodefplural{NSs}{neutron stars}
\acrodef{CO}{compact object}
\acrodefplural{COs}{compact objects}
\acrodef{DCO}{double compact object}
\acrodefplural{DCOs}{double compact objects}
\acrodef{RLOF}{Roche Lobe Overflow}

\acrodef{CE}{Common Envelope}

\acrodef{MW}{Milky Way}

\acrodef{SSE}{\texttt{SSE}}
\acrodef{BSE}{\texttt{BSE}}
\acrodef{MOCCA}{\texttt{MOCCA}}

\makeatletter
\def\l@subsubsection#1#2{}
\makeatother

\headwidth=\textwidth

\usepackage{adjustbox}
\usepackage{multirow}

\begin{document}

\title{\Large {\huge M}assive {\huge B}lack {\huge H}oles in {\huge G}alactic {\huge N}uclei}

\author{David Izquierdo-Villalba}
\email{david.izquierdovillalba@unimib.it}
\affiliation{Dipartimento di Fisica ``G. Occhialini'', Universit\`a degli Studi di Milano-Bicocca, Piazza della Scienza 3, I-20126 Milano, Italy}
\affiliation{INFN, Sezione di Milano-Bicocca, Piazza della Scienza 3, I-20126 Milano, Italy}
\author{Alessandro Lupi}
\email{alessandro.lupi@uninsubria.it}
\affiliation{Dipartimento di Scienza e Alta Tecnologia, Università degli Studi dell'Insubria, Via Valleggio 11, I-22100 Como, Italy}
\affiliation{INFN, Sezione di Milano-Bicocca, Piazza della Scienza 3, I-20126 Milano, Italy}
\affiliation{Dipartimento di Fisica ``G. Occhialini'', Universit\`a degli Studi di Milano-Bicocca, Piazza della Scienza 3, I-20126 Milano, Italy}
\author{John Regan}
\email{john.regan@mu.ie}
\affiliation{Centre for Astrophysics and Space Science Maynooth, Department of Theoretical Physics, Maynooth University, Ireland}
\author{Matteo Bonetti}
\email{matteo.bonetti@unimib.it}
\affiliation{Dipartimento di Fisica ``G. Occhialini'', Universit\`a degli Studi di Milano-Bicocca, Piazza della Scienza 3, I-20126 Milano, Italy}
\affiliation{INFN, Sezione di Milano-Bicocca, Piazza della Scienza 3, I-20126 Milano, Italy}
\author{Alessia Franchini}
\email{alessia.franchini@unimib.it}
\affiliation{Dipartimento di Fisica ``G. Occhialini'', Universit\`a degli Studi di Milano-Bicocca, Piazza della Scienza 3, I-20126 Milano, Italy}
\affiliation{INFN, Sezione di Milano-Bicocca, Piazza della Scienza 3, I-20126 Milano, Italy}

\date{\formatdate{\day}{\month}{\year}, \currenttime}

\begin{abstract}
\vs{4mm}
\begin{tcolorbox}
Massive black holes are key inhabitants of the nuclei of galaxies. Moreover, their astrophysical relevance has gained significant traction in recent years, thanks especially to the amazing results that are being (or will be) delivered by instruments such as the James Webb Space Telescope, Pulsar Timing Array projects and LISA. In this Chapter, we aim to detail a broad set of aspects related to the astrophysical nature of massive black holes embedded in galactic nuclei, with a particular focus on recent and upcoming advances in the field. In particular, we will address questions such as: What shapes the relations connecting the mass of massive black holes with the properties of their host galaxies? How do massive black holes form in the early Universe? What mechanisms keep on feeding them so that they can attain very large masses at $z=0$? How do \textit{binaries} composed of two massive black holes form and coalesce into a single, larger black hole? Here we present these topics from a mainly theoretical viewpoint and discuss how present and upcoming facilities may enhance our understanding of massive black holes in the near future.
\end{tcolorbox}
\end{abstract}

\begin{center}
    \phantom{\fontsize{50}{50}\selectfont I}\\
    {\fontsize{40}{10}\selectfont {\fontsize{50}{20}\selectfont M}assive {\fontsize{50}{20}\selectfont B}lack {\fontsize{50}{20}\selectfont H}oles\\[14mm]
    in {\fontsize{50}{20}\selectfont G}alactic {\fontsize{50}{20}\selectfont N}uclei}\\[80mm]
    {\noindent\makebox[\linewidth]{\resizebox{0.3333\linewidth}{1pt}{$\bullet$}}\bigskip}\\[3mm]
    {\fontsize{18}{5}\selectfont David Izquierdo-Villalba$^*$}\\ [6mm] {\fontsize{18}{5}\selectfont Alessandro Lupi$^*$}\\[6mm]        
    {\fontsize{18}{5}\selectfont John Regan$^*$} \\ [6mm]
    {\fontsize{18}{5}\selectfont Matteo Bonetti$^*$}\\[6mm]
    {\fontsize{18}{5}\selectfont Alessia Franchini$^*$}\\[4mm]
    {*all authors equally contributed to the chapter}
\end{center}
\newpage
\maketitle

\newpage
\tableofcontents
\newpage

\part{Supermassive black holes and their environments \\ \Large{Izquierdo-Villalba}}

\section{Introduction}
\acp{SMBH} inhabit the nuclei of many massive galaxies and are among the most fascinating objects in our Universe. Their typical masses span $10^5-10^{11} \msun{}$, even if there is increasing evidence for the existence of smaller \ac{IMBH},  as discussed in the previous Chapter. \acp{SMBH} appear to inhabit the majority of galaxies with stellar mass above $10^{11}\msun{}$.
Although their sphere of influence\footnote{The sphere of influence is a sphere centered onto the \ac{SMBH} where the motion of stars is significantly affected by the \ac{SMBH} potential; here we define it as the sphere containing twice the \ac{SMBH} mass in stars.} is much smaller ($10^{-3}-10^{-2}$ times) compared with the spatial extent of the galactic stellar population, \acp{SMBH} seem to correlate with several properties of their host galaxies - for example the bulge total stellar mass, the central luminosity and the velocity dispersion. This has led to the interpretation that the \ac{SMBH}  growth via accretion and the associated feedback may have had a relevant influence on the star formation in their host galaxies.

In spite of their relevance in the cosmic evolution of the galactic population, the origin of \acp{SMBH} remains debated. Several formation scenarios have been proposed in the last few decades, with each having their own advantages and disadvantages. Most of these processes can only occur effectively in the early Universe, where present day telescopes  have very limited capabilities. In the near future, upcoming \ac{GW} facilities may shed new light on this aspect and revolutionize our understanding in the formation of \acp{SMBH}.

Galaxies harbouring \acp{SMBH} grow more massive over cosmic time through repeated mergers; their central \acp{SMBH} may thus find themselves in the center of a galaxy merger remnant. The \ac{SMBH} pair can then shrink down to very small separations, form a binary system and eventually  merge via the emission of low-frequency \acp{GW}. This entire process has received great  attention in the last decade as the emitted \acp{GW} are potentially observable by future  \ac{GW} detectors such as the \ac{LISA}. The promise of future \acp{GW} detectors to pierce into the dark cosmos down to the earliest epochs of galaxy and \ac{SMBH} formation may allow us to shed new light on the origin of \acp{SMBH}, as well as their co-evolution with their host galaxies.

\acp{SMBH} will be the focus of this entire Chapter. We will start with an historical overview of their discovery; we will then detail the current knowledge about the coupled evolution between \acp{SMBH} and their host galaxies; we will cover the proposed formation process of  \acp{SMBH} and finally we will  describe the path to coalescence of \ac{SMBH} binaries.

\section{Historical overview}
Thanks to the advances in radio astronomy in the 1950's several works described the existence of a population of compact radio sources in which the optical counterpart was absent or, in some cases, only a faint star-like object was emerging. These new sources were classified as stars given their small angular size.
Interested in the new discovery, Maarten Schmidt took the spectra of one of these radio sources (3C 273). The results showed five broad emission lines which were incompatible with a spectrum generated by a star. In his work \cite{1963Natur.197.1040S}, he argued that the most plausible nature of 3C 273 was the nuclear region of a galaxy with a cosmological redshift of $z{\sim}\,0.158$. Over the next few years following this publication, similar objects (other than 3C 273) were reported by different studies, bringing more weight to the idea that these objects were of cosmological origin \citep[][]{1963ApJ...138...30M,1965AJ.....70..384W,1966AuJPh..19..471B,1966AuJPh..19..559B,1966ApJ...145..668O}. Despite the stellar nature of these objects being conclusively ruled out, their point-like shape similar to a star caused them to become known as \textit{quasi-stellar radio sources} or just \textit{quasars}.\\

The physical understanding of quasars started with \cite{1964ApJ...140..796S} and \cite{1969Natur.223..690L} which argued that quasars are the manifestation of gas accretion processes onto \acp{SMBH}. Several years later, important advances in this direction took place when  \cite{1973A&A....24..337S} presented a theory to describe the accretion of matter onto a stellar \acp{BH}. Even though it was originally proposed to explain the observational appearance of stellar \acp{BH} in binary systems, it could be perfectly extended to gas accretion onto \acp{SMBH}. The theory assumed that the matter which flows towards the \ac{SMBH} needs to have considerable angular momentum to prevent free fall accretion and be observable. As matter approaches  the \ac{SMBH}, the centrifugal forces become comparable to the gravitational ones, thus  the matter rotates on circular orbits and leads to the formation of a disc-like structure. In the theory, \cite{1973A&A....24..337S} showed that this configuration is not static but  matter loses angular momentum and spirals towards the \ac{SMBH} thanks to the viscous friction between adjacent disc layers. This inspiralling process lasts until the matter reaches the \ac{SMBH} where it is finally accreted. During the accretion event, gravitational energy is released and transformed into thermal energy which is eventually radiated away from the disc surface. \cite{1973A&A....24..337S} proposed that the total outgoing energy is fully determined by the  matter inflow rate onto the accretion disc, $\dot{M_{\rm d}}$:
\begin{equation} \label{eq:luminosity}
L\,{=}\, \epsilon \,  \dot{ M_{\rm d}} \,  {c^2},
\end{equation}
where  $L$ is the emitted luminosity, $c$ is the speed of light in vacuum, and $\epsilon$ is the so-called radiative efficiency (i.e. the fraction of the rest energy of the accreted matter which is released in the form of energy, $\sim0.1$) and depends on the \ac{SMBH} spin.\\

The development of theories able to link quasars with accretion events onto \acp{SMBH} sparked the interest in extragalactic objects that could be  powered in the same way. The first candidate host galaxies were reported by \cite{1943ApJ....97...28S}: the sample consisted of  nearby spiral galaxies whose nuclear spectral energy distribution displayed  broad emission lines  similar to those seen in quasars; galaxies with such properties became thus known as \textit{Seyfert galaxies}. Many more local galaxies showing similar behavior to the Seyfert population were reported by subsequent studies \citep[see][]{1976ApJ...206..898O,1980ApJ...240..429W,1985ApJ...288..205A}. Despite all these systems being somewhat similar to the prototypical quasar spectrum, their diversity in the line shapes (broad and narrow), extra types of emission lines, and in some cases the absence of radio emission made it too difficult to unequivocally conclude that these local galaxies were quasar-like objects at lower redshift. Thanks to the large efforts carried out in the '90s, it was possible to postulate the \textit{unification} model \citep{1987PASP...99..309L,1993ARA&A..31..473A} which established that all the galaxies containing an accreting nuclear \ac{SMBH} were in fact \textit{active galactic nuclei} (AGN) and could be brought under the same umbrella term. The specific features of both radio emission and line emission strongly depend on the accretion rate onto the \ac{SMBH}  and the matter distribution between the \ac{SMBH} and the observer. Seyfert I and II galaxies, radio galaxies, LINERs, or quasars are different manifestations of AGNs as defined through the unification model. Taking this unification model as a foundation, \cite{1982MNRAS.200..115S} tried to shed light on the expected \ac{SMBH} demographics in the nearby Universe. 
The line of thinking proposed by \cite{1982MNRAS.200..115S} was based on the fact that the mass density of \acp{SMBH} in nearby
massive galaxies should be the result of gas accretion of quasars and active galaxies at high redshift. By using available data of quasar bolometric luminosity functions,  \cite{1982MNRAS.200..115S} concluded that the total \ac{SMBH}  mass density in the local Universe would reach up to $\rm 4.7{\times}10^{4}\, \Ms/Mpc^{3}$ and, on average, every local giant galaxy should contain a $\rm 10^8\,{-}\,10^9\, \Ms$ \ac{SMBH}.  Nevertheless the results could only be considered as lower limits as a consequence of selection biases in the observations.  This became known as the ``Soltan argument'', and provided support to the idea that quiescent \acp{SMBH} should exist at the centre of many local galaxies.

Thanks to this previous work, the existence of \acp{SMBH} at the centre of galaxies became increasingly evident. However, direct confirmations were still pending. The first steps in this direction were done in the '70s and '80s \citep[see the review of][]{1995ARA&A..33..581K}. One of the pioneering works on this line was \cite{1978ApJ...221..731S} which studied the kinematics of the galaxy M87 by using spectroscopic data. The analysis showed a sharp increase in the velocity dispersion in the inner parts of the galaxy (${\sim}\,110\,\rm \pc$), inconsistent with the expected mass-to-light ratio. To explain such discrepancy, the authors hypothesised the existence of a compact massive object with $5\,{\times}\,10^9\,\rm \Ms$ at the center of M87. \cite{1988ApJ...324..701D} were able to report similar pieces of evidence for the galaxy M31 and the dwarf elliptical M32. Regardless of the geometry and the stellar dynamics included in their models, the authors were not able to reasonably reproduce the spectroscopic observations of these nearby galaxies. Indeed, similar issues to the ones of \cite{1978ApJ...221..731S} were reported. The mass-to-light ratios of these galaxies were too high and supported the idea that both M31 and M32  harboured \acp{SMBH} of $3\,{-}\,7\,{\times}\,10^7\, \Ms$ and ${\sim}\,8\,{\times}\,10^6\,\Ms$, respectively \cite[a similar conclusion was reached by][]{1984ApJ...283L..27T,1987ApJ...322..632T}. Analogous analyses were performed for NGC 3115, NGC 4594, and NGC 3377, reporting \ac{SMBH} masses of $10^6\,{-}\,10^7\, \rm \Ms$ \cite[see][]{1988ApJ...335...40K,1992ApJ...393..559K,1995ARA&A..33..581K}. 

All the previous works were fundamental for the development of our current theory of \acp{SMBH} and galaxy evolution, providing important hints about the fact that most of the local and nearby massive galaxies are the host of \acp{SMBH} with masses exceeding the $10^6\,\Ms$.

\section{The coupled  evolution of galaxies and SMBHs}

All the exciting results about \acp{SMBH} achieved in the '70s and '80s were the foundations of ambitious works about the study of the demography of \acp{SMBH}  and the properties of their host galaxies. The very first project addressing this was \cite{1998AJ....115.2285M} which reported for the first time a comprehensive analysis of the census of \acp{SMBH}  at the galactic centers of nearby galaxies. Using high-quality \textit{Hubble Space Telescope} (HST) photometry and ground-based spectroscopy, the authors were able to determine the bulge mass of 36 galaxies and constrain the mass of their central \acp{SMBH}. The analysis of \cite{1998AJ....115.2285M} reported two important relations between \acp{SMBH}  and galaxies: the mass of the \acp{SMBH} appeared to strongly correlate with the  luminosity and mass of their galactic bulges. These results were a milestone in the study of \acp{SMBH}, hinting for the first time at a possible co-evolution between the galaxy and the \ac{SMBH}  population. Further investigations on this topic were done by \cite{2000ApJ...539L...9F} and \cite{2000ApJ...539L..13G} who analysed for the first time the relation between the \ac{SMBH}  mass and the stellar velocity dispersion of its host galaxy ($\sigma$). These studies showed that the \ac{SMBH}  mass displayed a power-law correlation with $\sigma$ (with  slope ${\sim}\,3.75\,{-}\,4.8$) tighter than the ones previously reported for the mass and luminosity of the bulge. Indeed, \cite{2000ApJ...539L..13G} quantified that this newly discovered relation could feature up to ${\sim}\,40\%$ less intrinsic scatter than the other relations found by \cite{1998AJ....115.2285M}. Besides bulge mass, luminosity, and velocity dispersion, other correlations with galaxy structural properties were reported. By analyzing a sample of galaxies with  well-constrained \ac{SMBH}  masses, \cite{2001ApJ...563L..11G} found a correspondence between the \ac{SMBH}  mass and the central bulge concentration index \citep[defined as in][]{2001MNRAS.326..869T}. Initially it was suggested that this new correlation was marginally stronger than the one with the velocity dispersion and displayed a smaller (or comparable) scatter. Nevertheless, these results have been under debate and contradictory results have been found. 

\begin{table}[]
\begin{adjustbox}{width=1\textwidth}
\begin{tabular}{|c||l|}
\hline
\multirow{3}{*}{\textbf{Bulge mass}}                & \multirow{3}{*}{\begin{tabular}[c]{@{}l@{}} \protect{\cite{1998AJ....115.2285M}};  \protect{\cite{2003ApJ...589L..21M}}; \\ \protect{\cite{2004ApJ...604L..89H}}; \protect{\cite{2013ApJ...764..184M}};\\ \protect{\cite{2013ARA&A..51..511K}}; \protect{\cite{2019ApJ...873...85D}} \end{tabular}} \\
                                                    &                                                                                                                                                     \\
                                                    &                                                                                                                                                     \\ \hline
\multirow{3}{*}{\textbf{Bulge luminosity}}          & \multirow{3}{*}{\begin{tabular}[c]{@{}l@{}} \protect{\cite{2000ApJ...539L...9F}}; \protect{\cite{1998AJ....115.2285M}};\\ \protect{\protect{\cite{2003ApJ...589L..21M}}}; \protect{\cite{2008MNRAS.386.2242H}} \protect{\cite{2007MNRAS.379..711G}};\end{tabular}}                    \\
                                                    &                                                                                                                                                     \\
                                                    &                                                                                                                                                     \\ \hline
\multirow{3}{*}{\textbf{Bulge velocity dispersion}} & \multirow{3}{*}{\begin{tabular}[c]{@{}l@{}} \protect{\cite{2000ApJ...539L..13G}}, \protect{\cite{2013ApJ...764..184M}};\\ \protect{\cite{2013ARA&A..51..511K}}; \protect{\cite{2013ARA&A..51..511K}}; \\ \protect{\cite{2018Natur.553..307M}}\end{tabular}}         \\
                                                    &                                                                                                                                                     \\
                                                    &                                                                                                                                                     \\ \hline
\multirow{3}{*}{\textbf{Bulge effective radius}}    & \multirow{3}{*}{\protect{\cite{2003ApJ...589L..21M}}}                                                                                                                \\
                                                    &                                                                                                                                                     \\
                                                    &                                                                                                                                                     \\ \hline
\multirow{3}{*}{\textbf{Bulge Sérsic index}}        & \multirow{3}{*}{\begin{tabular}[c]{@{}l@{}} \protect{\cite{2001ApJ...563L..11G}}; \protect{\cite{2007ApJ...655...77G}};\\ \protect{\cite{2016ApJ...821...88S}} \end{tabular}}              \\
                                                    &                                                                                                                                                     \\
                                                    &                                                                                                                                                     \\ \hline
\multirow{3}{*}{\textbf{Galaxy stellar mass}}       & \multirow{3}{*}{\begin{tabular}[c]{@{}l@{}} \protect{\cite{2015ApJ...813...82R}}; \protect{\cite{2016ApJ...830L..12T}};\\ \protect{\cite{2012AdAst2012E...4E}};\\ \protect{\cite{2017MNRAS.472.4013C}}\end{tabular}}   \\
                                                    &                                                                                                                                                     \\
                                                    &                                                                                                                                                     \\ \hline
\end{tabular}
\end{adjustbox}
\caption{List of the works reporting correlations between the \ac{SMBH} mass and galaxy structural properties.}
\label{tab:Relations_Bulge_BH}
\end{table}

\begin{figure*}
\centering  
\includegraphics[width=1\textwidth]{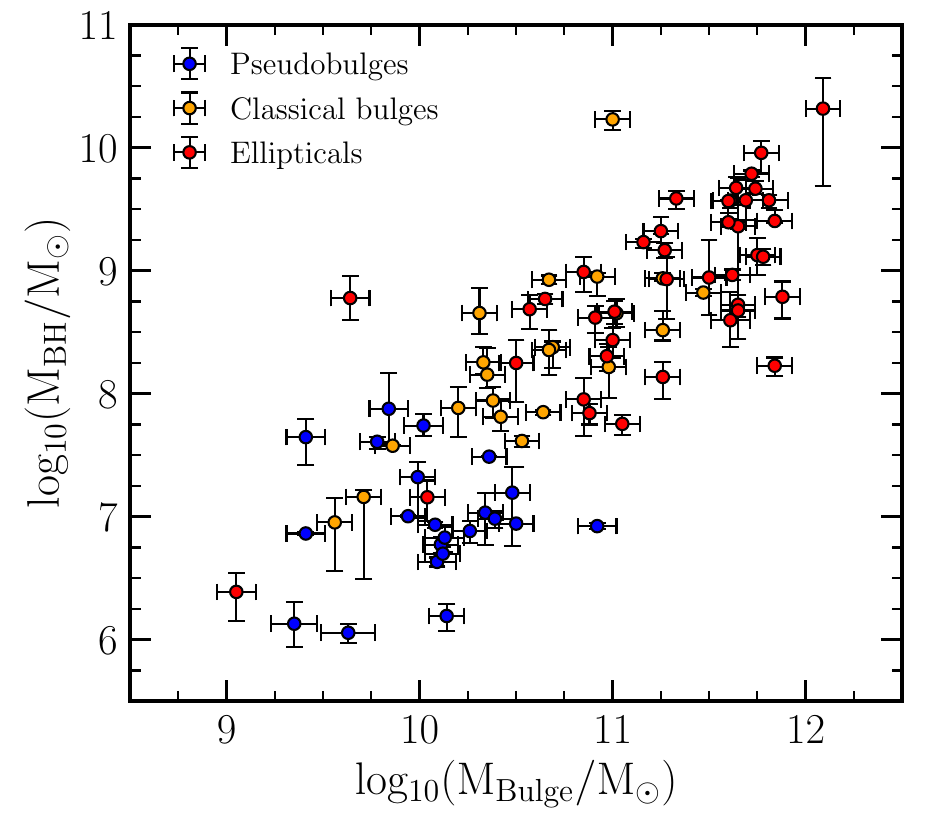}
\caption[]{\ac{SMBH} mass ($M_{\rm BH}$) as a function of the mass of its host galactic bulge  ($M_{\rm Bulge}$) as obtained from  local galaxies (adapted from \protect{\cite{2013ARA&A..51..511K}}). Data, taken from \protect{\cite{2013ARA&A..51..511K}}, represent galaxies featuring an elliptical (red), classical  (orange) and pseudo- (blue) bulge morphology.}
\label{fig:BHBulgeCorreltion_Kormendy}
\end{figure*}

During the last years, all the original correlations proposed by \cite{1998AJ....115.2285M} have been revisited and updated by several works. Today, these relations are commonly known as ``\textit{scaling relations}''. Among all the works that revisited them, we can highlight \cite{2004ApJ...604L..89H} and \cite{2013ApJ...764..184M} which showed that the galaxy-\ac{SMBH}  correlations might be slightly tighter than originally reported. For instance, \cite{2013ApJ...764..184M} showed that the best-fit for the relation between \ac{SMBH}  and bulge mass reads:

\begin{equation}
 \log_{10}\left(\frac{M_{BH}}{\Ms}\right)\,{=}\, 8.46 \,{+}\, 1.05\, \log_{10}\left(\frac{M_{Bulge}}{10^{11}\Ms}\right)
\end{equation}
where $M_{BH}$ and $M_{Bulge}$ correspond to the \ac{SMBH}  and bulge mass, respectively. In addition to these studies, \cite{2013ARA&A..51..511K} showed that such correlations are not universal but they depend on the galaxy morphology, with \acp{SMBH}  placed in massive bulges more correlated with the galaxy properties than the ones living in disc-dominated galaxies \cite[see also][]{2008MNRAS.386.2242H,2008ApJ...680..143G,2016ApJ...817...21S}.  Specifically, the authors presented their results by dividing the galaxy morphology in three different classes: elliptical galaxies, classical bulges and pseudobulges. Elliptical galaxies are objects that lack a stellar disc component and the whole morphology is characterized by a shperoidal structure composed by an old stellar population dynamically supported by their velocity dispersion (random motions). Classical bulges are elliptical-like spheroids surrounded by a stellar disc component. Finally, pseudobulges correspond to ellipsoidal structures at the centre of spiral galaxies whose stellar population is dominated by young stars and whose dynamics is supported by rotation rather than random motions.  The results of \cite{2013ARA&A..51..511K} are presented in Figure~\ref{fig:BHBulgeCorreltion_Kormendy}. Elliptical galaxies and classical bulge structures display tighter correlations with their central \ac{SMBH} than galaxies featuring a pseudobulge. The motivation behind these differences remains debated and further investigations are required \cite[see e.g.][]{2009MNRAS.399..621G,2013CQGra..30x4001S}.\\

All of the results shown by the previous works emphasize that \acp{SMBH} strongly correlate with the properties of their host galaxies, regardless of the specific amplitude of the correlation and its intrinsic scatter. In Table~\ref{tab:Relations_Bulge_BH} we report the main works that analyzed the  correlations between the \ac{SMBH} masses and galaxy structural properties.

To motivate all these relationships, it was proposed that \acp{SMBH}  evolve together with their host galaxies: any internal or external process underwent by a galaxy along its lifetime must have a repercussion on the growth of its nuclear  \ac{SMBH} (i.e active phase such as AGN/quasar). To corroborate this idea  and identify which specific mechanisms undergone by galaxies are the drivers of \acp{SMBH} growth, different observational and theoretical works were performed. In the following, we summarize their main findings.

\subsubsection{An observational point of view on the physics behind the scaling relations}

The search for an explanation about the co-evolution between galaxies and \acp{SMBH} and how these \acp{SMBH} reached such large masses has challenged astrophysicists during the last 50 years. 
Given that galaxies aggregate  hierarchically through the merger of smaller structures \citep{1998ApJ...499..542S,1999MNRAS.303..188K,2009ApJ...702.1005S,2010ApJ...719..229G}, galaxy interactions were the first process  being considered as a possible responsible for shaping the properties of galaxies, giving rise to bulge structures and driving the observed \ac{SMBH}-galaxy relations.  In fact, according to different studies, galaxy interaction can  lead to different outcomes depending on the baryonic mass ratio of the merging galaxies. On one hand, mergers between galaxies of similar baryonic content (called major mergers, indicatively for which the mass ratio between the involved mass is larger than $\sim 1:4$) lead to the destruction of galactic discs, trigger nuclear starbusts and concentrate the whole stellar population of the remnant galaxy in an elliptical structure. On the other hand, mergers between galaxies of different baryonic content (called minor mergers, indicatively for which the mass ratio between the involved mass is smaller than $\sim 1:4$) have a less destructive effect, leaving  the galactic disc of the most massive galaxy almost untouched and leading  the growth of its bulge component.\\ 

One of the pioneering works that shed light on the link between galaxy mergers and \ac{SMBH} activity was \cite{1988ApJ...325...74S}. By analyzing 10 bright infrared galaxies (ultra-luminous infrared galaxies, ULIRGs) the authors found that almost all the samples displayed AGN signatures. Given that ULIRGs are expected to represent advanced mergers of gas-rich disc galaxies \citep{1992ApJ...390L..53K,2002ApJS..143..315V}, these results were interpreted as the evidence for an evolutionary connection between \ac{SMBH} activity and galaxy mergers. \cite{1988AJ.....96.1575H} supported this scenario with the analysis of five more quasars placed in galaxies with strong tidal interaction features. During the following years, this merger-induced hypothesis gained a large relevance. 
By using optical observations of 20 nearby ($z\,{<}\,0.3$) luminous quasars, \cite{1997ApJ...479..642B}  showed that these quasars were placed in regions with a large galaxy density; this, coupled with the evidence of merger signatures in the sample, supported the idea that gravitational interactions play an important role in triggering  quasars. Similar conclusions were reached by \cite{2006ApJ...643..707V} who showed that the sizes and luminosities of quasar hosts and ULIRGs were statistically indistinguishable. The investigation of radio-loud AGNs was also fundamental for supporting the merger-induced scenario. For instance, \cite{2012MNRAS.419..687R} compared the population of quiescent and radio-loud AGN ellipticals. Even though both samples showed signatures of disturbed morphologies, likely the result of past or ongoing galaxy interactions, these features were brighter in the AGN population.\\

Despite being highly appealing  to explain the triggering of AGN activity and the observed correlations, the merger-induced scenario was questioned by several observational works. One of these studies  \cite{2012ApJ...758L..39T}  explored the existence of major merger signatures in a large sample of AGNs across luminosity and redshift. The results showed that, independently of redshift, the most luminous AGNs were the unique systems connected with major mergers. On the contrary, the less luminous AGNs lacked this signature and had to be triggered by different processes. The study of the galaxy morphology was also an alternative method to challenge the merger scenario. For instance, the study of the brightness profiles of high-$z$ heavily obscured quasar hosts suggested they had a disc-like morphology, contradictory with the expected elliptical structure, formed after a major merger. \cite{2012ApJ...744..148K} supported these results by finding that, at $z\,{\sim}\,2$, half of the AGNs with moderate X-ray luminosity ($10^{42\,{-}\,44}\,\rm erg/s$) were preferentially placed galaxies featuring a morphology dominated by the disc. At $z\,{<}\,1$, \cite{2011ApJ...726...57C} found larger fractions (${\sim}\,80\%$), concluding that the bulk of the \ac{SMBH} growth at low redshift occurred trough secular processes\footnote{{ In this chapter we define \textit{secular process} all the mechanisms that causes a slow rearrangement of galaxy mass, energy and angular momentum. Continuous star formation events or phenomena such as bar/spiral structure formation are the most representative secular processes in galaxy evolution. In this way, it is stated that the galaxy evolves secularly when  its evolution is ruled by any of these process.}}. These results were completed by \cite{2014MNRAS.439.3342V} who compared the morphology of the AGN host with a matched quiescent control sample. The analysis showed that both populations displayed comparable asymmetries, Sérsic indices, and ellipticities, ruling out that major mergers were unique reason for the MBH growth \cite[see also][]{2003ApJ...595..685G,2005ApJ...627L..97G,2007ApJ...660L..19P,2012ApJ...744..148K}.\\

Putting together the results of all the previous works, it was evident that mergers could explain the presence of AGN activity (at least the most luminous ones) but further processes were required to explain the presence of \ac{SMBH} activity in galaxies without  apparent  merger signatures. Several studies have explored the role of secular evolution in triggering the AGN phase, focusing for that on the galaxy morphology. \cite{2009MNRAS.397..623G} studied the morphology of X-ray AGN hosts at $z\,{=}\,1$ with HST photometry. The results showed that about 30\% of the sample was placed in disc galaxies, incompatible with the established scenario of major mergers destroying disc galaxies and triggering AGNs. The authors argued that internal instabilities undergone by the high-$z$ galaxies could lead to nuclear gas inflows, providing a favorable environment for the \ac{SMBH} growth. \cite{2011MNRAS.417.2721O} performed a similar study but focusing on a specific class of AGNs (Narrow-Line Seyfert 1), expected to be powered by Eddington-limited low mass MBHs. The results showed that the hosts of these objects displayed pseudobulge structures, expected to form through secular processes happening in galactic discs. \cite{2012ApJ...757...81B} provided further proofs about internal processes triggering AGN by comparing a low-$z$ galaxy sample of clumpy (unstable) discs with a similar one but with smoother (stable) discs. Based on line diagnostic methods and X-ray stacking, the results showed that the majority of clumpy discs had a high probability of containing an AGN. In addition to morphological analysis, some works have focused on the angular correlation function\footnote{{In cosmology, the angular correlation function measures how galaxies or AGNs cluster in the right ascension and declination plane. To this end, the method compares the distribution of galaxy or AGN pairs relative to that of a random distribution.}} of AGNs to further investigate the triggering mechanism of an AGN phase. Among these works, \cite{2011ApJ...736...99A} showed that the halo mass associated with the observed correlation function of $z\,{<}\,4$ X-ray-detected AGNs would be incompatible with these systems being triggered by major mergers. Instead, the authors argued that allowing secular evolution processes such as galactic tidal disruptions or disc instabilities would help in reducing the discrepancies between observations and theoretical models.\\

\subsubsection{A theoretical point of view on the physics behind the scaling relations} \label{sec:Theroety_BH_Galaxy_Correlation}

Parallel to observations, theoretical studies were carried out with the aim of understanding which processes underwent by the galaxy population trigger the growth of \acp{SMBH} and, ultimately, lead to the observed scaling relations. To this end, most of the works relied on analytical and numerical simulations, where different physical processes can be simulated and studied in detail. One of the first works in addressing the rise of the \ac{SMBH} activity was \cite{1989NYASA.571..190H}. The authors simulated the merger of spiral galaxies combining  N-body and hydrodynamics simulations. The simulations showed that such kind of encounters are capable of provoking gas inflows which accumulate $\rm {\sim}\,10^8\,{-}10^9\, \Ms$ near the nucleus of the primary galaxy and sustain the growth of the central \ac{SMBH} over ${\sim}\,10^8\, \rm yr$. The work of \cite{1992ApJ...393..484B} confirmed these results and highlighted that tail features and bar structures, born through the violent forces triggered by the merger, were a fundamental ingredient to cause a fraction of the gas disc to lose most of its angular momentum and fall into the center of the galaxy. Several years later, the sophisticated simulations of galaxy mergers presented in \cite{2005Natur.433..604D} were still supporting these scenarios. The results pointed out that besides triggering an intense burst of star formation, the merger gave rise to a luminous quasar, fuelled by a massive black hole accreting at $1\, \rm \Ms/yr$. The work of \cite{2005Natur.433..604D} also reported that the energy released by such accretion process was able to expel enough gas to stall the \ac{SMBH} growth and prevent further star formation events. These results were a milestone in the theoretical field of \acp{SMBH} since it was shown that, besides gas inflows, the energy released during \ac{SMBH} accretion events was fundamental to self-regulate the growth of both the galaxy and the \ac{SMBH} and thus, reproduce the observed correlation between \ac{SMBH} and galaxy velocity dispersion. {To have an idea about the typical energy released by an accreting \ac{SMBH} and how this one compares with the energy of a galactic bulge we can make some simple computations. By using Eq.~\ref{eq:luminosity}, the energy released by an accreating \ac{SMBH} in a given time, $dt$, is determine as:
\begin{equation}\label{eq:EnergyMBH}
\dot{E}_{\rm BH} \,{=}\,\epsilon \dot{M}_{\rm BH} {\rm d} t c^2/ (1-\epsilon)
\end{equation}
where $c$ is the speed of light and $ \dot{M}_{\rm BH}$ the accretion rate onto the \ac{SMBH}. On the other hand, the binding energy of a galactic bulge (i.e the minimum energy required to desegregate the structure) can be estimated as:
\begin{equation}\label{eq:EnergyBulge}
E_{\rm Bulge} \,{=}\, \frac{G M_{\rm Bulge}^2}{R_{\rm Bulge}}
\end{equation}
where $G$ is the gravitational constant and $M_{\rm Bulge}$, $R_{\rm Bulge}$ the mass and radius of the bulge, respectively. Taking into account these equations and assuming the reasonable values of $M_{Bulge}\,{=}\,10^{10} \Ms$, $R_{Bulge}\,{=}\,2\, \rm kpc$, $\dot{M}_{BH}\,{=}\,1\, \Ms/yr$, $dt\,{=}\,10^7 yr$ and $\epsilon\,{=}\,0.1$, we can draw the conclusion that:
\begin{equation}\label{eq:RatioEnergy}
\frac{E_{\rm BH}}{E_{B\rm ulge}} \,{\sim}\, 500.
\end{equation}
Thus, this simple calculation corroborates the theoretical results of \cite{2005Natur.433..604D}, stressing that the energy released by the MBH is large enough to affect the assembly of the galactic bulge, halt possible gas inflows towards the galactic nucleus and drive the observed correlations between the galaxy and the \ac{SMBH}.\\}

The possibility that the observed scaling relations and AGN activity could also arise in isolated galaxies with a self-regulated unstable disc was explored in \cite{2011ApJ...741L..33B} by running six idealized hydrodynamical simulations. As already seen in the merger case, the simulations showed that gravitational torques produced by disc instabilities are capable to drive  angular momentum out and drive  mass in, causing clump migration and strong gas inflow towards the galactic nucleus. The rate at which  the gas  reaches the galactic nucleus was shown to scale with the disk mass as $25\,(M_{\rm disc}/10^{11})(1\,{+}\,z)^{3/2}$, with values of $\rm 10 \, \Ms/yr$ for disc-dominated galaxes of disc mass $M_{\rm disc}\,{\sim}\,10^{11}\, \Ms$. Interestingly, the simulations of \cite{2011ApJ...741L..33B} showed that, in order to obey the observed bulge and \ac{SMBH} relation, the vast majority of the infalling gas should be used to trigger the bulge growth and only a small fraction of it (${\lesssim}\,10^{-3}$) would be able to reach the central \ac{SMBH}. All the previous analyses were extended by \cite{2008ApJ...686..815Y} who studied the scaling relations arising in simulations of both galaxy mergers and isolated galaxies with unstable discs. The authors reported that the inflows produced by both mergers and disc instabilities were enough to feed the nuclear \ac{SMBH} with gas. However, the thermal energy released by the \ac{SMBH} during the process of gas accretion was able to unbind the gas surrounding the \ac{SMBH}, terminating the inflow and stalling its growth. Caused by this self–regulation, the final mass of the \ac{SMBH} was not correlated with the galaxy gas fraction but rather with the binding energy of its host bulge. Besides this, \cite{2008ApJ...686..815Y} showed that the final scaling relation produced by unstable discs was systematically below the one generated by mergers, in agreement with the observational studies of \cite{2008ApJ...680..143G}.
Thus, hydrodynamics simulations showed that secular and merging processes could give rise to scaling relations characterized by different normalizations. Numerical simulations were not the unique tool used to explore the role of mergers and secular processes in regulating the \ac{SMBH} activity. For instance, \cite{2008ApJS..175..356H} used an analytical approach: By combining the well-known dark matter subhalo mass function at $z\,{<}\,6$ with dark matter halo occupation models, the authors showed that major mergers were enough to reproduce the observed quasar luminosity functions. Thus, gas inflows triggered by secular processes such as  bars or other disc instabilities would have a small contribution to the quasar activity at $z\,{>}\,1$.\\

\begin{figure*}
\hspace*{-2.5cm} \centering  
\includegraphics[angle=0,height=0.4\textheight]{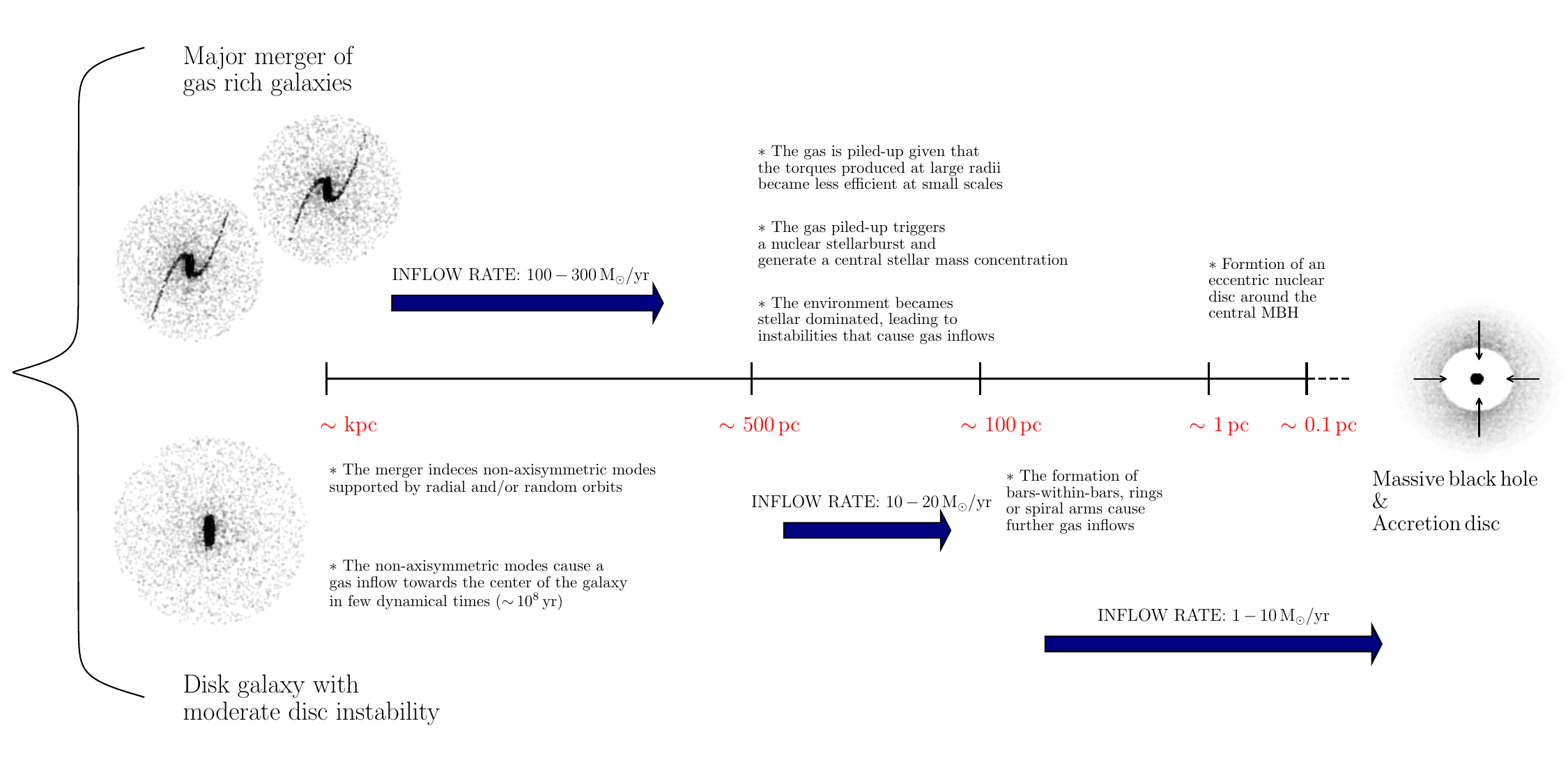}
\caption[]{Illustrative scheme of how the gas at kpc scales looses its angular momentum and reaches the vicinity of the nuclear massive black hole (${\sim}\,0.1\, \rm pc$) after a merger or a disc instability. The scheme is based on the work of \protect{\cite{2010MNRAS.407.1529H}}.}
\label{fig:Inflow_BH}
\end{figure*}

All these pioneering works pointed out that mergers and disc instabilities are efficient in funnelling gas towards the nuclear regions and trigger AGN activity. However, it remained unclear how gas looses angular momentum (at $\rm {\sim}\,kpc$ scales) and flows down to the \ac{SMBH} surroundings (${\sim}\,0.1\,\rm \pc$). To shed light on that, \cite{2010MNRAS.407.1529H} performed a suite of multi-scale hydrodynamic simulations that could follow the gas inflow from galactic scale down to sub-parsec scales. The authors focused on gas rich galaxy-galaxy mergers and isolated simulations of bar-unstable disc-dominated galaxies. We summarize  the results reported by  \cite{2010MNRAS.407.1529H} in Figure~\ref{fig:Inflow_BH}. In brief,  global gravitational instabilities produced by mergers and bars were enough to cause strong torques that efficiently leads part of the galaxy gas to flow towards the galaxy center. This migration was shown to happen on time-scales similar to a few galaxy dynamical times (${\sim}\,10^8\, \rm yr$). During the infall, the gas accumulates at ${\sim}\,0.1\,{-}\,1\, \rm k\pc$ since torques from large scales disturbances are less efficient in removing gas angular momentum at small scales \cite[see e.g.][]{2009MNRAS.398..303H}. To overpass this stalling and reach the vicinity of the nuclear \ac{SMBH} ($\rm {\sim}\,pc$ scales), the gas develops unstable and non-axisymmetric modes such as bars-within-bars, spirals arms or rings. The combination of all these multi-scale processes  are able to sustain accretion rates of $\rm {\sim}\,10 \, \Ms/yr$  onto \acp{SMBH}, compatible with the expectations for the most luminous quasars.  \\

All the theoretical works mentioned above focused on specific cases, where idealized galaxy mergers and galactic discs were simulated. In order to obtain a more complete theoretical picture of how galaxy mergers and disc instabilities ultimately lead to the observed scaling relations, simulations able to capture the cosmological context were required. To this end, semi-analytical models and cosmological hydrodynamical simulations have been the preferred tools. On the  one hand, a cosmological hydrodynamical simulation is in essence the same as the simulations described before used to study idealized galaxy mergers or isolated disc galaxies: both of them evolve galaxies and dark matter halos by solving self-consistently the gravity and fluid equations. However, cosmological simulations resolve these equations over cosmological times and large volumes (several Mpc) allowing for the emergence of the so-called \textit{large-scale structure}. Here the population of dark matter halos and galaxies interact over  wide scales and evolve through different physical processes from the cosmic dawn until the present day \citep{2020NatRP...2...42V}. Simulating cosmological volumes is computationally expensive and the development of cosmological hydrodynamical simulations requires a compromise between the simulated volume and the smallest (both in mass and length) structure that the simulation can self-consistently track. This limit is usually known as the \textit{simulation resolution}. All the physical processes happening below this limit are tracked by simple analytical recipes that capture the general behaviour of the physical processes occurring at scales that cannot be resolved. On the other hand, a semi-analytical model (SAM) is a tool that simulates the evolution of the galaxy population as a whole in a self-consistent and physically motivated manner. Unlike hydrodynamical comsological simulations, it does not solve the gravity and fluid equations since it does not include gas and stellar particles. Instead, galaxy properties such as star formation rate, stellar mass or magnitudes are a result of simple analytical recipes that simultaneously model multiple physical processes, which typically include gas cooling, star formation, AGN and supernova feedback, metal enrichment, massive black hole growth, and galaxy mergers \citep{2006MNRAS.370..645B}. All these processes are implemented through a system of coupled differential equations solved along the mass assembly history of dark matter halos, given by their respective merger tree, extracted either from N-body simulations or with  Monte Carlo simulations based on the Press-Schechter formalism \citep{1974ApJ...187..425P}.\\

One of the first works addressing the evolution of \acp{SMBH} in a full cosmological framework was \cite{1999MNRAS.308...77C}.  By using a simplistic semi-analytical model built on the  top of dark matter merger trees generated by Monte Carlo realizations, the authors explored how efficient have to be galaxy mergers to sustain accretion events onto the \acp{SMBH} and generate the known scaling relations. \cite{1999MNRAS.308...77C} showed that a linear correlation between the bulge and \ac{SMBH} mass required a linear link between the gas accretion onto the \ac{SMBH} and the amount of stars newly formed after the major merger. Non-linear relations between these two quantities led to a non-linear co-evolution between \ac{SMBH} and bulge masses.   
A few years later, \cite{2000MNRAS.311..576K} showed that joining the galaxy formation framework provided by semi-analytical models with the idea of \ac{SMBH} activity triggered by major merger events it was  possible to reproduce the observed B-band luminosity function of $z\,{<}\,2$ quasars (see also \citealt{2000ApJ...543..599C}). Besides this, the authors were able to prove that the observed decline in the number density of $z\,{<}\,2$ quasars is likely caused by the decrease in the galaxy merging rate, as well as the drop in the amount of cold gas available to fuel \ac{SMBH} activity, and the increase in the time-scale for gas accretion. Similar conclusions were reached by \cite{2003ApJ...587L..63M} when including in a different semi-analytical model the major-merger scenario as the main driver for triggering the growth of \acp{SMBH}. The results presented by \cite{2003ApJ...587L..63M} also pointed out that the merger scenario should be enough to reproduce the observed scaling relation and the redshift evolution of the number density of bright quasars. These results were reinforced by other works using semi-analytical models (see e.g \citealt{2003PASJ...55..133E} and \citealt{2003ApJ...593..661V}) that stressed the importance of merger events in triggering the accretion of gas onto \acp{SMBH} and causing the observed correlation between galaxies and \acp{SMBH}. Thanks to the improvements in cosmological dark-matter only simulations and the refinement of the galaxy formation paradigm, an increasing number of works explored the co-evolution between galaxies and \acp{SMBH} by using semi-analytical models specifically designed to be run on top of the merger trees extracted from the state-of-the-art of dark matter N-body simulations. One of the pioneering works on this new methodology was \cite{2006MNRAS.369.1808C} which used a semi-analytical model based on the merger trees of the Millennium dark matter simulation \citep{2005MNRAS.364.1105S} to explore the assembly of \acp{SMBH} and bulges. The work of \cite{2006MNRAS.369.1808C} concluded that if major mergers are the primary mechanisms triggering the growth of both bulges and \acp{SMBH}, the observed local correlation between these structures should display a redshift evolution primarily driven by the assembly of the bulge component. 
\cite{2005MNRAS.364..407C} explored the cosmological evolution of bulges and \acp{SMBH} using a similar semi-analytical approach. Unlike \cite{2006MNRAS.369.1808C}, the authors included both major mergers and disc instabilities as mechanisms that led to the bulge assembly but only galaxy major mergers as triggers for the \ac{SMBH} growth. The results showed the same trends reported by \cite{2006MNRAS.369.1808C} where the bulge assembly is the driver of the redshift evolution of the bulge-\ac{SMBH} relation. However, the inclusion of disc instabilities as an extra channel for bulge growth caused a bi-modality in the scaling relations at high-$z$, vanishing towards $z\,{=}\,0$. These novel results were fundamental to show that measurements of the bulge and \ac{SMBH} masses at high-$z$ could be used as a powerful tools to explore the mechanisms through which bulge growth may occur. To test if the major-merger scenario was successful in reproducing the cosmological evolution of the quasar bolometric luminosity function and the clustering of quasars\footnote{The clustering of AGNs and quasars provides information about the spatial distribution of active \acp{SMBH}. This can be used to investigate the typical environments where AGNs/quasars are placed. The knowledge about these environments can provide new insights onto the mechanisms triggering the accretion of \acp{SMBH} and thus, the processes leading to the observed relations.}, \cite{2008MNRAS.385.1846M} and \cite{2009MNRAS.396..423B} extended and improved the semi-analytical model presented in \cite{2006MNRAS.369.1808C}.  In their implementation, \acp{SMBH} did not accrete gas instantaneously after a galaxy merger. Instead, the accretion was coupled with a light curve model consisting of two different phases. One in which the \ac{SMBH} growth is Eddington limited, while the other is a quiescent regime in which the \ac{SMBH} grows at low Eddington rates. With these modifications, the new version of the semi-analytical model was still able to reproduce the local correlations between bulges and \acp{SMBH}. However, at low bulge masses (${<}\,10^9\, \Ms$) the correlations diverged from the liner fit, being more compatible with a power-law behaviour. Regarding the activity of \acp{SMBH}, the model showed a good performance in reproducing the evolution of the faint end of the quasar bolometric luminosity function and the clustering of $z\,{<}\,3$ quasars. However, the merger-induced scenario included in the semi-analytical model struggled in reproducing the number of bright quasars at $z\,{\geq}\,3$, suggesting that this population required an extra channel of grow in addition to mergers.\\

Motivated by the observational studies suggesting that secular evolution might play a role in triggering the active phase of \acp{SMBH}, many different semi-analytical models explored  the role of such kind of processes in the co-evolution between galaxies and massive black holes. \cite{2012MNRAS.426..237H} used a similar methodology than \cite{2008MNRAS.385.1846M} but including  the possibility that disc instabilities could trigger AGN/quasar phase on top of the merger-induced scenario. The results showed that the addition of this extra mechanism was enough to reproduce the evolution of the bolometric luminosity function up to $z\,{=}\,5$ down to a luminosity of $10^{44}$ erg/s. Thus, disc instabilities appeared to be another important channel to reproduce the cosmological evolution of the quasar/AGN number density. Despite these important results, \cite{2012MNRAS.426..237H} included \textit{ad hoc} the assumption that disc instabilities only trigger very sub-Eddington \ac{SMBH} accretion rates, failing to prove wether disc instabilities can be more important than mergers for the activity of \acp{SMBH}. \cite{2012MNRAS.419.2797F} shed more light on  topic by including  a comprehensive model of \ac{SMBH} growth in a semi-analytical model, in which gas accretion takes place after both mergers and disc instabilities, on a time scale given by the dynamical time of the hosting bulge. The results showed that disc instabilities taking place in self-gravitating discs play a very important role in the mass evolution of \acp{SMBH}. Specifically, the model predicted that \acp{SMBH} accrete typically   100 times more gas through disc instabilities than via major and minor mergers, regardless of the explored redshift. The work of \cite{2014A&A...569A..37M} also provided important results on the role of disc instabilities in the growth of \acp{SMBH}. Unlike other models, \cite{2014A&A...569A..37M} included  a sophisticated prescription for the gas mass inflow after disc perturbations based on the physical description of the \cite{2011MNRAS.415.1027H} numerical simulations. The work showed that while at $z\,{<}\,4.5$ the bright end of the AGN luminosity in the B-band can only be explained thanks to disc instabilities, at $z\,{>}\,4.5$ the abundance of bright quasars is underestimated if disc instabilities are the only mechanisms for \ac{SMBH} feeding. On the top of the luminosity functions, \cite{2014A&A...569A..37M} studied the behaviour of the \ac{SMBH}-galaxy mass relation when running two versions of the semi-analytical model. In the first (second) run, galaxy interactions (disc instabilities) were the unique drivers of the \ac{SMBH} evolution. In both cases, the results showed to be consistent with the observed local scaling relations, but they displayed important differences in the slope and scatter. The merger-induced scenario produced a less scattered and steeper correlation than the disc instability one. Besides this, the model showed that the correlation between the AGN luminosity and the star formation rate could be another important property to distinguish which mechanisms taking place in galaxies drive the evolution of \acp{SMBH}. In particular, \cite{2014A&A...569A..37M} showed that a scenario where the disc instabilities take the major role would display a tighter correlation between AGN luminosity and the star formation rate  than a merger-induced scenario. \cite{2016MNRAS.456.1073G} performed a follow up of \cite{2014A&A...569A..37M} work but studying whether mergers and disc instabilities leave any distinctive imprint in the clustering strength of AGNs. The results showed that the merger-induced model better reproduced the observed abundance of active, central galaxies than the disc instability one. Further differences were found in the number of satellite AGN inside groups and clusters, being systematically larger in the disc instability scenario. Indeed, \cite{2016MNRAS.456.1073G} suggested that the low number of satellite AGNs predicted by the merger-induced scenario might indicate that different feeding modes could simultaneously contribute to the triggering of satellite AGN. A similar conclusion was reached by the semi-analytical model presented in \cite{2018MNRAS.474.1995R}. The results showed that the relation between the \ac{SMBH} mass and bulge velocity dispersion would be underestimated at $M_{\rm BH}\,{\lesssim}\,10^7 \, \Ms$ if no extra growth models are accounted on top of the merger-induced scenario. Based on all these results, the state-of-the-art semi-analytical models include both mergers and disc instabilities as fundamental mechanisms to build-up the population of massive black hole population across cosmic time and reproduce the observed correlation between \acp{SMBH} and galaxies (see for instance, \citealt{2019MNRAS.482.4846S,2019MNRAS.487..198G,2020MNRAS.495.4681I,2020MNRAS.494.2747M}).\\

Parallel to the results provided by semi-analytical models, cosmological hydrodynamical simulations further helped in shedding light on the redshift evolution of \acp{SMBH} and on the observed scaling relations. 
One of the first works that addressed the assembly of \acp{SMBH} in a cosmological hydrodynamical simulation was \cite{2015MNRAS.452..575S}; by using the   \texttt{Illustris} simulation \citep{2015A&C....13...12N}, the authors showed that \acp{SMBH} of $M_{BH}\,{>}\,10^7 \, \Ms$ tightly correlate with the galaxy properties, in agreement with observations. However, at low masses, the correlation displayed a large scatter consequence of the supernovae feedback affecting the galaxy assembly. On top of this, the authors reported a redshift evolution in the normalization of the $M_{\rm BH}\,{-}\,M_{\rm stellar}$. The faster assembly of \acp{SMBH} compared to the one of galaxies caused that at fixed bulge mass, \acp{SMBH} were more massive at high-$z$. Thanks to the good resolution of the \texttt{Illustris} simulation, enough to well resolve thousands of galaxies, \cite{2015MNRAS.452..575S} proved that the position of \acp{SMBH} in the scaling relation was an important piece of information to determine the galaxy properties. In particular, \acp{SMBH} residing above the median $M_{\rm BH}\,{-}\,M_{\rm stellar}$ (or $ M_{\rm BH}\,{-}\,\sigma$) relation were hosted in quiescent galaxies, with an old stellar population and low gas content. On the contrary, \acp{SMBH} below the median relation were placed in star-forming galaxies. These results were important given that cosmological simulations corroborated previous theoretical works pointing out that the assembly of \acp{SMBH} leaves an imprint on the properties of their hosting galaxies (the so-called \textit{AGN-feedback}). Similar results were also found by \cite{2015MNRAS.454..913D} when analyzing the \texttt{MassiveBlack-II} cosmological simulation \citep{2015MNRAS.450.1349K}. Specifically, the authors showed that \acp{SMBH} above the $M_{\rm BH}\,{-}\,M_{\rm stellar}$ relation tend to grow faster than their galaxy hosts but once their mass exceeds a certain fraction of the galaxy stellar mass, \acp{SMBH} slow down their growth and the secular assembly of the galaxy brings it  to the median of the scaling relation. The work of  \cite{2015MNRAS.454..913D} also corroborated the results about the redshift evolution of the  $M_{\rm BH}\,{-}\,M_{\rm stellar}$ and $ M_{\rm BH}\,{-}\,\sigma$ relations. To this end, the authors quantified the steepens of the relations. While at $z\,{<}\,1$ the slope remained almost constant, at higher redshift it can undergo an increase of 50\%.  For instance, at $z\,{\sim}\,0$ the $M_{\rm BH}\,{-}\,M_{\rm stellar}$  relation showed a slope of ${\sim}\,1.2$ while at $z\,{\sim}\,4$ it raised up to $1.8$. Interestingly, \cite{2015MNRAS.454..913D} proved that the evolution of the slope was also influenced by selection effects, being shallower for randomly selected samples than for the ones  chosen through specific cuts in stellar/\ac{SMBH} mass. These effects could in principle bias our understanding of the evolution in the scaling relations. The \texttt{SIMBA} cosmological simulation \citep{2019MNRAS.486.2827D} shared the same behavior as \texttt{Illustris} and \texttt{MassiveBlack-II}: the $ M_{\rm BH}\,{-}\,\sigma$  relation evolved in such a way that galaxies with a given \ac{SMBH} mass have higher $\sigma$ values at higher redshifts \citep{2019MNRAS.487.5764T}. However, \texttt{SIMBA} showed a no redshift dependence in the $M_{\rm BH}\,{-}\,M_{\rm stellar}$. Finally, \cite{2021MNRAS.503.1940H} explored how the $M_{\rm BH}\,{-}\,M_{\rm stellar}$  behaves in the \texttt{TNG} simulations \citep{2019ComAC...6....2N}. Interestingly, the results showed a no linear correlation at low stellar masses ($\rm {<}\,10^{10}\, \Ms$) and high redshift,  consequence of a too-efficient supernova feedback regulating the star formation rate of the galaxy. On the other hand, at low redshift the correlation changes to a linear behavior result of the supernovae feedback being less efficient.\\

Besides providing a valuable piece of information about how the scaling relation behaves at different epochs of the Universe, cosmological hydrodynamical simulations have been shown to be important tools to understand the assembly of the \ac{SMBH} population. For instance, using the \texttt{EAGLE} simulation, the authors of reference \cite{2016MNRAS.462..190R} corroborated the existence of a switch in the nature of the \ac{SMBH} accretion with cosmic time, also known as \textit{downsizing}. While at $z\,{>}\,2$ active massive black holes are the main population of \acp{SMBH}, at lower redshift the number of inactive \acp{SMBH} increases until becoming the main type. Indeed, at $z\,{=}\,0$ only a tiny fraction of \acp{SMBH} (${\leq}\,1\%$) are undergoing strong accretion processes. \cite{2016MNRAS.462..190R} reported that such behavior was the main responsible of shaping the evolution of the \ac{SMBH} mass function, i.e the number density of \acp{SMBH} per logarithmic bin of mass. Specifically, the amplitude of the \ac{SMBH} mass function rapidly increased from $z\,{>}\,5$ down to $z\,{\sim}\,2$. On the contrary, at $z\,{<}\,2$ the vast majority of $\rm M_{BH}\,{>}\,10^7\, \Ms$ were already formed and the massive end of the \ac{SMBH} mass function barely changed. Similar results were also found by other simulations such as \texttt{Illustris} and \texttt{SIMBA} \citep{2015MNRAS.452..575S,2019MNRAS.487.5764T}. Despite the consensus about the \ac{SMBH} downsizing process, the fuelling mechanism of \acp{SMBH} in cosmological simulations are still under debate. For instance, \cite{2018MNRAS.481..341S} studied  how galaxy mergers impact the growth of \acp{SMBH} in the \texttt{MAGNETICUM} simulation. The results showed that major and minor merger events can increase by a factor of two the probability of triggering the accretion process onto \acp{SMBH}. However, at $z\,{>}\,2$ almost all the galaxies contained an \ac{SMBH} with a certain level of gas accretion, irrespective of the merger activity. Indeed, such activity was triggered thanks to the large content of dense and cold gas present in high redshift galaxies. On this line we can find the results of \cite{2020MNRAS.494.5713M} which reported that mergers provoke important rapid gas accretion  onto \acp{SMBH} of ${\sim}\,10^7 \Ms$ hosted in galaxies of ${\sim}\,10^{10} \Ms$ at $z\,{<}\,1$. Despite this, the authors showed that the majority of the mass accreted onto the \acp{SMBH} occurs outside the periods of mergers, and is  related to galaxy secular evolution. The association between \ac{SMBH} activity and the presence of nearby galaxies was explored by \cite{2020ApJ...904..150B} in the \texttt{TNG} simulations. The authors showed that $z\,{<}\,1$ \acp{SMBH} hosted in galaxies with companion satellites at distance $\rm {\leq}\,0.1\,Mpc$ display an enhancement of the accretion rates compared with random \acp{SMBH}. However, the number of objects with these characteristics was subdominant (${<}\,40\%$). In this way, the results suggested that mergers play a role in triggering AGNs, but they have a relatively minor role in fuelling the \ac{SMBH} population as a whole.\\

\subsubsection{Processes impacting the co-evolution between galaxies and SMBHs: gravitational recoils}

As discussed in previous sections, the scaling relations observed between galaxies and \acp{SMBH} rise thanks to the nuclear gas inflows provoked by galactic mergers/disc instabilities and the subsequent energy released by the \ac{SMBH} during the gas accretion process. However, some processes are able to blur these correlations. One of them we would like to focus on here is the gravitational recoil. Galaxy-galaxy mergers lead to the formation of \ac{SMBH} binaries. Owing to a series of processes better discussed in the next Sections, these \acp{SMBH} are able to reach sub-parsec scales and reach the final coalescence via \ac{GW} emission. During the \ac{SMBH} merger, the linear momentum of the system has to be conserved. As a result,  the remnant \ac{SMBH} formed after the coalescence may experience a kick or recoil. The specific magnitude of the speed of these kicks can be established via numerical simulations of general relativity. For instance, \cite{2012PhRvD..85h4015L} showed that the recoil velocity undergone by a remnant \ac{SMBH} formed after the coalescence of two \acp{SMBH} of spin $a_1$ and $a_2$ and mass $M_{\rm BH,1}$ and $M_{\rm BH,1}$ can be parametrized as:
\begin{equation} \label{eq:recoil_velocity}
\begin{split}
& \vec{V}_{\rm recoil} \,{=}\, v_m\boldsymbol{\hat{e}}_1 + v_{\perp}\left( \cos \zeta \boldsymbol{\hat{e}}_1 + \sin \zeta \boldsymbol{\hat{e}}_2 \right) + v_{\parallel} \boldsymbol{\hat{n}}_{\parallel} \, , \\
& v_{m} \,{=}\, A_m \nu^2 \frac{(1-q)}{(1+q)} \left[ 1 + B_m \nu \right] \, , \\
&  v_{\perp} \,{=}\, H \frac{\nu ^2 }{(1+q)} \left( a^{\parallel}_2 - q a^{\parallel}_ 1 \right) \, , \\
& v_{\parallel} \,{=}\, \frac{16 \nu^2}{(1+q)} \left[ V_{1,1} + V_{A} \tilde{S}_{\parallel} + V_{B}\tilde{S}^2_{\parallel} + V_{C} \tilde{S}^3_{\parallel}\right] |\vec{a}^{\perp}_2  - q\vec{a}^{\perp}_1| \cos(\phi_{\Delta} - \phi_1) \, , 
\end{split}
\end{equation}
where, $v_m$ is the velocity due to the \textit{mass-asymmetry contribution} and, $v_{\parallel}$, $v_{\perp}$ are \textit{spin contribution} producing kicks parallel and perpendicular to the orbital angular momentum respectively. $q \,{=}\, \rm M_{BH,2}/M_{BH,1} \,{\leq}\, 1$ and $\nu \,{=}\, q/(1\,{+}\,q)^2$ are the binary mass ratio and the symmetric mass ratio, respectively. $|a_i|$ is  the initial spin magnitude of the massive black hole $i$, $\parallel$ and $\perp$ refer respectively to components parallel and perpendicular to the orbital angular momentum. $\boldsymbol{\hat{e}}_1$  and $\boldsymbol{\hat{e}}_2$ are orthogonal unit vectors in the orbital plane. $\zeta = 145^{o}$ is the angle between the unequal mass and spin contribution \citep{2007PhRvL..98i1101G,2008PhRvD..77d4028L}, $\vec{\tilde{S}} \,{=}\, 2 (\vec{a}_2 \,{+}\, q^2\vec{a}_1)/(1\,{+}\,q)^2$. $\phi_{\Delta}$ is the angle between the in-plane component $\vec{\Delta}^{\perp} \,{=}\, (M_{\rm BH,1} \,{+}\, M_{\rm BH,2})(\vec{S}_2^{\perp}/M_{\rm BH,2} \,{-}\, \vec{S}^{\perp}_ 1/M_{\rm BH,1})$ and the infall direction at merger. The values of the other coefficients are obtained numerically: $A_m \,{=}\, 1.2 \times 10^4$, $B_m \,{=}\, -0.93$, $H \,{=}\, 6.9\times 10^3$ \citep{2007PhRvL..98i1101G,2008PhRvD..77d4028L} and $V_{1,1}\,{=}\, 3677.76\, \rm km/s$, $V_A \,{=}\, 2481.21 \,\rm  km/s$, $V_B \,{=}\, 1792.45\, \rm km/s$ and $V_C \,{=}\, 1506.52 \, \rm km/s$.\\

By using different combinations of masses and spins in Eq.~\ref{eq:recoil_velocity}, it is easy to get an idea of the typical order of magnitude for the recoil velocity. Specifically, the values can vary from few to thousands of km/s (see e.g \cite{2008MNRAS.390.1311B} or \cite{2010MNRAS.402..682D}). Another noticeable thing is that the magnitude of the kick is independent on the two \ac{SMBH} masses, while it grows for larger $q$ and spin magnitudes. For the specific cases in which the kick is larger than the escape velocity of its host galaxy, the newborn \ac{SMBH} is ejected from the galaxy. If these kind of ejections are common in the cosmological assembly of galaxies and \acp{SMBH}, it is likely that they will cause important effects in the \ac{SMBH}-galaxy scaling relations. To explore these consequences, \cite{2011MNRAS.412.2154B} run a suite of hydrodynamical simulations of gaseous galaxy mergers where merged \acp{SMBH} receive a recoil. The analysis of these simulations showed that the depletion of \acp{SMBH} at the centre of galaxies after gravitational recoils leave an important imprint in the local scaling relations. Specifically, the normalization of the \ac{SMBH} and bulge velocity dispersion relation decreased and its intrinsic scatter increased by  ${\sim}\,0.13\,{-}\,0.3 \, \rm dex$. \cite{2015MNRAS.446...38G} explored as well the effect of gravitational recoils in the \ac{SMBH}-galaxy co-evolution. By using a different analytical approach to treat mergers the authors found that gravitational recoils can reduce down to ${\sim}\,85\%$  the occupation fraction of \acp{SMBH}\footnote{The \ac{SMBH} occupation fraction is a quantity used to express the fraction of galaxies  that host a nuclear \ac{SMBH} over the total number of galaxies.} placed in the brightest cluster galaxies (i.e the most massive galaxies in the Universe). In the line with previous findings, \cite{2015MNRAS.446...38G} showed that the ejection of \acp{SMBH} enhances the scatter of the \ac{SMBH}-bulge mass correlation of brightest cluster galaxies. Finally, the work of \cite{2020MNRAS.495.4681I} used a semi-analytical model to explore the effect of gravitational recoils on the whole galaxy population. The results showed similar effects in the \ac{SMBH} occupation fraction than the ones reported by \cite{2015MNRAS.446...38G}: gravitational recoils cause a progressive depletion of nuclear \acp{SMBH}  with decreasing redshift and stellar mass. For instance, the authors reported that the ejection of \acp{SMBH}  can reduce up to ${\sim}\,40\%$ the occupation fraction of \acp{SMBH} placed in dwarf galaxies (${<}\,10^{9}\,\Ms$), that necessarily feature a smaller escape velocity. Besides, \cite{2020MNRAS.495.4681I} showed that these \ac{SMBH}  ejections  lead to changes in the predicted local \ac{SMBH}-bulge relation. These modifications were more evident in pseudobulge structures, for which the relation was flattened at bulge masses ${>}\,2\,{\times}\,10^{10} \, \Ms$ and the scatter increased up to three orders of magnitude. The authors proved that these changes were caused by the fact that after the \ac{SMBH}  ejection, the galaxy can undergo multiple galaxy mergers. These interactions deposit in the empty galaxy (in terms of \acp{SMBH}) new \acp{SMBH}  whose mass is not bound to be correlated at all with its new host galaxy.\\

In this Section, we have extensively shown that all the observational and theoretical works carried out until now corroborate the idea that \acp{SMBH} and galaxies have a tight relationship: the cosmological assembly of  galaxy influences the one of \acp{SMBH}  and vice-versa. Thanks to theoretical works, the reason for this co-evolution seems to be more clear: nuclear gas inflows provoked by violent (galaxy mergers) or secular (disc instabilities) processes and the energy released by accreting \acp{SMBH}  self-regulate concurrently the growth of galaxies and \acp{SMBH}. Future works will unveil even further information about this co-evolution and will shed light on the intrinsic scatter of the observed \ac{SMBH}-galaxy scaling relations. Part of the unknown aspects about the aforementioned co-evolution and scaling relations are necessarily linked with the fact that it remains unclear how \acp{SMBH} formed in the early Universe, what was their typical mass, number density and host environment. For this, in the next Section we  describe the processes that may have led in the first place to the formation of \acp{SMBH}.


\part{The formation scenarios of SMBHs \\ \Large{Lupi and Regan}}

\section{Introduction}

While observations point to an occupation fraction very close to 100\% for \acp{SMBH} in massive galaxies, the origin of \acp{SMBH} remains 
unknown \cite{Kormendy2013, Faber1997, Magorrian1998}. Over the past four decades many theoretical proposals have been put forward to address the origin 
of massive black holes (MBHs) in galaxy centres \cite{Rees1978, Rees1984}. Note that here and in the next sections we use the term MBHs rather than SMBHs; we believe this term is more appropriate as it encompasses the populations of both SMBHs and IMBHs, which may have shared the same formation process;  IMBHs could constitute the (partially  ungrown) \textit{seeds}\footnote{The term `seed' is used generically to refer to the original progenitor black hole.} that eventually became SMBHs in the present-day Universe.

\begin{figure}[!t]
\centering
\includegraphics[width=16.0cm, height=10cm]{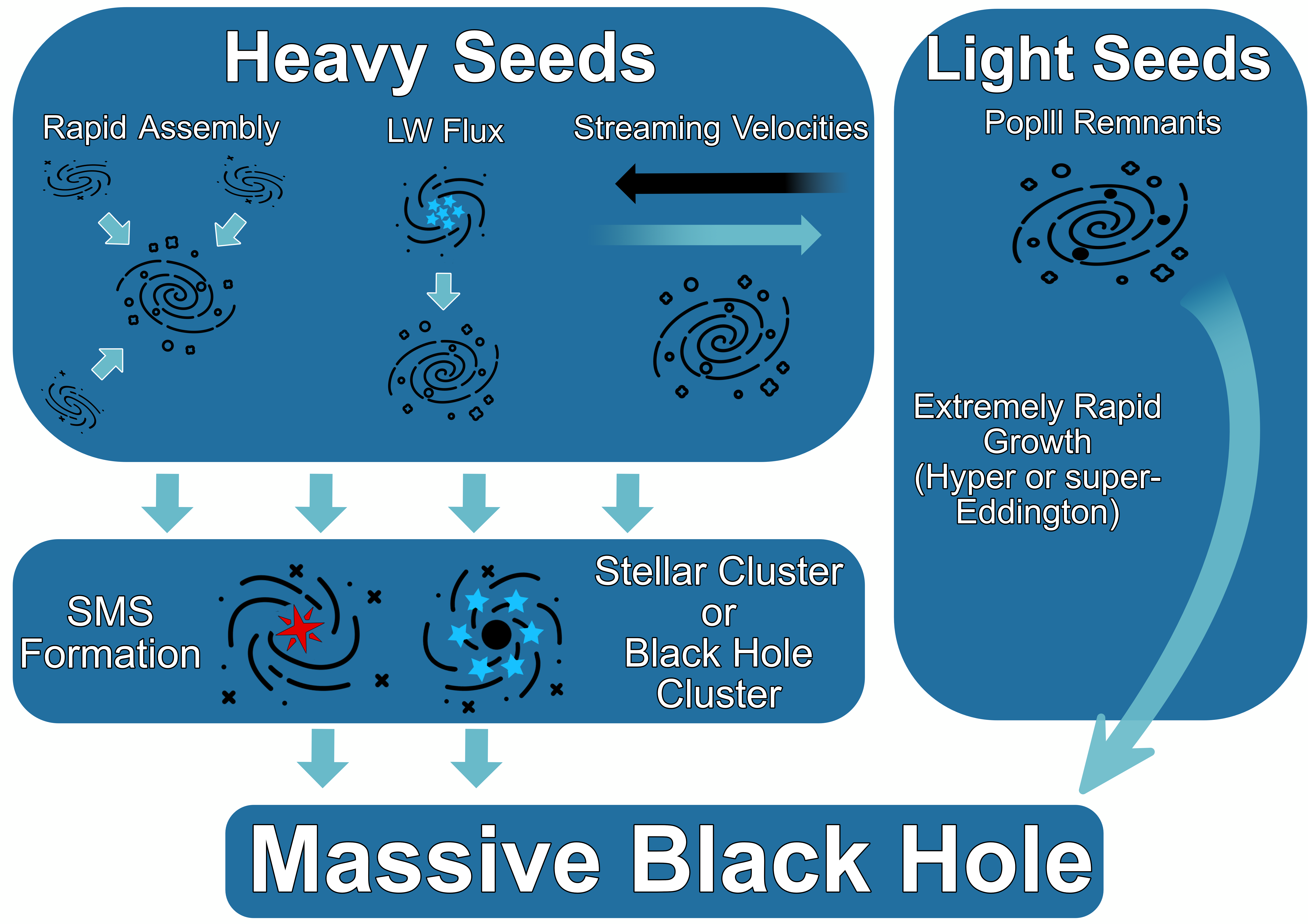}
\caption{The formation pathways leading to both the creation of heavy and light seeds. Heavy seed pathways can lead to either the realisation of a monolithic collapse and a SMS or in the event of significant fragmentation the realisation of a dense stellar cluster or subsequently a black hole cluster. Light seeds are born from the remnants of the first generation of stars and are expected to have typical masses in the range M$_{BH} \sim 10^1 - 10^3 \ \Ms$. Heavy seeds are
expected to cover the range of M$_{BH} \sim 10^3 - 10^6 \ \Ms$.}
\label{Fig:Seeds}
\end{figure}

In Figure \ref{Fig:Seeds} we illustrate the most up-to-date theoretical proposals for the formation pathways of MBH seeds. 
MBHs are expected to be seeded at high $z$ ($z \gtrsim 10$) in either metal-free or metal-poor environments. The demographics of MBHs inside early galaxies is 
currently unknown with quantitative estimates of the number densities of MBHs as a function of redshift varying by several orders of magnitude \cite{Agarwal_2012, Agarwal_2014b, Dijkstra2014a, Habouzit_2016, Wise2019}. The 
variation can be partly explained by our lack of understanding of both the exact processes responsible and the exact environmental conditions required to seed MBHs. We now 
explore these issues in detail. \\
\indent The (astrophysical) seeds from which MBHs grow can be broken down into \textit{light} and \textit{heavy} 
seeds. Light seeds are born from the remnants of the first generation of stars and are expected to have typical masses in the range 
M$_{\rm BH} \sim 10^1 - 10^3 \ \Ms$. Heavy seeds are
expected to cover the range of M$_{BH} \sim 10^3 - 10^6 \ \Ms$. In the following sections we will 
describe in detail both the formation pathways and growth prospects of \textit{light} and 
\textit{heavy} seeds. We will break the \textit{heavy} seed formation scenario into two 
complementary parts. The first section will deal with the dynamical formation of a MBH seed within a dense, gaseous or stellar, environment
(e.g. a dense stellar cluster, a nuclear star cluster\footnote{Nuclear star clusters are aggregate groups of stars with very small
separations between the stars, with the centre of mass of the cluster typically located
very close to the centre of the galaxy} or within a dense black hole cluster) and the subsequent formation of 
a MBH(s) at the cluster centre. The second section will explore the related idea of supermassive star (SMS) formation in a similar environment and the subsequent transition of the system into one with one or more MBHs at the centre. Both scenarios may indeed overlap potentially distinguished only by the degree of fragmentation at the outset \citep[e.g][]{Chon_2020}. \\
\indent Another potential pathway to form MBHs is through the formation of primordial black holes  in the very early universe ($z \gg 1000$) with MBH seeds forming due to phase transitions as the 
Universe cooled following the Big Bang. These MBH seeds are not astrophysical in origin and are therefore distinct to the \textit{light} and \textit{heavy} seeds discussed above. Primordial black holes have previously and continue to this 
day to be postulated to be possible dark matter candidates. The theory that primordial black holes make up some, if not all of, dark matter is reinforced by the continued non-detection  of particle dark matter candidates in the mass interaction cross-section parameter space and energy ranges where they have been predicted \cite{Baudis2012, Boveria2018}. If primordial black holes are created during the early Universe phase transitions with a broad range of masses then they may account for the 
existence of SMBH at early epochs \cite{Cappelluti2022}. We refer the interested reader to Chapter~4 of this same book, which focuses specifically on primordial black holes.\\
\indent Observational signatures of the seeding process responsible for the formation of MBHs is still beyond the range of current (and indeed near future) telescopes. With the seeding process existing at small mass scales at very high redshift direct, clear, evidence is likely to be very difficult to achieve in practice. Nonetheless, there are a number of avenues that can be explored, that together can build up sufficient circumstantial evidence to point towards
certain seeding lines. In the final section, we will explore how high-$z$ observations of quasars, low-$z$ observations of fossil dwarf galaxies 
and future \ac{GW} measurements can all contribute to understanding the origin of 
MBHs.

\subsection{Light Seeds} \label{Sec:LightSeeds}
\begin{figure}
    \centering
    \includegraphics[width=\textwidth]{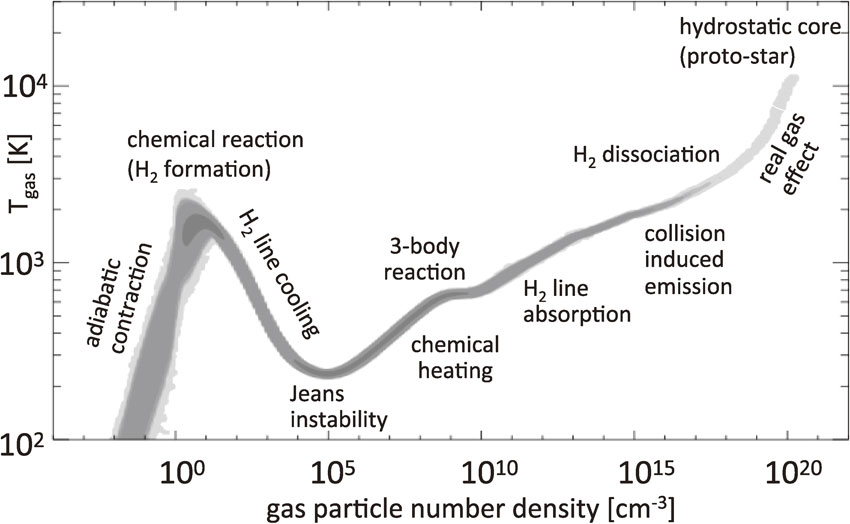}
    \caption{Density‐temperature relation for the collapse of primordial clouds. The different evolutionary phases and the description of the most relevant processes in each phase are also annotated \citep{Yoshida2019}.}
    \label{fig:firststars_collapse}
\end{figure}

After the seminal work by \cite{Rees1978}, the main idea behind SMBHs forming from a population of light seeds has been put forward by \cite{Madau_2001}, who suggested that these seeds might be the remnants of the core collapse of the first generation of stars, characterised by an initial pristine composition (formed out of primarily Hydrogen and Helium, without metals, and commonly dubbed PopIII stars). According to the currently accepted hierarchical growth  paradigm, galaxies assemble starting from smaller entities via smooth accretion and mergers during the cosmic evolution. When baryons fall within the potential well of the underlying dark matter distribution, they heat up to about the virial temperature of the halo
\begin{equation} \label{Eqn:Tvir}
    T_{\rm vir} = 1.98\times 10^4\left(\frac{\mu}{0.6}\right)\left(\frac{\Omega_{\rm m}}{\Omega^z_{\rm m}}\frac{\Delta_{\rm vir}}{18\pi^2}\right)^{1/3}\left(\frac{hM_{\rm vir}}{10^8\Ms}\right)^{2/3}\frac{1+z}{10}\rm K,
\end{equation}
where $\mu$ is the mean molecular weight, $\Omega_{\rm m}$ is the cosmological mass density parameter at $z=0$, $\Omega^z_{\rm m} = \Omega_{\rm m}(1+z)^3/[\Omega_{\rm m}(1+z)^3+\Omega_\Lambda]$, $\Delta_{\rm vir}= 18\pi^2+82d-39d^2$, $d=\Omega_{\rm m}^z-1$, $h$ is the local value of Hubble constant in units of $100\rm km/s$, and $M_{\rm vir}$ is the halo virial mass~\citep{Barkana2001}.
In order to form stars, baryons need to contract to very high densities, and this can only be achieved if they are allowed to efficiently cool down via radiative emission, contract, and fragment in less than a Hubble time. Given the primordial composition, the only available cooling mechanisms are Hydrogen and Helium lines (for $T\gtrsim 8000$~K), and the inefficient molecular hydrogen (H$_2$) cooling at lower temperatures. The latter obviously depends on the formation of H$_2$, which in the primordial Universe occurs via the associative detachment process of H$^-$ \citep{Galli1998} in moderately dense gas with $n_{\rm H}\gtrsim 100\rm\, cm^{-3}$. These requirements set a minimum mass to the baryonic structures that can form in the Universe, as early as $z\sim 30$, at $M_{\rm vir}\gtrsim 10^5 \ \Ms$ \citep{Trenti2009}: the so-called \textit{minihaloes}.
As the gas contracts, the chemical and thermodynamical state of the gas evolves according to a complex balance between different cooling and heating processes, which has been originally studied by \cite{Palla1983} and further refined over time, resulting in the evolution in Fig.~\ref{fig:firststars_collapse}. 

As the gas gets denser, the Jeans mass\footnote{The Jeans mass is defined as the enclosed mass at which the gas becomes gravitationally unstable and starts to collapse, as pressure is no longer able to sustain it from its gravitational collapse.} decreases down to about $0.1 \Ms$, which corresponds to the typical protostellar core mass, independent of the metallicity conditions. In metal-rich conditions typical of the local Universe, metal cooling quickly drives the temperature to very low values already at moderate densities, making H$_2$ cooling irrelevant \citep[see, e.g,][]{Omukai_2001,Krumholz2011}, and this results in widespread fragmentation within the halo and the formation of low-mass stars well described by a Kroupa/Chabrier initial mass function \citep{Kroupa_2001,Chabrier_2003}. In the absence of metals, instead, the fragmentation is less efficient, and this leads to the almost monolithic collapse of relatively massive clumps that are directly accreted onto the protostellar core, finally ending up in the formation of more massive stars. This crucial difference has led to the idea that PopIII stars are characterised by a more top-heavy initial mass function extending up to $M\sim 10^3 \ \Ms$ \citep{Bromm_1999,Bromm_2002}. In the last decade, the increase in computational power and the inclusion of more detailed physics (for example radiation and jets from the protostar) in numerical simulations has allowed a more detailed exploation of the initial growth of PopIII stars, with conflicting results. In particular, some studies suggest that the first stars might be less massive than originally thought \citep[e.g.][]{Greif_2012,Stacy_2012} with a characteristic mass close to $40 \Ms$, whereas others found that protostellar fragments could merge together efficiently and build a top-heavy initial mass function \citep{Hirano_2015}. While a final consensus on the PopIII initial mass function has not been reached yet, due to the complexity of the problem and the limitations in the physical modelling \citep{Hirano_2017}, the general agreement is that stars with initial masses well above $100\Ms$ are more likely to form in pristine environments. 

After entering the Main Sequence, stars evolve following a relatively well known evolutionary path, first burning Hydrogen in their core to form Helium, and then, if massive enough, moving to the Helium burning phase to produce Carbon and Oxygen (the Carbon-Oxygen core), commonly called the `Horizontal branch'. After this second stable phase, massive stars enter into a phase of rapid evolution, where elements are quickly converted into heavier ones, up to Iron. After the formation of an Iron core, nuclear fusion becomes energetically inconvenient, and the equilibrium of the stars is broken. At this stage, the gravitational pull of the core triggers its collapse, producing a neutron star  or a black hole, and the envelope can be dispersed in a supernova event. The final fate of massive stars is highly dependent on the Carbon-Oxygen core properties after the Helium burning phase, and this requires accurate modelling of the different evolutionary phases of the star, that takes into account many aspects such as initial metallicity,  rotation,  magnetic fields, as well as the mass loss of the star during its evolution. 

Even using simple arguments, one can determine that the typical lifetime of a star depends on its mass, with massive stars having stellar lifetimes significantly shorter than their low mass counterparts. With more detailed calculations, one can also demonstrate that very massive PopIII stars would evolve very quickly, with typical lifetimes of no more than 1~Myr, leaving behind a compact object. The actual nature of this compact object is still debated, and is affected by the amount of mass lost via stellar winds. For zero metallicity stars, \cite{Marigo_2001} showed that stellar winds are weak, thus resulting in a black hole with a mass comparable to that of the stellar progenitor $(10-1000 \ \Ms)$, although recent studies including stellar rotation suggested that large mass losses would occur in fast-rotating stars, leading to a supernova explosion, thus preventing the formation of a black hole \citep{Yoon_2012,Stacy_2013,Murphy_2021}.

\begin{figure}
    \centering
    \includegraphics[width=\textwidth]{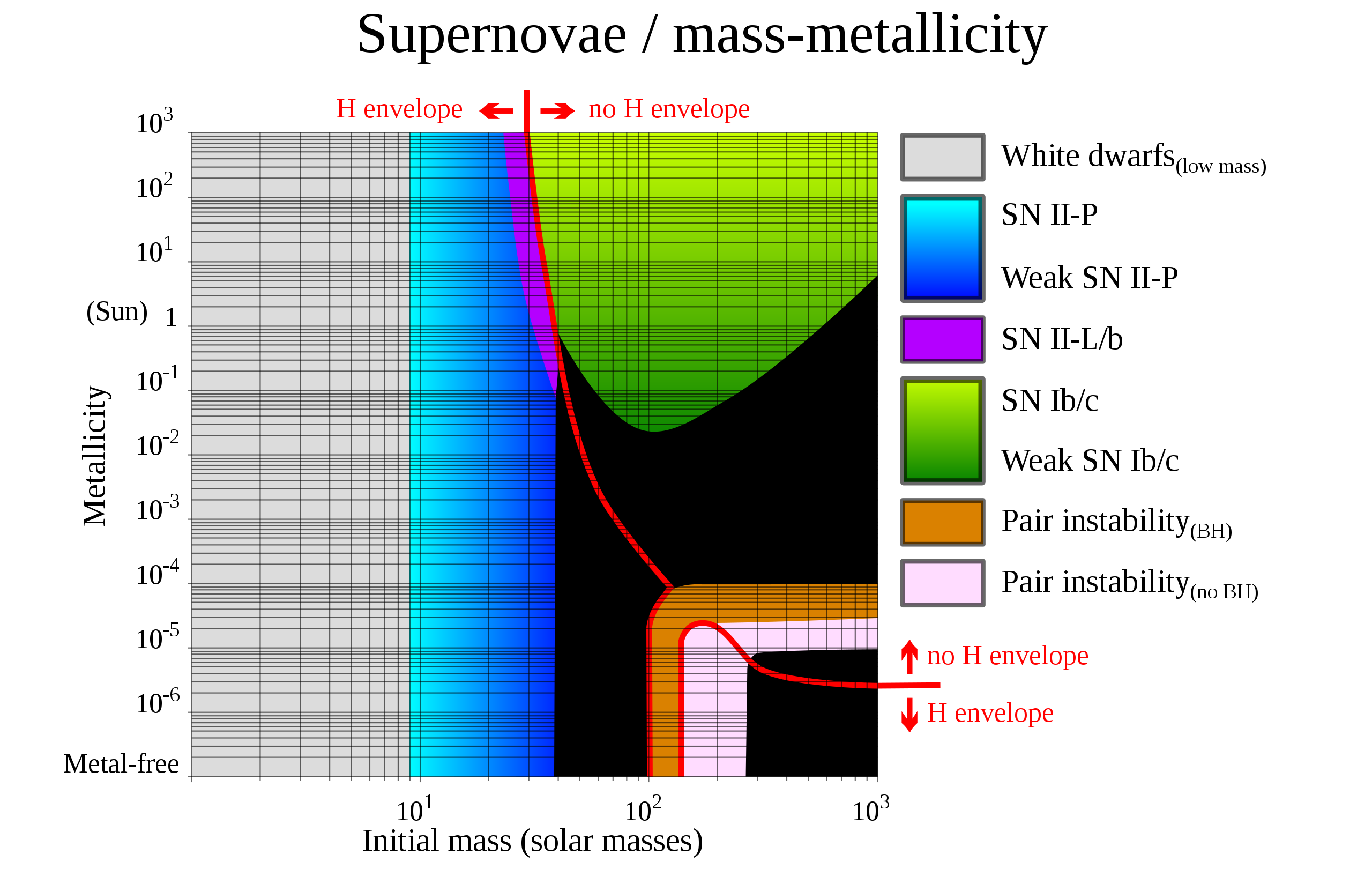}
    \caption{Mass-metallicity diagram of the final fate of stars, highlighting different end-products \citep{Heger2003}. The grey region corresponds to stars smaller than $8\Ms$, which end their life relatively quietly as white dwarfs, independent of metallicity. For more massive stars, metallicity starts to play a crucial role, with stars at Solar (or super-Solar) metallicities exploding as type II (for stars up to $\sim 40\Ms$) or type Ib/c supernovae (for more massive stars), leaving a neutron star behind. At lower metallicities, stars above $40\Ms$ typically do not explode at all, but collapse directly into a black hole. The only exception is for stars above $100\Ms$ and $260\Ms$ and metallicities below $10^{-5}\Zs$, which most of the time become unstable to pair production, which carries energy away, producing a rapid contraction and the subsequent thermonuclear explosion of the entire star, leaving no remnant.}
    \label{fig:stellar_evolution}
\end{figure}

To give a general idea of the different fates of stars, Fig.~\ref{fig:stellar_evolution} shows the results reported by \cite{Heger2003}, in particular the pair instability region (pink) where stars are expected to explode leaving no remnants, the grey region of low-mass stars (similar to our Sun) ending their life as white dwarfs, and the massive stars which explode as supernovae leaving a neutron star behind or that collapse into black holes (as a black region). It is easy to notice that in metal-free conditions, all stars above $40\Ms$ (outside the pair instability region) are expected to form black holes. This, combined with the typically higher mass of PopIII stars, suggests that black holes as massive as $260\Ms$ or more are likely to form via stellar evolution in the young Universe, thus representing potential seeds (light) for the MBH population observed. 

Because of their early formation and the common conditions under which they form, light seeds seem the most natural path and might in theory explain the entire MBH population, in particular when we consider that PopIII stars are expected to form in most (if not all) haloes in the high-redshift Universe. 
However, simple calculations show that, while they can easily explain the majority of MBHs observed in the local Universe, they have to accrete continuously at the Eddington limit  in order to account for the MBH population observed at high redshift \citep{Johnson_2013}, a requirement that is not very likely, especially in the light of recent results about the suppressing effect of stellar feedback on the MBH growth in low-mass galaxies \citep[$M_{\rm star}\lesssim 10^{10} \Ms$;][]{Dubois_2015,AnglesAlcazar_2017,Lupi_2019,Trebitsch_2020}. 

A potential solution to this issue has been recently proposed by \cite{Madau_2014} and \cite{Volonteri_2015}, and relies on intermittent accretion at super-critical rates, i.e., above the Eddington limit rates \cite{Lupi_2016,Pezzulli_2016}.\footnote{The Eddington limit corresponds to the maximum luminosity reached via spherical accretion above which radiation pressure overcomes the MBH gravitational pull.} In a seminal work, \cite{Abramowicz_1988} showed that under specific conditions (high densities and large inflow rates), the radiation produced in accretion discs can be trapped within the flow and only partially released or not released at all, resulting in the so-called slim discs. The transition from the standard geometrically thin optically thick \cite{Shakura_1973} accretion disc, where the heat produced is locally released, to the slim disc has been found to occur when the luminosity exceeds a critical value $L \simeq  0.3 \ L_{\rm Edd}$. Because of radiation trapping, the slim disc solution remains moderately luminous ($L \gtrsim L_{\rm Edd}$), even at super-Eddington accretion rates, as the radiative efficiency decreases. The plausibility of this accretion regime has  been demonstrated observationally in tidal disruption events \citep{Lin_2017} and ultra-luminous X-ray sources \citep{Bachetti_2014}. Recent studies \citep{Sadowski_2016a,Sadowski_2016b,Jiang_2019} have showed, however, that the total feedback efficiency (the fraction of rest-mass energy released) during super-Eddington accretion phases can be as high as $\sim 30\%$ for highly spinning black holes, mostly released as kinetic winds/jets rather than radiation. Depending on specific environmental conditions and the coupling efficiency of the jet with the surrounding gas, this high efficiency can be sufficient to suppress the MBH growth, resulting in sub-Eddington average accretion rates \cite{Regan_2019,Massonneau_2022}, or allow accretion to proceed almost unimpeded. The latter can occur in the so-called hyper-Eddington accretion regime \citep{Inayoshi_2016,Shi_2022}, where the feedback is almost suppressed, or in very dense environments (e.g. nuclear star clusters) \citep{Alexander_2014,Natarajan_2021}. To date, a final consensus is still missing, due to many limitations of the current simulations, as the limited resolution, the poorly constrained coupling efficiency of the jet, and the environmental conditions in which these seeds are embedded \cite{Regan_2019}.
 
\subsection{Heavy Seeds Generated in Nuclear Star Clusters}
\begin{figure}
    \centering
    \includegraphics[width=\textwidth]{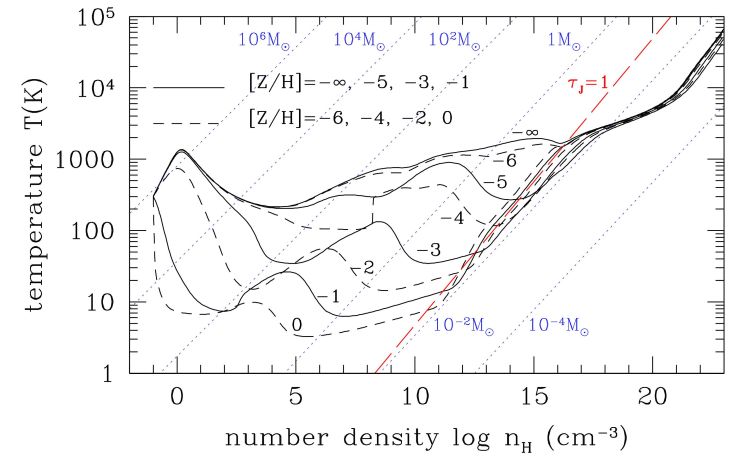}
    \caption{Density‐temperature relation for the collapse of clouds with different metallicities \citep{Omukai_2001}, defined by the logarithmic ratio between the metal mass fraction $Z$ and the hydrogen mass fraction $H$. The red dashed line corresponds to the conditions at which the cloud becomes optically thick ($\tau_{\rm J}=1$), and the dotted lines to different values of the Jeans mass. The different curves highlight how metal cooling (and correspondingly the cooling by dust grains that form more efficiently in metal-richer environments) brings the gas temperature down more efficiently for increasing metallicity, bringing the gas to lower  Jeans masses, which favour widespread fragmentation and the formation of lower mass stars.}
    \label{fig:Omukai2005}
\end{figure}

When massive stars explode as supernovae, they inject heavy elements into their surroundings, increasing the gas metallicity. Metals (and dust, which forms from the aggregation of carbon and silicon composites) provide additional  cooling to the gas, which more easily reaches low temperatures, as shown in Fig.~\ref{fig:Omukai2005}, where the evolutionary tracks of clouds at different metallicities are reported in the  density--temperature plane \citep{Omukai_2005}. A faster decrease in temperature allows the gas to more easily fragment into small clumps, thus ending with the formation of less massive stars (with an initial mass function more similar to the present one). Unlike PopIII stars, these lower-mass stars are commonly organised in clusters with typical sizes of a few pc. 

While stellar clusters can in general form everywhere within galaxies, observations of galactic nuclei often evidence the presence of massive and dense stellar complexes surrounding MBHs \citep{Seth2008,Graham2009,Georgiev2016,Nguyen2018}, called nuclear stellar clusters (NSCs). Detailed observational analysis also shows that the mass of the NSC relative to that of the MBH scales with the stellar mass of the galaxy host \citep{Neumayer2020}, with the NSC becoming less and less important as the host and the central MBH grow in mass. Our Galaxy is not different, as it hosts both a NSC of mass $10^7 \ \Ms$ \citep{Launhardt2002, Schodel2009} and a MBH of mass $4.3\times 10^6 \ \Ms$ \citep{Genzel2003,Ghez2008,Gillessen2009}. This particular evidence might indicate the possibility that the MBH formed in-situ inside the NSC, hence the need to understand how this can occur in practice.

The idea of forming MBHs in stellar clusters dates back to the 60s, with the pioneering work of \cite{zeldovich1965}, then refined and expanded at the end of the 80s by \cite{Quinlan1987,Quinlan1989,Quinlan1990}, which highlighted how gravitational interactions among the cluster members would lead to the catastrophic collapse of the system and the formation of a single massive object.
Inside a NSC, stellar mass black hole remnants of massive stars, being more massive than the remaining stellar objects, tend to segregate towards the centre of the cluster forming a core. Then, the energy exchange between the core and the rest of the system would lead the core to contract more and more up to the point at which a general relativistic instability sets in, finally driving the stellar black holes to collapse into a single object. Despite the validity of the conclusions, the conditions required for this process to occur were rather extreme, hence unlikely, i.e. the NSC had to be very massive and dense compared to typically observed NSCs, so that stellar binaries would not survive in such systems. Indeed, hard binaries (those with a binding energy larger than the typical kinetic energy of the other objects in the NSC) would act as a heating source for the system, injecting energy via three-body interactions, but also removing mass via the gravitational wave-induced recoil after binary mergers \citep{Sigurdsson1993,Lousto2010}. This could lead to the complete evaporation of the NSC, preventing the formation of a massive seed black hole, although the actual result also depends on the initial mass of the heaviest black hole in the cluster \citep{Miller2002}.

After being put aside for about two decades, this idea was revived by \cite{Davies2011} in a cosmological context, motivated by the results of numerical simulations that showed how very large gas inflows towards the nuclear region of galaxies might occur during mergers at high-redshift \citep{Mayer2010}. If no fragmentation occurs and these inflows reach the galaxy central pc, they can significantly deepen the potential well of the NSC at a level sufficient to ensure that hard binaries quickly coalesce via gravitational wave emission and the recoiling black holes are efficiently retained \citep{Miller2012,Antonini2019}. The absence of heating sources in the cluster would help the core of stellar mass black holes to contract and reach the density conditions at which rapid mergers would occur, ending with the formation of a massive seed black hole. In order to constrain the typical properties of these seeds, \citep{Lupi2014} explored the relevance of this mechanism using semi-analytical models of cosmic structure formation, finding that it would produce a population of seed black holes comparable to that of the light-seed scenario, but i) with typically larger masses compared to PopIII remnants, of the order of $10^3\Ms$, and ii) at lower redshifts and under less stringent metallicity conditions. 

Another possible path to form seed black holes in NSCs is via runaway collisions of stars well before massive ones evolved forming black holes (or neutron stars). This idea was originally proposed by \cite{PortegiesZwart2002} and requires a low (but not zero) metallicity, which results in a relatively inefficient fragmentation, hence in the formation of a compact NSC with a core collapse timescale shorter than the evolutionary time of its most massive stars (around a few Myr). The motivation behind the requirement of a short core collapse timescale is to guarantee a high rate of collisions among massive stars, which are expected to have the largest radii, before they turn into compact objects which, by having a smaller cross section, would inhibit the mass build-up. The fate of the collision product, a more massive star, would then depend on its mass and spin, according to the stellar evolutionary tracks also considered in the case of PopIII remnants. In general, if the star formed reaches about $10^3 \ \Ms$ or more, the most likely fate would be the direct collapse into a black hole heavier than that formed by PopIII remnants. As the actual conditions for this process to occur were not clear from the idealised study of \cite{PortegiesZwart2002}, several authors tried to include this model in semi-analytic cosmological models \citep{Devecchi2009,Devecchi2012} or hydrodynamical simulations \citep{Katz2015}, finding that a very massive star  of about $500-10^3 \ \Ms$ was likely to form. 
Despite the apparent simplicity of this formation scenario, several uncertainties are present. One of them is gas dynamics. In particular, before massive stars explode, the NSC might be still embedded within a massive gas cloud, and gas might be accreted by the orbiting stars on timescales comparable to those of stellar collisions. This idea has been recently explored using N-body simulations including static gas distributions \citep{Boekholt2018,Reinoso2018,Das2021}, showing that gas accretion can boost the mass of the colliding stars (or protostars), ending up with a very massive star of $10^{4-5} \ \Ms$, although the effect of mass losses via stellar winds might inhibit the efficient growth \citep{Das2021b}. 
Another possibility is instead that, if the metallicity is lower than $10^{-4} \ \Zs$, fragmentation might occur only at very high densities, producing a single massive protostar surrounded by many smaller objects. In these conditions, dynamical friction on gas and stars would bring the stellar objects in the NSC towards the centre, producing a `bombardment' of stellar collisions with the central massive protostar, and the final formation of a very massive star of $\sim 10^4 \ \Ms$ \citep{Tagawa2020}.

\subsection{Heavy Seeds generated from (Super-)Massive Stars}
\begin{figure}[!t]
\centering   
\includegraphics[width=\textwidth]{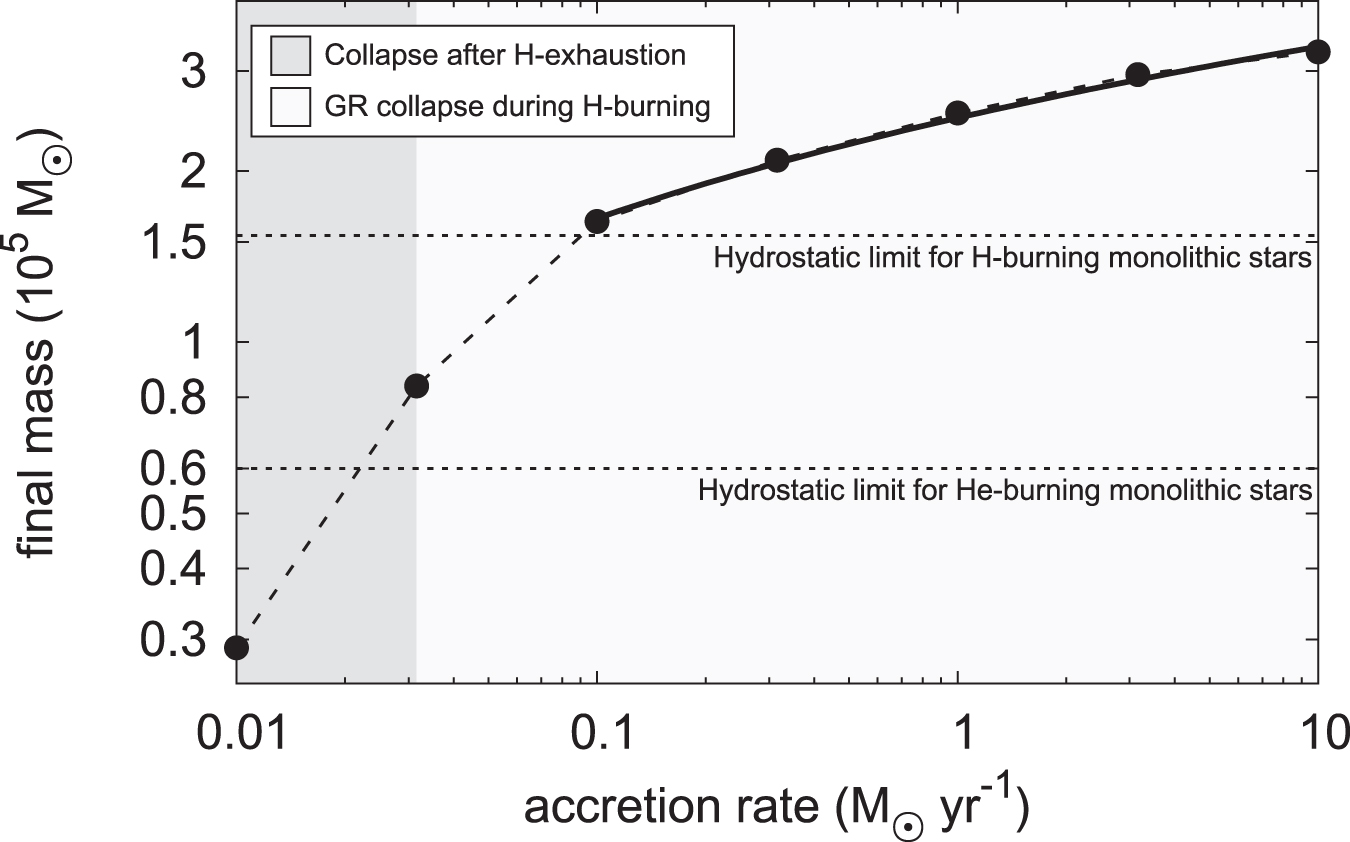}
\caption{Final masses as a function of (continuous) accretion rate. Black circles denote individual models undergoing constant accretion, with the trend given by the black dashed
line. The solid black line plots the fit formula of Equation \ref{Eqn:Woods17}. Also shown are the limiting masses for hydrostatic hydrogen and core helium burning for monolithically
formed SMSs.  The maximum masses for hydrostatic hydrogen and helium burning are for monolithically formed (as opposed to
accreting) SMSs. Figure reproduced with permission from \cite{Woods2017}.}
\label{Fig:Woods17}
\end{figure}

As discussed in the previous section, heavy seeds can be formed in very dense environments where a 
sufficient gas inflow can either result in the formation of a NSC or can feed an existing cluster of stars. In a similar vein SMSs can be formed when accretion onto a protostar exceeds a critical rate 
\cite{Hosokawa2012, Hosokawa2013, Sakurai2016, Haemmerle2017b, Haemmerle2018}. This critical accretion rate onto the stellar surface is somewhat uncertain with some dependence on the episodic nature 
of the accretion. However, different stellar evolution codes and treatments agree on 
a critical accretion rate of $ 5 \times 10^{-3} \lesssim \rm{\dot{M}_{crit}} /(\Ms/{\rm yr})  \lesssim 5 \times 10^{-2}$.
Surface accretion rates exceeding $\rm{\dot{M}_{crit}}$ cause the star to bloat as high entropy gas is
accreted onto the star preventing the star from achieving thermal relaxation. As the star bloats and
the surface area increases the star cools and achieves an effective temperature, 
$\rm{T_{eff}} \sim 5000 \ \rm{K}$. Such stars will therefore appear red, emitting the vast majority of their radiation in the infrared and hence their
feedback is somewhat `weaker' than comparable mass PopIII stars whose spectral energy distribution peaks in the far ultraviolet. \\
\indent Depending on the sustained accretion rate onto the star, the star can reach a maximum
mass of approximately $1.5 \times 10^5 - 3.3 \times 10^5 \ \Ms$ \cite{Woods2017}. As the mass approaches 
this maximum mass value the general relativistic instability is reached \cite{Chandrasekhar_1964b} causing the SMS to collapse directly into a black hole. A star of this mass collapsing
directly into a black hole with little or no mass loss will leave behind a MBH with a mass of 
$\rm{M_{MBH}} \gtrsim 10^5 \ \Ms$. \citet{Woods2017} provide the following fitting function to model the maximum mass of a rapidly accreting SMS as a function of the accretion rate

\begin{equation} \label{Eqn:Woods17}
    \rm{M_{SMS, final}} \approx \Big [ 0.83 \ \log_{10} \Big ( \frac{\rm{\dot{M}}}{\Ms \ \rm{yr^{-1}}} \Big) + 2.48 \Big ] \times 10^5 \ \Ms
\end{equation}

where $\rm{M_{SMS, final}}$ is the final SMS mass and $\rm{\dot{M}}$ is the stellar accretion rate.

Figure \ref{Fig:Woods17} shows the maximum attainable masses from which Equation \ref{Eqn:Woods17} is derived. Independent treatments and calculations by both \cite{Haemmerle2018} and \cite{Umeda2016} give consistent 
results verifying, at least theoretically, that SMSs are both stable and can grow to masses of approximately $10^5 \Ms$ before collapsing directly into a MBH. \\
\indent Creating the conditions necessary for SMS formation has been the subject of numerous detailed calculations over the past decade. In that regard the 
focus of the community has been on so-called atomic cooling haloes. These are dark matter halos that have attained a virial temperature of approximately 
8000 K with the primary gas coolant being atomic Hydrogen. This virial temperature can be related to the halo virial mass through the following equation \cite{Fernandez_2014} (cf. Eqn \ref{Eqn:Tvir}):
\begin{equation}
    {M_{\rm halo} = 2 \times 10^7 \Big(\frac{T_{\rm vir}}{10^4} \frac{21}{1+z} \Big)^{3/2} M_{\odot}}
\end{equation}
where ${M_{\rm halo}}$ is the halo virial mass and ${T_{\rm vir}}$ is the halo virial temperature.
This gives the expected minimum mass that must be attained in order to achieve sustained accretion at a rate sufficient to drive 
SMS formation. In addition to a minimum mass threshold there is also a restriction on metal enrichment of the halo. Metals offer a number of cooling emission lines enabling the gas to cool to tens of degrees Kelvin, which result in star formation with masses similar to those (if perhaps somewhat higher, see also \S \ref{Sec:LightSeeds}) of present day star formation. As a result, in addition to the minimum mass restriction, a further restriction is that haloes either remain pristine or very metal poor, to avoid excessive fragmentation of the gas due to cooling instabilities. Three main mechanisms have been identified that may provide the correct conditions to allow pristine (or near-pristine) atomic cooling haloes to be realised. \\

\subsubsection{Lyman-Werner Irradiation}
\indent In order to achieve the correct conditions to enable SMS formation we need a process that allows a sufficiently large Jeans mass to develop, by-passing the formation of `normal' PopIII stars. Lyman-Werner  radiation corresponds to photon energies in the range 11.8 eV - 13.6 eV and so while below the Hydrogen ionisation edge, Lyman-Werner radiation is energetic enough to disassociate H$_2$ \cite{Field_1966}. By dissociating H$_2$ from pristine haloes,  PopIII star formation is effectively stalled and instead the halo continues to accrete mass without forming stars. Intense local Lyman-Werner radiation, in combination with a mild Lyman-Werner background, from a (very) nearby galaxy is likely required to enable the halo to reach the atomic cooling limit without initiating star formation \cite{Regan_2017}. In this case ideal conditions for forming SMSs can be created. The intense Lyman-Werner background can be found in the so-called synchronised pair scenario where two haloes, closely separated in space and time, 
can produce the desired effect. One of the two neighbouring haloes begins to form stars as it crosses the atomic cooling limit. Indeed it may be possible that not only two, but a number of nearby intense sources may occur -- a scenario dubbed synchronised multiplets \cite{Lupi_2021}. The intense Lyman-Werner radiation given off
by this initial starburst is effective at dissociating much of the H$_2$ from its neighbour. This allows the neighbour to exceed the atomic cooling limit without forming stars, resulting instead in the formation of a SMS. 

\subsubsection{Rapid Assembly}\label{sec:rapid_assembly_BHseeds}

\indent The second mainstream mechanism to produce SMSs is via the hierarchical formation of structures itself. As a halo grows in mass, the rapid accumulation of matter may give rise to the birth of massive stars as has been shown by \citet{Wise_2019, Lupi_2021} via the rapid assembly paradigm and by \citet{Latif_2022} in the rarer case of a turbulence driven scenario. Both mechanisms are governed by the same underlying principles and in fact their theoretical foundations can be traced back to papers by \citet{Yoshida_2003a, Fernandez_2014}. \\ 
\indent In the rapid assembly case advocated by \citet{Wise_2019} and \citet{Lupi_2021}, the rapid assembly results in dynamical heating of the halo through repeated (minor) mergers. If the 
dynamical heating rate exceeds the cooling rate (driven almost exclusively by H$_2$ in mini-haloes) then the halo cannot form PopIII stars. Instead the halo continues to grow until it exceeds the atomic cooling threshold (in a similar way to above with the Lyman-Werner pathway) and atomic Hydrogen acts as an efficient coolant. In this case again,  a halo can remain pristine and accumulates a substantial baryonic core which can result in large accretion rates onto any collapsing object(s) and 
therefore the formation of a SMS or indeed multiple SMSs. Simulations by \citet{Regan_2020} have shown that in such a scenario, where a halo is growing very rapidly and augmented by a mild ($J_{LW} \sim 5 \ J_{21}$) Lyman-Werner field that stars with masses in the range M$_* = 1000 \ \Ms - 10,000 \ \Ms$ can form. \\
\indent \citet{Latif_2022} modelled a rarer manifestation of the above scenario where a rare halo, found at the nexus of a number of converging 
flows, experienced large turbulent support which exceeded the support from rotation. The result was the formation of stars with masses 
in the range M$_* = 10,000 \ \Ms - 100,000 \ \Ms$. Whether more massive stars can form, in this or related ways, and be truly super-massive (i.e. M$_* \gtrsim 10^5 \ \Ms$) is a subject of ongoing research.

\begin{figure} [!t]
\centering
\includegraphics[width=\textwidth]{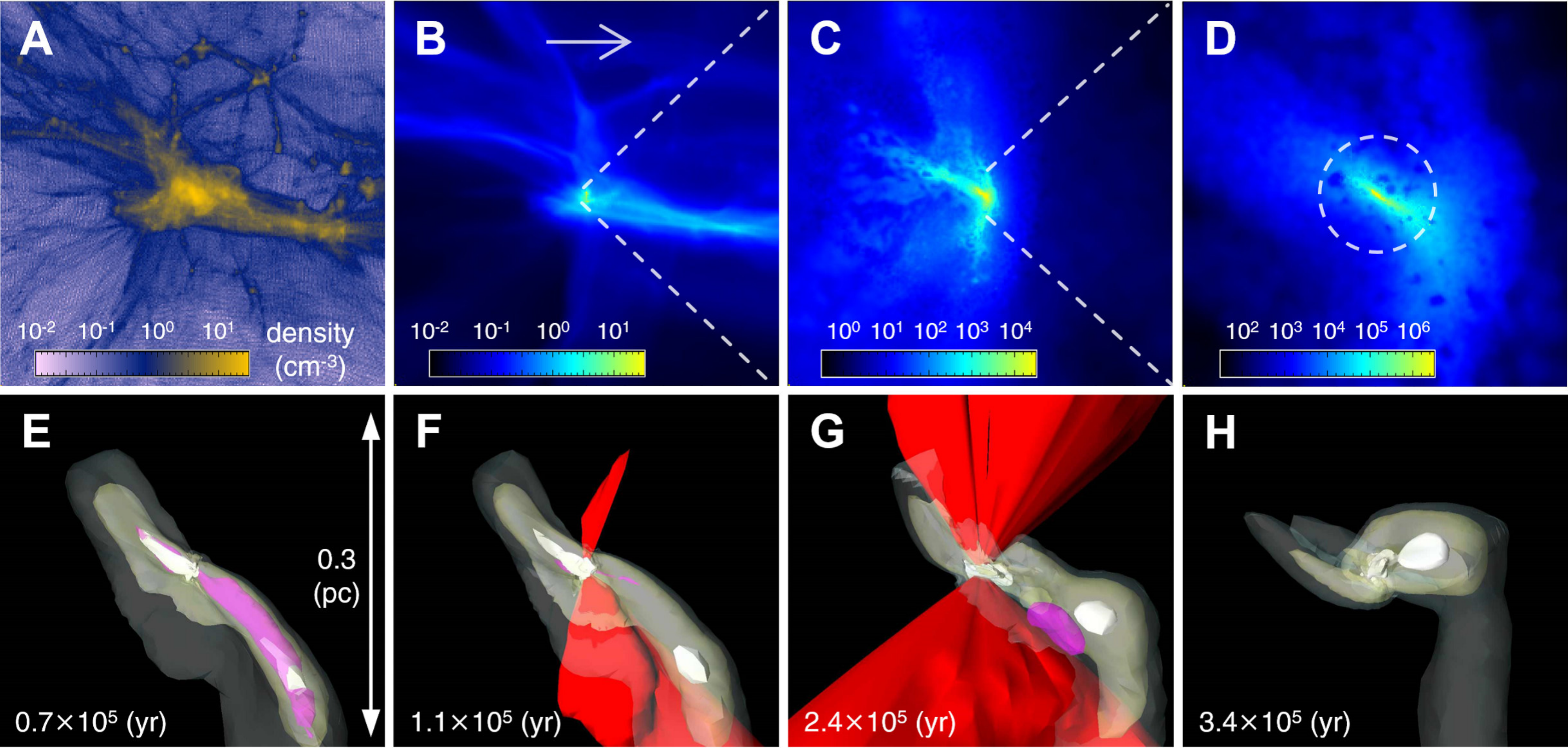}
\caption{Large-scale density distribution and the structure around an accreting protostar.
(A) Projected density distribution of dark-matter component around the star-forming cloud at
$z$ = 30.5. The box size is 2500 pc on a side. The virial mass of
the main dark matter halo located at the centre is $2.2 \times 10^7 \Ms$. (B to D) Projected density
distribution of the gas component in regions of 2500, 100, and 10 pc on a side, from left to
right. The horizontal arrow in (B) shows the direction of the initial supersonic gas stream. The
dashed circle in (D) indicates the Jeans length, within which the cloud is gravitationally unstable
given its mass of $26,000 \Ms$ . (E to H) Evolution of the temperature and density structure in the
protostellar accretion phase after the protostar formation. Coloured in white, red, and magenta
are the iso-contours of gas density (at $10^6, 10^5$ and $ 3 \times 10^4$ cm$^{-3}$), photoionized hydrogen
abundance with $\ge$ 50\% (H II region), and the number fraction of hydrogen molecules with
$\ge$ 0.2 \%. Adapted from Hirano et al. (2017) \cite{Hirano_2017}}
\label{Fig:Hirano}
\end{figure}
\subsubsection{Baryonic Streaming Motions}
\indent Finally, baryonic streaming velocities \cite{Tseliakhovich_2010} can lead to the formation of pristine atomic cooling haloes -- and in a similar vein to the above pathways the 
formation of a SMS. Baryonic streaming velocities arise at recombination due to relative velocity differences between the baryons and the dark matter. The relative differences have a mean of zero but larger fluctuations (up to tens of km/s) are possible in rare regions. A number of authors \cite{Tanaka_2014, Latif_2014c, Hirano_2017, Schauer_2017}  have studied the effect that streaming velocities can have on the formation of MBH seeds at early times. In particular \cite{Hirano_2017} found that for relative velocity fluctuations at three times the root-mean-squared value they were able to form a SMS with a mass of approximately $34,000 \ \Ms$ (see Figure \ref{Fig:Hirano}). Such a massive star would be an ideal 
seed from which to realise a MBH seed. \\

\indent Each of the pathways described above have advantages and disadvantages and in particular contain inherently different number density predictions. It is also possible that two (or more) of the 
pathways work in unison \cite{Wise2019, Schauer_2017, Kulkarni_2021} possibly forming the most massive seeds which go on to become the quasars observed at $z \gtrsim 6$. A number of high resolution cosmological (radiation)-hydrodynamical simulations have attempted to probe each of the above scenarios using a variety of techniques to extract the seed black hole mass formed in each case. As noted above, \cite{Hirano_2017} looked at the impact of baryonic streaming velocities on MBH seed formation and found that MBH seeds of approximately $34,000 \Ms$ can form. 
In addition to this, simulations by \cite{Regan_2020} found SMSs with masses of more than $6,000 \ \Ms$ formed in their simulations examining the rapid assembly channel
while \citet{Latif_2022} formed stars with masses of up to $100,000 \ \Ms$ as mentioned in the end of Sec.~\ref{sec:rapid_assembly_BHseeds}. \cite{Regan_2018} examined the Lyman-Werner formation channel (using somewhat idealised initial conditions by imposing Lyman-Werner background fields) and formed SMSs with masses upwards of $100,000 \ \Ms$. Further work examining each of these pathways (and ideally with pathways working in tandem organically) are now required although the computational task for this remains formidable. In addition to larger sample sizes, increased resolution is, at a minimum, needed to examine the thorny issue of gas instability and fragmentation in such systems as this may lead to a lower than expected initial mass function for very massive star formation \cite{Chon_2020}.\\
\indent Related to SMSs as potential progenitors for MBHs, quasi-stars have also been postulated as possible progenitors \cite{Begelman2008}.  Quasi-stars are 
defined as rapidly accreting stars in which the core of the star has collapsed to form a black hole which continues to accrete the surrounding stellar envelope.
Such objects are predicted to form if the infall of gas 
into the centre of a halo exceeds about 1 $\Ms \ \rm{yr^{-1}}$. The collapsing gas traps its own radiation and forms a radiation pressure-supported SMS. When the core of the SMS
collapses, the resulting system becomes a quasi-star. The quasi-star can then acculate mass at a rate commensurate with the mass of the surrounding gaseous envelope and hence the core effectively accretes at rates greatly exceeding its Eddington rate. Similar to a SMS, a quasi-star displays an outwardly cool envelope and appears red. \cite{Volonteri2010} used merger trees to track gas rich mergers which may be a trigger for quasi-star formation at $z \gtrsim 10$,
furthermore they restrict their models to gas inflows of low angular momentum gas -- which are needed to achieve the very high accretion rates into the centre of the halo. Modelling quasi-stars as a blackbody with $T_{QS} = 4000$ K, \cite{Volonteri2010}
predicted number counts of suasi-star hosting haloes for JWST\footnote{The James Webb Space Telescope.} in the 2–10 $\mu$m band. Assuming a sensitivity of 10 nJy at 2 $\mu$m, they found that JWST could detect up to several quasi-stars per field at $z \sim 5–10$. However, it should also be noted that standard stellar modelling of such high accretion rates show no evidence of quasi-star formation \cite{Woods2018} and therefore their existence must await observational confirmation \cite{Woods2017}.

\subsection{Exotic Formation Channels}
In addition to the three main scenarios, more exotic physics has also been considered to explain the formation of MBH seeds. Among them:
\begin{itemize}
\item The most famous model, and probably also the most studied to date, is inherited from theoretical studies about dark matter, and corresponds to primordial black holes forming from the collapse of primordial baryonic perturbations in the Universe. Depending on the horizon scale at the time of the collapse, black holes with different masses can be formed, from tiny ones that have already evaporated due to Hawking radiation (formed just after inflation) up to potentially millions of solar masses (after the time of $e^+ e^-$ annihilation). Constraints on the abundance of primordial black holes can be obtained using different observational tracers that would be affected by their presence, as the cosmic X-ray background or the effects on the cosmic microwave background \citep{Carr2020}. The detection of \acp{GW} from LIGO/VIRGO of merging black holes with $\sim 30\Ms$ was used to cast doubts on the tight constraints existing in this mass interval, triggering new interest in this potential formation mechanism (see, e.g., \citep{Carr2018});
\item Another possible mechanism is based on cosmic strings, topological defects appearing during phase transitions in the Universe, which can intersect, creating loops. If the loop angular momentum is low enough, the contraction proceeds up to the point of black holes formation. The typical masses of these seed black holes can reach $\sim 10^5\Ms$ at $z\sim 30$ \citep{Brandenberger2015};
\item Recent results provided by the Large Hadron Collider on supersymmetric particles\footnote{Supersimmetry is an extension of the Standard Model of particle physics, where each fermion has a supersimmetric boson partner, and vice versa.} dark matter candidates, together with the `too big to fail'\footnote{Haloes deemed "too big to fail" are anomolously low-luminosity haloes which should have on-going star formation but for some reason do not.} and the `cusp-core'\footnote{The cusp-core problem relates as to whether a dark matter halo should have a core (constant density) or a cusp (peak of density) at its center.} problem in astrophysics, have triggered a great deal of research aimed at extending the standard model of particle physics, resulting in theories of self-interacting dark matter\footnote{ In these theories, dark matter particles become collisional below a specific interaction scale, and start behaving as a pressure-dominated fluid.} and mirror sectors, the latter being almost identical `dark' replicas of the standard model \citep{Berezhiani2005}. According to this mirror model, dark particles might dissipate energy in the same way as baryons, resulting in the potential rapid collapse of structures in the early Universe. In particular, \citep{Damico2018} and \citep{Latif2019} showed that in a Universe made of axions and a small fraction of mirror dark matter, the initially low ionisation fraction of dark electrons and the low temperature of dissipative dark matter might suppress fragmentation, favouring the collapse of intermediate mass black holes, which would be then able to accrete mass from both the dark and baryonic sectors, more easily growing to the mass of the most massive quasars to date.  
\end{itemize}

\subsection{Differentiating Seed Formation Channels through Observations}

Differentiating between seed formation channels is likely to be a difficult and challenging undertaking even in the medium term with upcoming space-based and ground-based observations. The fundamental reason being that in order to categorically distinguish seed formation channels, the seed formation processes would need to be observed at exquisite precision and likely across a range of epochs and environments. The epoch in question 
is the very early Universe at  $z\gtrsim 10$ and the mass scale is $M_{\rm BH} \lesssim 10^5 \ \Ms$. By any metric this is a challenging proposition.  However, there are other observations that could give promising hints (if not direct observation) of the seeding process. 
Dwarf galaxies may for instance host fossilised MBHs that formed at high-$z$ but remained essentially unaltered since their formation epoch. In that way these fossilised MBH are true representations of the seeding mechanism with their mass today being the same as their seed mass. A full determination of the seeding problem may ultimately require a multi-messenger (i.e.  electromagnetic plus \ac{GW}) approach. This would be made up of observations in the GW wavebands as well as the traditional electromagnetic wavebands. 
We therefore begin by exploring the benefits of a multi-messenger approach and the potential for a direct observation of the MBH seeding process. \\
\indent Accreting MBHs are expected to be X-ray bright; due to the strong penetrating capacity of X-rays this offers a potentially very clear signature of active MBHs. Catching the on-going formation of a heavy seed MBH would be the most definitive proof of the heavy seed pathway \cite{Haiman_2019}. X-rays are likely the best way to detect high-$z$ MBH seeds.  In order to achieve this a next generation X-ray facility would be required. As it stands, the proposed Lynx\footnote{https://www.lynxobservatory.com/} mission is best placed to detect the smoking gun of heavy seed formation as it has the required sensitivity (flux limit of $10^{-19}$ erg s$^{-1}$ cm$^{-2}$). In addition to X-ray, detections in the optical and infrared would be complementary to diagnose the host galaxy characteristics. JWST, along with the Roman and Euclid missions, should have the required sensitivity to probe the host galaxy properties assuming that the 
host can be sufficiently well located in the X-ray. If a future X-ray telescope can identify the 
early stages of heavy seed formation then this can be used to constrain and differentiate between 
seed formation channels (particularly if evidence of SMS star formation can be sampled in the same galaxy \cite{Woods_2021}). \\
\begin{figure} [!t]
\centering
\includegraphics[width=\textwidth]{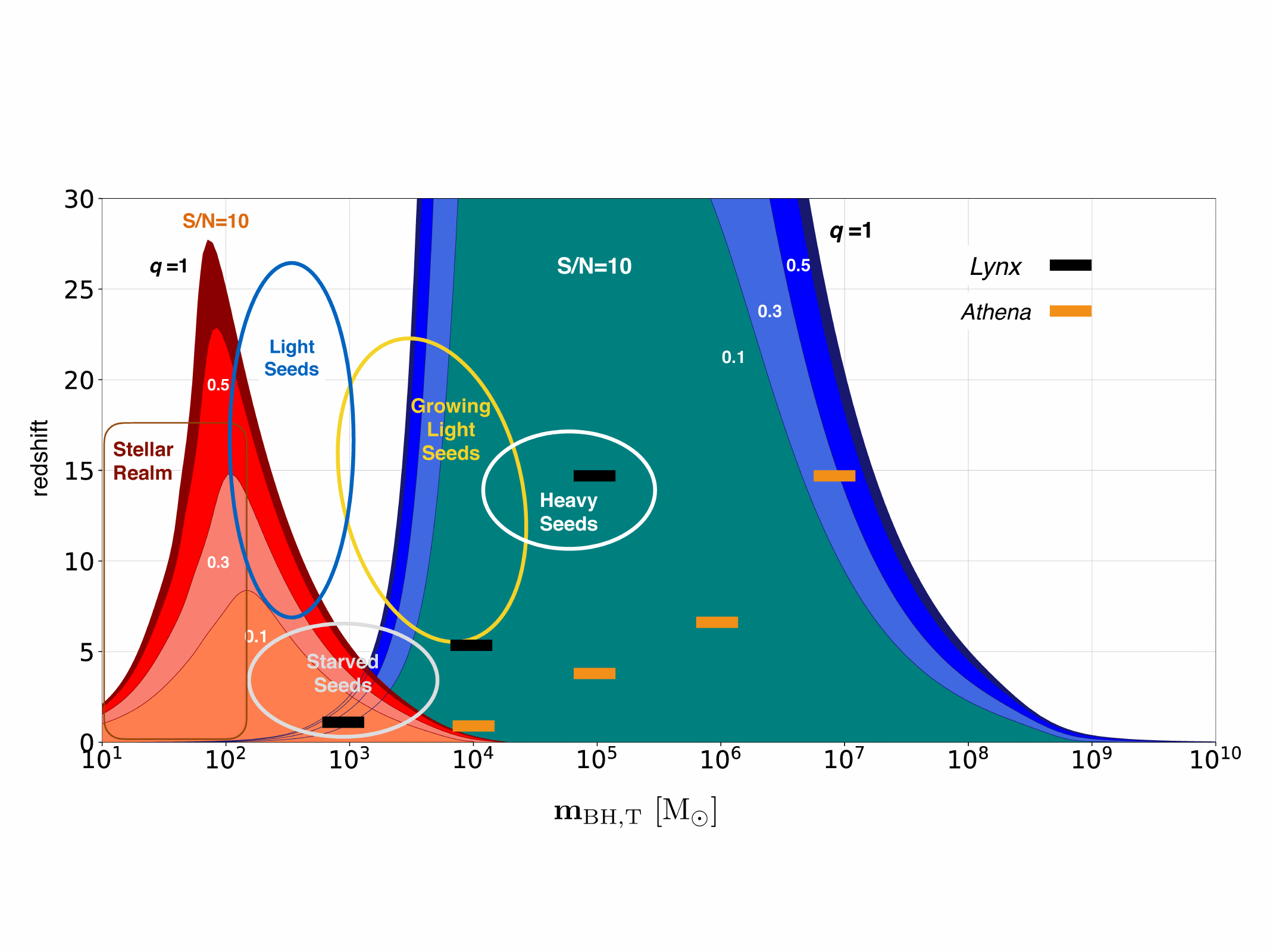}
\caption{The \ac{GW} and electromagnetic landscape in the plane with merging black hole binary mass versus redshift. Colour-coded areas give the average \ac{GW} horizon computed for a detection threshold equal to S/N = 10: contour lines
refer to binaries with mass ratios $q = 1, 0.5, 0.3, 0.1$ both in the ET and LISA bandwidth. Upper limits (shown as thick horizontal bars) indicate the sensitivity
of the deepest pointing, in the [0.5 - 2] keV observed band, by Athena (orange) and Lynx (black)
given the limiting fluxes of $2.4 \times 10^{-17}$ and $10^{-19}$ erg s$^{-1}$ cm$^{-2}$,
respectively. The upper limits are inferred assuming that black holes are emitting at the Eddington limit and adopting a bolometric correction (L$_X$ / L$_{bol}$) of
10\%. Ellipses highlight the islands in the plane where light (blue) and heavy (white) seeds are expected to form as well as where light seeds are
expected to grow via accretion and mergers (yellow). The transit to the SMBH domain covers the entire LISA area and electromagnetic observations are key to discover
the high-mass tail of the SMBH distribution. The light-grey ellipse below $z \sim 5$ marks the population of long-living "starved" seeds. Note that in this island,
coordinated multi-band observations are possible with LISA having the capability to first follow the early inspiral in IMBHs and ET the
merger phase, enhancing the ability to carry on precise measurements of the source parameters also 
at $z \sim 5$ \cite{Jani_2020}. The islands have overlap with
the GW horizon, but an empty inaccessible region is present between ET and LISA, corresponding to the Deci-Hz GW domain. The island corresponding to
the stellar realm is included, on the left, for comparison. Figure reproduced with permission from \cite{Valiante_2021}.}
\label{Fig:Valiante}
\end{figure}
\begin{figure}[!t]
\centering
\includegraphics[width=14.0cm, height=10cm]{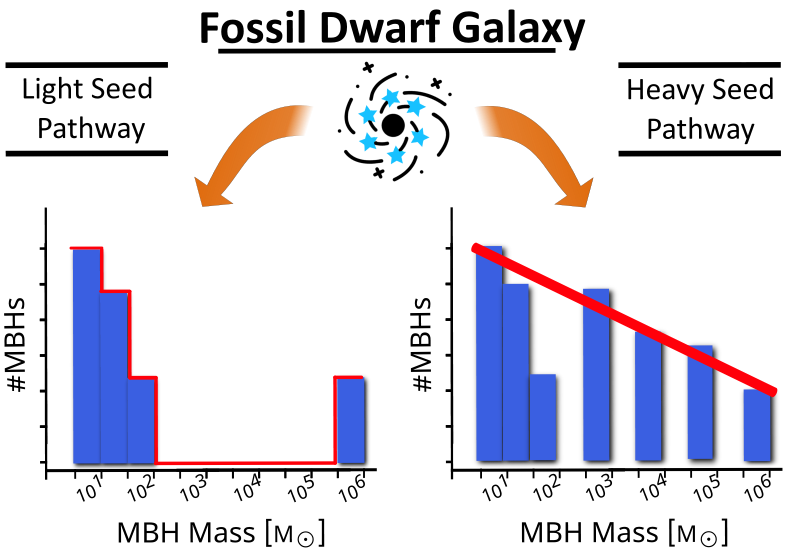}
\caption{Graphical illustration of the demographic signatures of both light (left hand side) and heavy (right hand side) seeds. Light seeds (born from PopIII star for example) are predicted to only grow in the rare cases where they experience successive bursts of accretion. Given the rarity of such events, at most one MBH is expected per dwarf galaxy with a light seed as progenitor. Heavy seeds born here from the SMS route may instead arise with many sibling MBHs. In this case this initial spread of heavy seeds is expected to persist across cosmic time. The discovery of multiple MBHs in a dwarf galaxy would then be a unique and powerful signature of a heavy seed formation channel. Figure reproduced with permission from \cite{Regan_2023}.}
\label{Fig:LeoI_MBH}
\end{figure}
\indent  \citet{Valiante_2021} used semi analytic models to track the growth 
of both light and heavy seeds across cosmic time and investigated the multi-messenger capacity of 
different \ac{GW} detectors on discriminating between seeding pathways. The results of their analysis are outlined in Figure \ref{Fig:Valiante}. The waterfall plots show the capacity of the Einstein Telescope \citep{Maggiore_2020} (ET) (orange) and LISA \citep{Klein_2016}(green) to detect massive black hole binaries with a given mass and redshift. The 
detectability of the same binaries (assuming accretion at the Eddington limit) is also shown. Marked
on their figure as ellipses are the regions in which different seeds are expected to reside and so 
by building up a sufficient statistical sample constraints on the relative populations of each can be 
determined. \\
\indent LISA, and to a lesser extent ET, will have the ability to detect the mergers of MBHs out to the seeding epoch at $z \gtrsim 10$. LISA's ability to detect MBH binary mergers out
to high-$z$, combined with an X-ray telescope (e.g. Lynx) ability to detect the electromagnetic signal from 
accreting MBHs with masses in the seeding mass range gives the potential of constraining the seeding pathway. The difficulty that remains is that the competing scenarios of a rapidly accreting light seed versus a rapidly accreting heavy seed (earlier in its evolution) must still be disentangled. To combat this complexity, a sufficient sample of \ac{GW} and electromagnetic detections at $z \gtrsim 10$ will be required. This statistical sample can then be compared against theoretical models to ascertain the seeding mechanism. The dynamics of light seeds is such that they are not expected to easily sink to the centre of galaxies \citep{Tremmel_2015, Pfister_2019, Beckmann_2022} and hence the prevalence of MBHs with masses greater than $10^5 \ \Ms$ may be more limited in the case where \textit{all} seeds originate from light seeds. Seeds born from heavy seeds, on the other hand, have a greater 
probability of accreting as they can more easily sink to the potential centre. Additionally, the relative fraction of electromagnetic signals versus merging signals may help to differentiate between models. In the case where both light and heavy seeds struggle to grow then the only signal of the existence of MBHs will come from \acp{GW} and, if this is the case, the heavy seed scenario will be the dominant pathway \citep{Hartwig_2018}. \\
\indent As discussed at the beginning of this Section, the low-$z$ environment may also provide strong circumstantial evidence of the seeding process. Dwarf galaxies present an excellent laboratory in which to investigate MBH seeds. As alluded to in the above paragraph the dynamics of MBHs with masses less than $10^5 \ \Ms$ (and perhaps as high as $10^6 \ \Ms$) means the MBHs find it difficult to sink to the centre of their host galaxies where the densities are highest. Dwarf galaxies are by definition low-mass galaxies, if they follow the
empirical $M-\sigma$ relation, then they should host central MBHs with masses close to or at least within the range of (heavy) seed mass black holes. In dwarf galaxies with low central densities and/or dwarf galaxies that have had relatively quiescent merger histories then any 
MBH could be viewed as carrying information on the seeding mass. This is particularly true for off-nuclear MBHs that are detected, since their accretion is likely to have been relatively light (relative to its initial mass). \\
\indent Over the last decade there has been an explosion in the detection of MBHs in dwarf galaxies \citep{2013ApJ...775..116R, 2017IJMPD..2630021M, 2018ApJ...863....1C, 2018MNRAS.478.2576M, 2019MNRAS.488..685M, 2020ApJ...898L..30M, Reines_2020}, as extensively presented in previous {\color{blue} Chapter \textbf{2}}, to the point where the existence of MBHs in dwarfs is no longer the question but has instead transformed into a determination of the occupation fraction of MBHs in dwarf galaxies. The occupation fraction of MBHs in dwarf galaxies is important in understanding MBH
seeding as the mass scale at which the occupation fraction drops to zero indicates the minimum galaxy mass that can form or host a MBH. Most MBHs in dwarf galaxies are currently found using optical emission line diagnostics which attempt to differentiate between star light and the emission from accreting MBHs. Initial work on using dwarf galaxies as laboratories in which to test 
seeding models was pioneered by \cite{Volonteri_2008} and \cite{VanWassenhove_2010}. Using semi-analytic models they used the predicted number density of different seeding models to predict the occupation fraction of MBHs in today's dwarf galaxies surrounding the Milky Way. Of course this model could be run in reverse as well if we were able to detect all MBH in surrounding dwarf galaxies. \\
\indent In this regard, a recent claim by \cite{Bustamante-Rosell_2021} of a MBH in the centre of the Leo I galaxies is intriguing. \citep{Bustamante-Rosell_2021} claim the black hole
mass is of the order of $10^6 \ \Ms$, making this dwarf galaxy extremely `obese' in terms of its black hole to stellar mass ratio. Interestingly, this obesity of black hole mass to
stellar ratio ratio is predicted by heavy seed models \citep{Agarwal_2013} and so must be taken seriously. If this claim holds then by itself this constitutes extremely strong evidence for a heavy seed model. Building on the Leo I claim, \citet{Regan_2023} argue that the detection of multiple MBHs in a dwarf galaxy would be near definitive proof of the heavy seed channel as there is no plausible way to populate a dwarf galaxy with multiple MBHs by means other than a heavy seed channel (most likely via the intermediate stage of very massive stars). This postulate is graphically illustrated in Figure \ref{Fig:LeoI_MBH}. On the left hand side, Figure 
\ref{Fig:LeoI_MBH} \cite{Regan_2023} illustrates what the population demographics should look like in a dwarf galaxy where the seeding is provided by a single light seed, allowed to grow  spectacularly by accretion, by interacting with a number of dense gas clouds or indeed a light or heavy seed that grows 
via dynamical interactions at the centre of a dense nuclear cluster. In both cases these models predict a single MBH should exist in the dwarf galaxy (perhaps but not necessarily) at the centre. On the right side is illustrated the case where a heavy seed model prevails. In this case, under the assumption that mild fragmentation results in a spread of initial heavy seeds (e.g. \citep{Regan_2020}), then the dwarf galaxy should contain two or more heavy seeds, i.e. two or more MBHs. Therefore, if a dwarf galaxy can be found with multiple MBHs, then this provides compelling evidence for the heavy seed pathway -- though the limiting factor of MBH deposited by halo mergers must also be disentangled. \\
\indent In summary, low-$z$ dwarf galaxies are a powerful complimentary tool to high-$z$ direct observations of the seeding process. Near-future, sensitive, observations of MBHs in dwarf galaxies are likely to be pivotal in our understanding of MBH demographics. \\

In this Section, we have extensively covered the formation pathways for MBHs, and we have highligted that upcoming \ac{GW} detections from facilities as LISA and ET may prove to be extremely valuable for disentangling among different formation scenarios. In the next Section, we focus on the pathways that can lead to the formation of MBH binaries that can eventually become detectable \ac{GW} sources and help us shedding light on the cosmological evolution of MBHs.

\part{Massive black hole binaries \\ \Large{Bonetti, Lupi  and Franchini}}
\section{Introduction}
According to the hierarchical formation scenario, galaxies aggregate via accretion of matter from large-scale filaments and repeated mergers of smaller structures \citep{2010MNRAS.406.2267F,2021MNRAS.501.3215O}, thus suggesting that, if MBHs are hosted in these systems, MBHs would definitely pair and coalesce, becoming sources of \ac{GW} radiation that might be observed by LISA, ET, or  Pulsar Timing Array experiments \citep[e.g][]{2016PhRvD..93b4003K,2019MNRAS.486.2336D,2020ApJ...904...16B,2021MNRAS.500.4095V}.
However, the exact number and properties of these sources depends on several poorly constrained parameters associated to the MBH population (as the initial mass or the occupation fraction), but also on the host and environment properties, which are the topics of ongoing research by the scientific community.

The current picture behind MBH pairing and coalescence dates back to the seminal work by \citet{Begelman80}, who first explored the dynamics of MBHs in merging galaxies, proposing the existence of three main phases: a large-scale orbital decay driven by dynamical friction against the different galaxy components (the dark matter halo on tens of kpc scales, gas and stars on kpc scales), a shrinking phase after a gravitationally bound binary has formed (on pc and sub-pc scales), driven by the interaction with gas, stars, or potentially other MBHs, and a final relativistic phase, when the MBHs coalesce thanks to the energy loss due to GW emission.

In the following subsections, we detail the dynamical evolution of MBH binaries across these different stages. We first  focus on the large-scale (dynamical-friction driven) stage of the evolution, which plays a crucial role in the definition of the final fate of the MBH pair. This stage starts at the onset of the galaxy merger and ends roughly when the two MBHs form a bound Keplerian system, typically at or below pc scales (depending on the MBH mass). Initially, this large-scale evolutionary phase was considered the simplest one to describe. However, over the years, a much more complex picture has emerged, due to the strong dependence of the orbital decay on the physical processes behind galaxy formation, which are still extremely uncertain.

\subsection{The large-scale inspiral}

\subsubsection{The simplest approximation: the collisionless isothermal sphere}
As already mentioned, galaxies interact and merge during the cosmic assembly history, followed by their central MBHs. While the two merging galaxies are still separate entities, the interaction between the two galaxy nuclei (and the central MBHs) can be approximated as the orbital encounter of a massive perturber (usually the nucleus of the smallest galaxy, containing the MBH) in a diffuse collisionless medium (the most massive galaxy in the pair). Since the nucleus/MBH is much more massive than the single particles (dark matter or stars) of the main galaxy, the dynamics can be easily described by dynamical friction \citep{1943ApJ....97..255C}, i.e. the drag resulting from particles that have been deflected by the perturber gravitational pull, creating a trail behind it. This drag results in a deceleration of the perturber, which loses angular momentum and spirals towards the centre of the main galaxy. In the ideal case of an infinite homogeneous medium characterised by a density $\rho$ and an isotropic Maxwellian velocity distribution of particles, the resulting drag force can be expressed as
\begin{equation}
    \mathbf{F}_{\rm DF} \propto -M_{\rm p}^2 \rho \frac{\mathbf{v}}{v^3} \ln{\Lambda} f(v/\sigma),
\end{equation}
where $M_{\rm p}$ is the perturber mass, $\mathbf{v}$ is the perturber relative velocity with respect to the homogeneous background, $v$ its speed, $\ln\Lambda\sim 10$ is the Coulomb logarithm, and $f(v/\sigma)$ is a function which depends on the actual background velocity distribution, and scales as $(v/\sigma)^3$ for $v/\sigma<<1$ and is 1 for larger ratios. Note here that the perturber mass entering this formula is not necessarily the MBH alone, but in principle can also be the entire galaxy, as long as it can be considered a compact bound system compared to the background. In the simplest case of a MBH on a circular orbit at radius $r$ within a singular isothermal sphere (where $\rho\propto r^{-2}$ and the velocity distribution is exactly Maxwellian), the orbital decay timescale in Gyr is \citep{1987gady.book.....B}
\begin{equation}
    \tau_{\rm DF,Gyr} \approx \frac{8}{\ln \Lambda} r_{\rm kpc}^2\sigma_{200} M_{\rm p,7}^{-1},
\end{equation}
where $r_{\rm kpc}$ is the radius in kpc, $\sigma_{200}$ is the velocity dispersion of the background in units of 200~km~s$^{-1}$, and $M_{\rm p,7}$ is the perturber mass in units of $10^7\rm\, \Ms$.
For instance, in a galaxy with $\sigma_{200}=1$, a MBH with the unit mass above at 10~kpc from the center would take about $80$~Gyr to reach the centre, a time much longer than the age of the Universe.

Obviously, such a simple analysis does not take into account the complexity behind galaxy formation and evolution, in particular the 
role of the mass growth of the system over cosmic time, the presence of density and velocity anisotropies, the role of gas, the mass loss of the perturber (in case it is not a naked MBH), and the interplay among gas, MBHs, and stars in shaping galaxies.
These effects are expected to play a crucial role in the evolution, significantly changing the orbital decay timescale, and have profound implications for the detectability of merging MBHs. In the following, we will discuss some of these additional effects, with the aim of giving a broad picture  of the large-scale evolution of MBH pairs.

\subsubsection{The impact of different density profiles}
Obviously, real galaxies feature much more complex profiles than a singular isothermal sphere, even if we only consider the dark matter halo, which is typically described by a Navarro-Frenk-White profile $\rho \propto r^{-1}(1+r)^{-2}$. More detailed analyses also found that some galaxies exhibit cored dark matter profiles  \citep{2015AJ....149..180O} which could be due to the nature of the dark matter itself \citep{2017PhRvD..95d3541H} or to baryonic effects \citep{2010Natur.463..203G}. Such differences can have crucial consequences on the pairing timescale, as recently shown by 
\citet{2018ApJ...864L..19T}, which simulated the merger between dwarf galaxies with different density profiles, finding that cored ones would cause the secondary MBH stalling at relatively large separations (50–100 pc), contrarily to non-cored density profiles.

\subsubsection{Axisymmetric distributions: the effect of discs}\label{sec:discs_large_scale}
A large fraction of galaxies in our Universe exhibit rotationally-suppoorted disc-like strucutres rather than dispersion-dominated, isotropic distributions. Even without considering gas, the stellar distribution in disc galaxies does not match the conditions assumed by \citet{1943ApJ....97..255C}, in particular because of the net rotational motion in the disc plane. Several studies have investigated the effect of galactic discs (~1–10 kpc) on the pairing of two MBHs into a binary system, suggesting that dynamical friction in rotating discs usually drive initially prograde and eccentric orbits towards circularization, and it reverses the angular momentum of counter-rotating trajectories, then again promoting circularization \citep[see, e.g][]{2006MNRAS.367..103D,2011ApJ...729...85C,2020MNRAS.494.3053B} and thus affecting the orbital decay timescale.

\subsubsection{Dynamical friction in gaseous environments}
Although most local galaxies are dominated by collisionless components, with gas being subdominant, at high-redshift galaxies were very rich in gas \citep[][]{2018ApJ...853..179T,2020ApJ...902..110D}. At first order, the orbital decay of a massive perturber into a gas-dominated main galaxy can be considered comparable to that in a collisionless system, described above. 
However, we have to consider that the wake lagging behind a massive perturber would be affected by collisional effects (e.g. pressure and coupling with radiation) which could alter it, indirectly affecting the impact of dynamical friction. In fact, for similar densities, the gas-driven dynamical friction can be weaker, stronger, or of similar strength compared to the stellar-driven case, depending on the Mach number of the perturber \citep{1999ApJ...513..252O}. The overall impact of gas is however still debated, as gas is typically consumed during galaxy mergers in rapid star formation bursts, suppressing it \citep{2017MNRAS.471.3646P}, but in the case of equal-mass gas-rich mergers, in which gas is able to overcome the stellar contribution to dynamical friction \cite{2013MNRAS.429.3114C}.
Another important effect to consider is the effect of feedback (by stars and MBHs).
Radiative feedback by an inspiralling MBH can have a fundamental role on the pairing timescale, as the ionizing radiation that emerges from the innermost parts of the MBHs’ accretion flows pushes the gas away from the MBH and creates an ionized region larger than the characteristic size of the gaseous dynamical friction wake around it, with a dense shell of gas in front of the MBH, as a consequence of the snowplow effect. In these conditions, the
dominant contribution to the MBH acceleration comes from the dense shell ahead, which acts as a "negative dynamical friction", accelerating the MBH and lengthening the inspiral time \citep[e.g.][]{2017ApJ...838..103P, 2020MNRAS.492.2755G, 2020MNRAS.496.1909T}. 
In low-mass galaxies, the impact of feedback becomes even worse, as supernova explosions combined with the radiation background coming from the Universe reionisation evaporate and expel the gas, suppressing the growth of the MBH and its orbital decay \citep{2019MNRAS.486.2336D}.
Finally, gas in galaxies can form clumps of different masses and sizes, which typically represent the large-scale structures in which molecular clouds, hence new stars will form. When a MBH approaches one of these clumps, its orbit can be severely altered, depending on the properties of the clumps. In observed galaxies, clumps have typical masses of $10^5-10^7\Ms$ (consistent with giant molecular clouds in the local Universe), although more massive ones (up to $10^8-9\Ms$) can potentially form at high redshift \citep{2017MNRAS.464.2952T,2017MNRAS.468.4792T,2019MNRAS.489.2792Z}. 
The role of clumps has been commonly addressed via numerical simulations, that showed that the clumpiness of the gaseous medium renders the orbital decay highly stochastic, with situations in which the MBH separation does not shrink \citep{2015MNRAS.449..494R}, and others in which the decay is promoted \citep{2013ApJ...777L..14F, 2015ApJ...811...59D, 2015MNRAS.453.3437L}.
In the first case, the main reasons behind a longer orbital decay are the scatterings out of the disc plane, in regions where the density is much lower and the dynamical friction less efficient, or the impact of feedback, which can open cavities that reduce the density around the MBH \citep{2017ApJ...838...13S}. 

\subsubsection{Time evolution of the main galaxy}

In the case of a long-lasting inspiral, i.e. when dynamical friction is inefficient, the galaxy itself might change its structure due to different processes as new star formation (and stellar feedback), gas inflows from large scales, or even multiple mergers. These changes can significantly alter the orbital decay of the secondary MBH, either accelerating it or slowing it down. Although simple analytical arguments can be made to infer the change in the MBH pairing timescale for very simple density distributions and evolutions \citep{2020MNRAS.498.2219V}, more realistic conditions necessarily need to be determined numerically. In particular, while the growth of a disc galaxy experiencing minor mergers can be easily modelled using the mass and scale radii evolution of its main components (Varisco et al. in preparation) using semi-analytic models, strong perturbations can only be followed using N-body simulations.

\subsubsection{Global asymmetries: galactic bars}
Globally asymmetric matter distributions like stellar bars \citep{1986ApJ...300...93W,1989MNRAS.239..549W} generate torques which become particularly important at resonances between the perturber’s orbital frequency and the orbital frequency of
the background matter. Bars are extremely common in low-redshift disc galaxies, contributing to about half of massive  ($M_*\gtrsim 10^{10} \, \Ms$) discs  
\citep[e.g.][and references therein]{2016A&A...595A..67C}. Although the fraction of observed barred galaxies at high redshift is still highly unconstrained  \citep[see, e.g.][]{2008ApJ...675.1141S, 2014MNRAS.438.2882M, 2014MNRAS.445.3466S}, the common idea (based on both simulations and observations) is that they can be hosted in massive galaxies at all redshifts, with an increasing mass threshold for their formation as a function of redshift \citep{2015A&A...580A.116G}. Due to their strong perturbation on the background distribution, bars are expected to significantly affect MBH pairing during mergers. In particular, this has been demonstrated using hydrodynamical simulations  \citep{2017MNRAS.467.4080F} in which a population of secondary MBHs was placed at different radii and at different angles with respect to a forming bar, and in semi-analytical models \citep{2022MNRAS.512.3365B}. In both cases, a stochastic behaviour in the pairing time-scale was found, with some of the secondary MBHs being pushed towards the centre of the main galaxy, and others being ejected by a slingshot with the bar. Noticeably, it was found that the orbital
decay of the secondary MBHs was dominated by the global torque provided by the bar rather than by the local effect of dynamical friction. 

The different aspects discussed in the previous sections clearly show how the pairing process of MBH binaries is extremely challenging and requires a detailed description of the background matter distribution, accounting for all components and their effect on the orbital evolution of the perturber. Unfortunately, such an endeavour cannot be made via analytic calculations, but it requires semi-analytical or numerical approaches, which are more computationally expensive, thus limiting the parameter space exploration. In addition, the limited resolution of current models does not allow us to properly describe the complexity of baryonic physics during galaxy formation, increasing the uncertainty on the results. Finally, another level of uncertainty that has to be mentioned comes from the fact that, in the first stages of galaxy mergers (when the MBHs are at kpc-scale distances), the secondary MBH could be still surrounded by its galaxy nucleus, which increases the total perturber mass and potentially accelerates the decay. As the pairing proceeds, however, this additional mass is stripped, altering the pairing timescale and leaving a naked MBH behind. Properly describing the stripping process in semi-analytical and numerical models is still challenging, and this further limits our ability to predict the pairing efficiency of MBHs.

\subsubsection{The transition to the small scale inspiral}

As long as dynamical friction or other processes are efficient enough to ensure both MBHs reach the center of the galaxy merger remnant, the self-gravity of the two MBHs is no longer negligible and they can form a bound pair. This happens at a separation of the order of $\sim G M/\sigma^2$ (roughly the influence radius of the binary system) where $M$ is the total binary mass and $\sigma$ the characteristic velocity dispersion of the stellar background \citep{1980Natur.287..307B,2001ApJ...563...34M}. 
The subsequent evolution of the newly-formed MBH binary is  determined by the specific features of the environment hosting it, but in general we can identify two main pathways. In galaxies with very little gas we expect that the MBH binary evolution is mainly driven by interaction with the stars in the vicinity of the binary (a process known as stellar hardening), while in gas-rich galaxies hydrodynamical torques can dominate the radial inspiral (or possibly outspiral!). Despite the evolution is generally analysed separately for each of the two viable channels, in realistic galaxies we expect that stellar and gaseous hardening processes operate at the same time, therefore their joint effect has also to be taken into account to ultimately understand the MBH binary path to coalescence. Finally, the binary evolution can be affected by one (or more!) further incoming MBH brought in by an additional merger; this is especially likely to occur if the  stellar or gaseous hardening process is very inefficient.
In the following, we detail the physics behind these different small scale processes that can drive the binary towards its GW-driven coalescence.

\subsection{The small-scale inspiral: hardening in stellar backgrounds}

When the density profile of a galactic nucleus is dominated by the stellar background, the main mechanism that can shrink MBH binaries involves the interaction of the binary with stars in its vicinity. This interaction is purely gravitational and proceeds mainly through a series of three-body encounters in which stars generally intersect the MBH binary orbit and exchange energy and angular momentum \citep{1992MNRAS.259..115M, 1996NewA....1...35Q, 2006ApJ...651..392S}. Here we describe the basics of how such process actually works and how it connects to the specific conditions of different galactic environments. We also give an overview of the tools and techniques used so far to gain insight into such  evolution.

\subsubsection{Binary-star interaction: energy and angular momentum exchanges}\label{sec:dyn_proc_MBHB}

Before considering the specific situation involving a MBH binary, we start by analysing the typical phenomenology of three-body encounters, for now setting no constraints on the masses involved. Moreover, without loss of generality, we assume that the three bodies are well represented by point particles within the framework of Newtonian dynamics.

We start from an initial state characterised by a well separated bound binary and a third body. In this framework, the interaction is driven by conservative gravitational forces only, therefore total energy and angular momentum are conserved, but nothing prevents an exchange of those quantities between the two sub-components of the systems. Under the above conditions, the ensamble of possible outcomes can be summarised as follows \citep[see e.g.][]{1983ApJ...268..319H, 1983AJ.....88.1549H}:
\begin{itemize}
    \item fly-by: the interaction resolves quickly during a single passage of the incoming third body, which after interacting with the binary returns to infinity. Due to the energy exchange, the binary orbital properties can be substantially altered. If $K$ is the kinetic energy at infinity in the centre of mass  frame, while $E_b = G m_1 m_2/(2a)$ is the binary binding energy (with $m_1,m_2,a$ being the binary masses and its semi-major axis), then energy conservation ensures
    \begin{equation}
        K_i - E_{b,i} = K_f - E_{b,f} \longrightarrow K_f - K_i = E_{b,f} - E_{b,i}.
    \end{equation}
    Therefore when $K_f>K_i$ (i.e. the third body gains energy), we have
    \begin{equation}
        E_{b,f} = \frac{G m_1 m_2}{2 a_f} > E_{b,i} = \frac{G m_1 m_2}{2 a_i},
    \end{equation}
    i.e. $a_f < a_i$, meaning that the binary has shrunk. The opposite happens if the interaction leads to $K_f<K_i$, i.e. if the third body gives energy to the binary.
    
    \item Exchange: the incoming third body swaps with one of the bodies originally bound in the binary. The probability of exchange is much higher when the third body is more massive than at least one of the binary components. Also in this case the binary generally increases its binding energy, i.e. if $m_2 < m_3 < m_1$ then
    \begin{equation}
        E_{b,f} = \frac{G m_1 m_3}{2 a} > E_{b,i} = \frac{G m_1 m_2}{2 a}.
    \end{equation}
    
    \item Resonant interaction: when the total energy is negative, the third body and the binary can form a meta-stable triplet showing an extremely complex dynamics. After 10-100 (or even more) dynamical times\footnote{The dynamical time of a system is a characteristic timescale for a particular change to take place for that specific system. There can be different dynamical times for every systems and they are generally important because they play a major role in determining stability. For the case of a MBH binary, the most important dynamical time is represented by the orbital period, i.e the time over which the two binary components complete an orbit.} the systems generally disrupts, again leaving a bound binary and an escaping body. The end state is not intrinsically different from what produced by the previous two outcomes (fly-by and exchange), but the remarkable difference is how this is attained; the strong and chaotic resonant interaction, being extremely sensitive to the initial condition, erases any memory of the initial state. This is why chaos makes essentially impossible any analytical mapping between initial and final states, making numerical methods the only feasible way to investigate the full problem.
    
    \item Ionisation: finally, when the total energy of the system is positive it may happen that all the three bodies escape in different directions leaving no bound components. This happens if the velocity of the incoming body at infinity (prior to the interaction) exceeds the critical velocity 
    \begin{equation}
        v_c^2 = \frac{G m_1 m_2 (m_1+m_2+m_3)}{m_3 (m_1+m_2) a},
    \end{equation}
    derived under the assumption that total energy at infinity equals zero, i.e.\footnote{At infinity there is no potential energy contribution between the third body and the binary, thus only kinetic energy is non zero. In Eq.~\ref{eq:energy_inf}, it is expressed as the kinetic energy of an effective two-body system with components $m_1+m_2$ and $m_3$ and reduced mass $\mu = (m_1+m_2)m_3 / (m_1+m_2+m_3)$.}
    \begin{equation}\label{eq:energy_inf}
        E_f = \frac{1}{2} \frac{(m_1+m_2) m_3}{m_1+m_2+m_3} v_c^2 - \frac{G m_1 m_2}{2 a} = 0.
    \end{equation}
    
\end{itemize}

If the above outcomes tell us what we can expect from a single three-body encounter, when considering real astrophysical binaries we need to remember that those live in galactic environments and not in vacuum. Therefore a binary generally undergoes many encounters with other objects (mainly stars) found in its vicinity. On a statistical base, binaries in astrophysical environments are separated in two categories: hard and soft binaries. The threshold diving them is set by the characteristics of the environment hosting them. Hard binaries are defined as those whose binding energy is larger than the average kinetic energy of the other components of the systems, i.e. when
\begin{equation}\label{eq:hard_bin_energy}
    E_b > \frac{1}{2} \langle m \rangle \sigma^2,
\end{equation}
where $\langle m \rangle$ is the typical mass of background objects and $\sigma$ their characteristic velocity dispersion. Conversely, soft binaries have less binding energy than the average energy of other stars. As stated by Heggie's law, hard binaries tend to become harder, while soft binaries generally further soften \citep{1975MNRAS.173..729H}. This means that, statistically speaking, as a consequence of many three-body encounters hard binaries tend to get tighter and tighter, while soft binaries loose binding energy and eventually dissolve. We are of course interested in binaries that live long, we therefore focus our attention on hard binaries and specifically how the ``hardening process'' shrinks them and how this depends on the environment.\footnote{A caveat for the soft-hard binary picture is represented by MBH binaries. Those binaries even when they are technically soft, they do not dissolve. This difference is due to the intrinsic mass of MBHs, which is orders of magnitude higher than stellar objects, making the ionisation of a MBH binary due to stellar scattering negligibly unlikely.}

\subsubsection{Binary hardening: analytical estimates and scattering experiments}

We consider a hard binary immersed in a field of stars. Turning Eq.~\ref{eq:hard_bin_energy} into a condition on the binary separation we get that binaries are hard when $a\leq a_h$, where $a_h = G \mu /(4\sigma^2)$ is known as the hardening radius. We can then express the interaction rate with the field stars as the product of their number density $n$ times their typical velocity $\sigma$ times the binary cross-section $\Sigma_{\rm cross}$, i.e. 

\begin{equation}\label{eq:interaction_rate}
    \frac{dN}{dt} = n \Sigma_{\rm cross}\sigma.
\end{equation}

The cross-section of the binary can be expressed as a geometrical cross-section, i.e. $\Sigma_{\rm cross} = \pi b_{\rm max}^2$, where $b_{\rm max}$ is a maximum impact parameter, which differs from the standard impact parameters one would define at infinity (say $b$). The difference is due to the gravitational focusing, i.e. the non-negligible deflection from a straight line of the trajectory of the approaching third body, which makes the effective pericentre $r_p$ much smaller than $b$ (at least for energetic encounters). The maximum impact parameter is related to the pericentre $r_p$ by 

\begin{equation}\label{eq:bmax}
    b_{\rm max}^2= r_p^2\left(1 + \frac{2G (m_1+m_2+m_3)}{r_p \sigma^2}\right).
\end{equation}

We can further assume for simplicity (but still describing relevant astrophysical situations) that the binary is hard, the mass of the third body is much smaller than $m_1,m_2$ and that the effective pericentre is $\lesssim 2a$ (i.e. just to focus on the most energetic, close encounters).
Under such assumptions $2G (m_1+m_2+m_3)/(r_p \sigma^2) \gg 1$, therefore for the interaction rate we get
\begin{equation}\label{eq:interaction_rate2}
    \frac{dN}{dt} = \frac{4\pi G (m_1+m_2) n a}{\sigma}.
\end{equation}

To actually get how a binary evolves due to three-body encounters we need then to couple the interaction rate with the amount of energy each encounter can exchange, on average. For each encounter, the average binding energy variation can be related to the binary energy through \citep{1992MNRAS.259..115M, 1996NewA....1...35Q}
\begin{equation}\label{eq:deltaE}
    \langle \Delta E_b \rangle = \eta \frac{m_3}{m_1+m_2} E_b, 
\end{equation}
where the dimensionless coefficient $\eta$ parametrises the energy flow and can be estimated through the numerical integration of the three-body equations of motion (see next).

Coupling Eq.~\ref{eq:interaction_rate2} and Eq.~\ref{eq:deltaE}, the rate of binding energy exchange is given by
\begin{equation}\label{eq:deltaE_rate}
    \frac{dE_b}{dt} = \frac{dN}{dt} \langle \Delta E_b \rangle = \frac{2\pi \eta G^2 m_1 m_2 \rho}{\sigma}, 
\end{equation}
where we have considered that $m_3 = \langle m \rangle$ is representative of the field star mass and therefore $\rho = n \langle m \rangle$. From Eq.~\ref{eq:deltaE_rate} we can note a very important feature of hard binaries: if the characteristics of the environment do not vary sensibly (i.e. $\rho$ and $\sigma$ are constant), \textit{hard binaries harden at a constant rate}. Finally, we can express the energy evolution in terms of the semi-major axis
\begin{equation}\label{eq:delta_a_inv_rate}
    \frac{d}{dt}\left(\frac{1}{a}\right) = \frac{2}{G m_1 m_2} \frac{dE_b}{dt} = \frac{4\pi \eta G \rho}{\sigma}, 
\end{equation}
or more explicitly
\begin{equation}\label{eq:da_dt}
    \frac{da}{dt} = - a^2\frac{G H \rho}{\sigma}, 
\end{equation}
where again $H = 4\pi\eta$ is a dimensionless coefficient that parametrise the energy exchange and can be measured numerically. We can note that the binary shrinking depends on the combination $\rho/\sigma$, i.e. the higher the density and the lower the velocity dispersion, the stronger is the hardening. Moreover, the hardening efficiency depends on $a^2$, meaning that as the binary gets tighter the efficiency of stellar hardening decreases accordingly.

Following similar calculations, also the eccentricity variation can be express in a form similar to Eq.~\ref{eq:da_dt}, specifically
\begin{equation}\label{eq:de_dt}
    \frac{de}{dt} = a\frac{G H K \rho}{\sigma},
\end{equation}
where the additional quantity $K$ is another dimensionless parameter that depends on the specific interactions between the MBH binaries and the stars.

To actually estimate how energy and angular momentum are drained from the MBH binary, we actually need to assess the numerical values of the parameters $H$ and $K$. This is generally attained through the scattering experiments, where a large number of single MBH binary-star interactions are simulated to properly cover all possible scattering configurations. The initial setup involves a MBH binary and a star that starts far away on an hyperbolic orbit.\footnote{The choice of hyperbolic orbits is simply motivated by the fact that stars initially are  not bound to the MBH binary. The velocities of stars are determined by the galactic potential well and their total energy is lower than zero only with respect to the galaxy system, but not with respect to the three-body system star-MBH binary.} When the star gets close to the binary, a complex three-body interaction starts, which results in the vast majority of cases in a quick ejection of the star, but in same cases can also produce metastable triples, eventually leading to the final ejection of the lighter body, i.e. the star. When this happens the fraction of energy and angular momentum drained from the MBH binary is recorded. 
Finally, by properly averaging over the whole ensable of simulations, and specifically over the velocity distribution of stars, the desired information about $H$ and $K$ is  extracted \citep[see][for a detailed description of the procedure]{1992MNRAS.259..115M,1996NewA....1...35Q,2006ApJ...651..392S}. 

\begin{figure}
    \centering
    \includegraphics[scale=0.37]{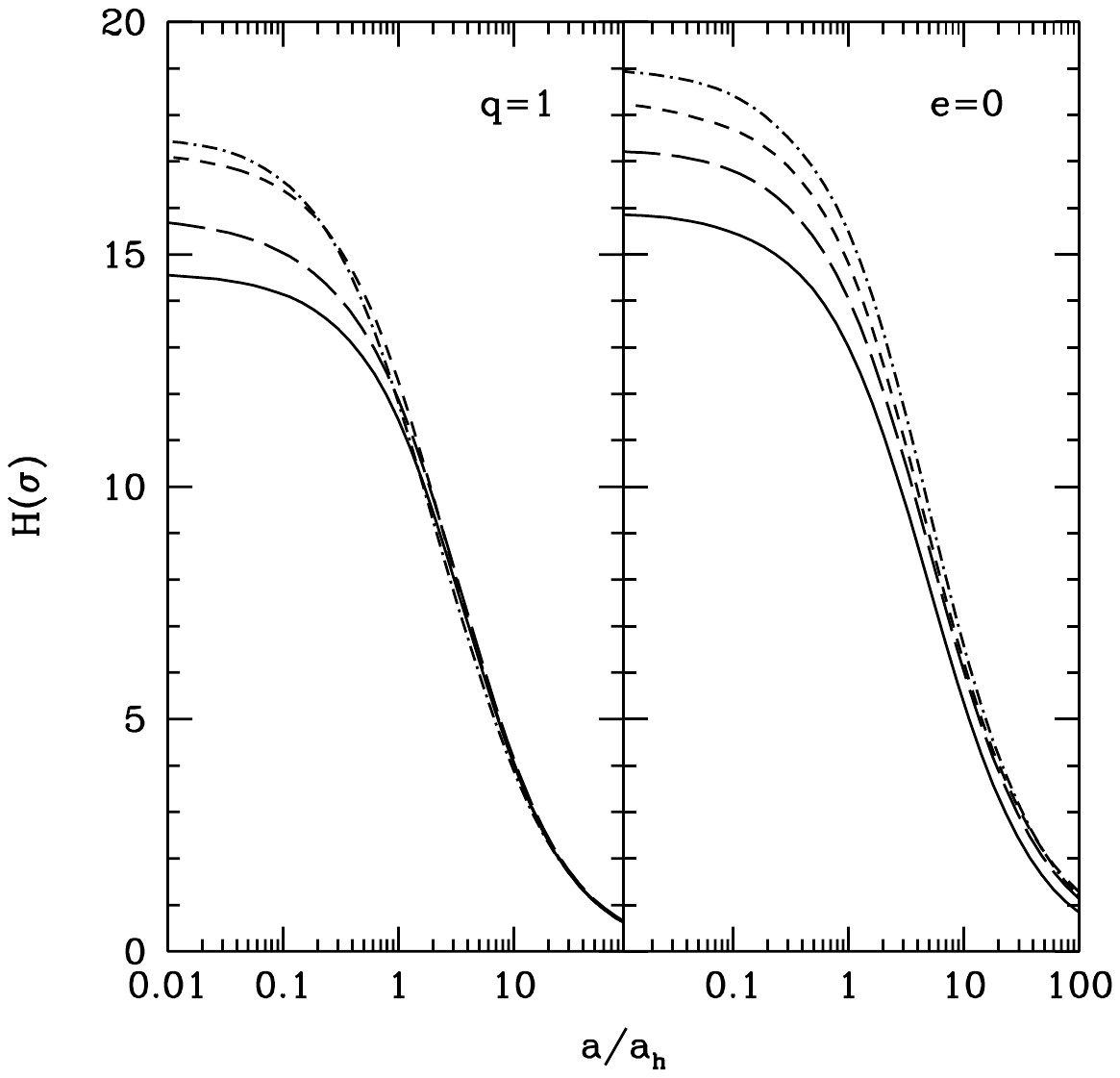}\includegraphics[scale=0.37]{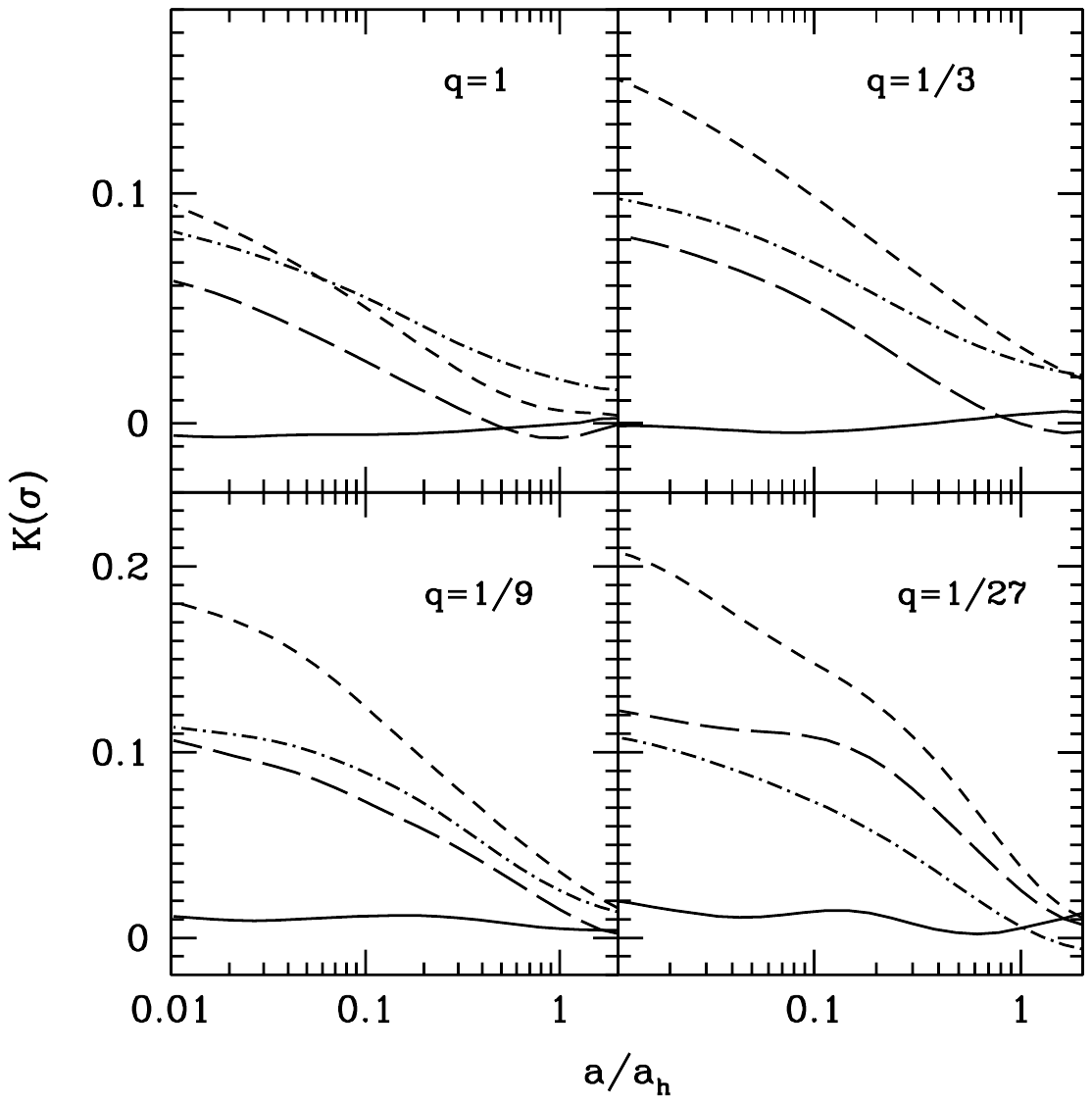}
    \caption{Binary hardening rate $H$ (left) and eccentricity growth rate $K$ (right) averaged over a Maxwellian velocity distribution as a function of binary hardness $a/a_h$. For $H$ different line styles refer either to different eccentricities $e = 0, 0.3, 0.6, 0.9$ (left) or to different mass-ratios $q = 1/3, 1/9, 1/27, 1/81$ (right). For $K$ different line styles denote various eccentricities $e = 0, 0.3, 0.6, 0.9$. Figure adapted from \citet[][see their Figures 3 and 4]{2006ApJ...651..392S}.}
    \label{fig:HK_rate}
\end{figure}

In Fig.~\ref{fig:HK_rate}, the values of $H,K$ as a function of the binary hardness are shown for different binary parameters. The hardening rate, parametrized through the $H$ parameter, is a rapidly growing function of the binary hardness, then when the binary is hard ($a<a_h$), $H$ saturates around numerical values between 15 and 20 depending on the mass-ratio of the binary. Concerning $K$, we can observe that for a circular binary (solid line in the plot), its value remains close to zero meaning that binaries with $e=0$ tend to stay circular during the hardening evolution. On the contrary, initially eccentric binaries tend to increase the eccentricity as they shrink and get harder.

\subsubsection{Evolution in realistic galactic environments}

The scattering experiments framework is designed such that the orbits of incoming stars have a pericentre within a factor of a few times the MBH binary separation. This choice is motivated by the fact that such technique aims at characterising how a MBH binary evolves when stars are efficiently scattered. However, when considering a realistic stellar distribution in a galaxy this is actually not the case. Indeed, only a very small fraction of the galactic stellar population has an angular momentum small enough to interact with the binary located at the very centre of the nucleus: only stars with angular momentum $J$ such that $J\leq J_{\rm lc} \approx \sqrt{2 G M_{\rm bin} a}$ can properly interact with the MBH binary and efficiently extract energy and angular momentum. Such stars (that again, are a small number compared to the total number of stars in galaxies) are known to belong to the so-called loss-cone\footnote{The loss-cone of a single MBH is defined as the region in the phase-space of energy and angular momentum, where the angular momentum of an incoming star is small enough that it can be captured by a MBH, either through direct collision or through the process of orbital decay. The loss-cone of a MBH binary is defined in the same way but is generally larger and depends on the binary separation.} of the binary \citep[see e.g.][]{2013degn.book.....M}.

As observed during scattering experiments, each star-binary interaction ultimately causes the ejection of the star, which leaves the binary neighbourhood with a high speed and generally without a way back. This implies that  many stars have to be ejected  to sensibly shrink a MBH binary, effectively lowering the density of stars that can further interact with the MBH binary and scouring a low density core \citep{2015ApJ...810...49V,2018MNRAS.477.2310B}. In other words, the binary hardening depletes the reservoir of stars in the loss-cone, jeopardising its own effectiveness. Therefore, in order to maintain the efficiency of the process (i.e. keep the loss-cone full), the loss-cone needs to be refilled somehow: with too few stars the binary shrinking cannot proceed on timescale shorter than the Hubble time (empty loss-cone regime).

The loss-cone repopulation can happen either through collisional or collisionless processes. The former involve a redistribution of energy and angular momentum among the objects that form a galaxy. This phenomenon is due to the fact that galaxies are not perfectly smooth systems, therefore gravitational encounters among the galaxy constituents can actually change the energy and angular momentum of each of them. Such changes, due to the endless small kicks that each object receives, happen on the timescale of two-body relaxation, 
\begin{equation} \label{eq:t2br}
    t_{\rm rel} \approx \frac{0.1 N}{\ln N} t_{\rm cross}
\end{equation}
which for the vast majority of massive galaxies (with a number of stars $N\sim 10^{11}$ and a typical crossing time $ t_{\rm cross} \sim 100$ Myr) is way longer than the Hubble time \citep{1987gady.book.....B}.\footnote{Still, though always quite long, relaxation can play a non-negligible role in determining the present structure of many stellar systems, such as globular clusters ($N \sim 10^5, t_{\rm cross} \sim 10^5$ yr, lifetime 10 Gyr), the central parsec of galaxies ($N \sim 10^6, t_{\rm cross} \sim 10^4$ yr, lifetime 10 Gyr), the centers of galaxy clusters ($N \sim 10^3, t_{\rm cross} \sim 1$ Gyr, lifetime 10 Gyr).} 
Therefore, an efficient repopulation of the loss-cone only relying on relaxation is generally not possible in large massive galactic systems, while it may be slightly more effective in less massive galaxies where  relaxation proceeds on a faster pace due to the combination of a smaller size of the host galaxy that makes the system more collisional and the fewer number of objects in the galaxy.
As a matter of fact, in spherically symmetric galaxies, where the motion of stars is dictated by the conservation of energy and all components of the angular momentum, loss-cone refilling can happen \textit{only} through the process of two-body relaxation, i.e. a collisional refilling whose timescale is typically quite long even exceeding the age of the Universe. Such feature lies behind the idea of the famous \textit{final parsec problem} \citep{2001ApJ...563...34M,2003ApJ...596..860M}. In its original theoretical statement, dating back to the seminal paper of \citet{1980Natur.287..307B}, bound MBH binaries, forming typically at parsec scales, cannot reach the GW-driven inspiral because of the severe loss cone depletion determined by the stellar hardening itself, which unavoidably make MBH binaries to stall around a fraction of the initial separation at formation. This was also confirmed numerically by early full N-body simulations of spherical galaxies, therefore providing a proof strengthening the actual existence of the problem \citep{2004ApJ...602...93M,2005ApJ...633..680B}.

\begin{figure}
    \centering
    \includegraphics[scale=0.4]{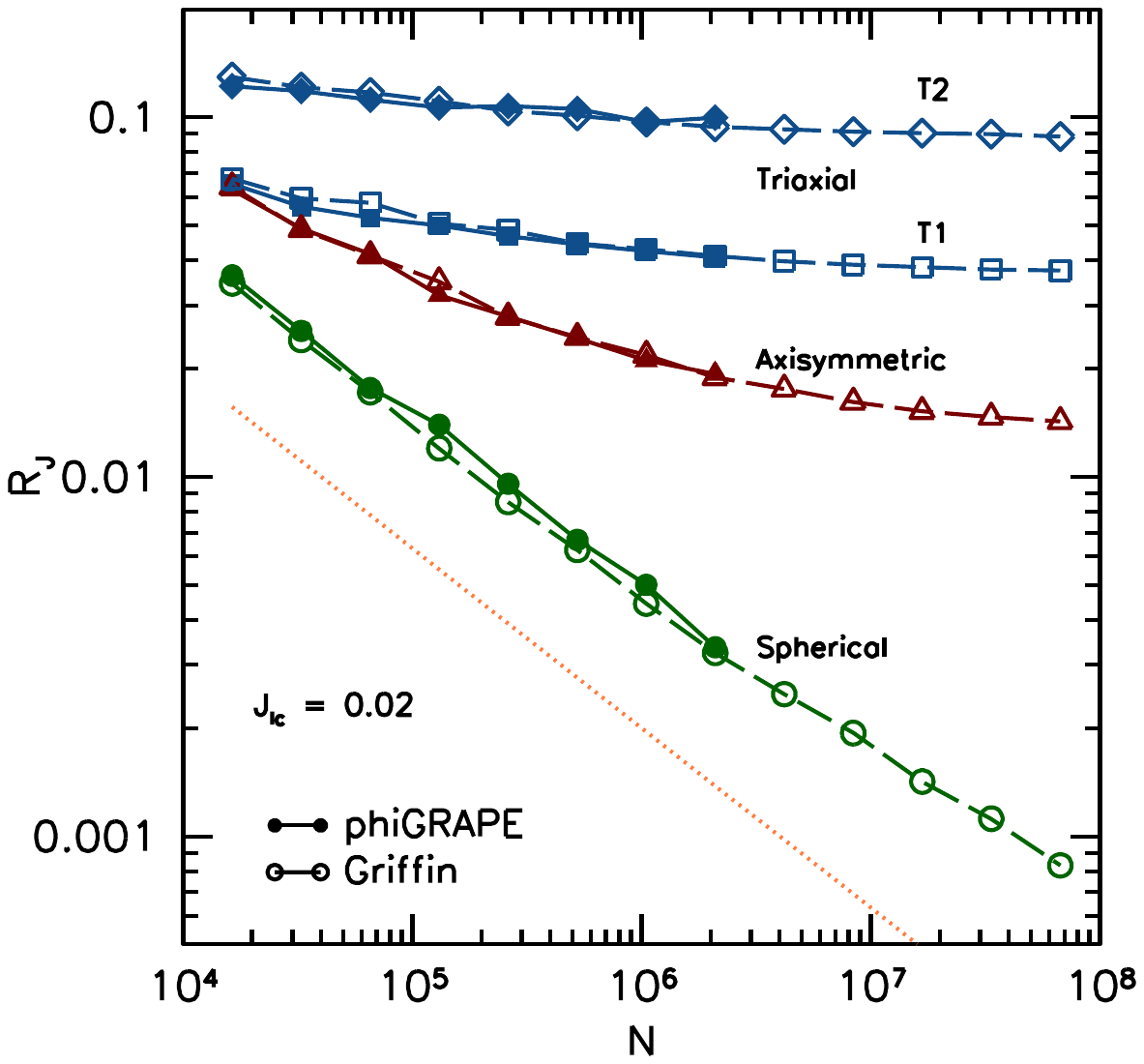}
    \caption{Refilling parameter $R_J = N(J\leq J_{\rm lc})/N$, defined as the number of stars with angular momentum lower than that of the loss-cone over total as a function of particle number for different models (spherical, axisymmetric and triaxial, as labelled). Different symbols refer to different state-of-the-art numerical codes, specifically phiGRAPE (direct N-body) and griffin (fast multipole method). The dotted line shows the clear $1/\sqrt{N}$ trend expected for collisional loss-cone repopulation solely due to two-body relaxation and typical of spherical profiles. Figure adapted from \citet{2017MNRAS.464.2301G} (their Figure 5).}
    \label{fig:refilling}
\end{figure}

Throughout the last decades, various evidences have been provided by the community concerning the fact that the final parsec problem actually would only occur in idealised, perfectly spherical galaxies, which actually do not exist in nature. Indeed, from the theory side, following the hierarchical build up of the cosmic structure, galaxies are expected to undergo merger events, yielding newly formed galaxies that naturally deviate from spherical symmetry \citep[see e.g.][]{2011ApJ...732...89K,2011ApJ...732L..26P}. Moreover, from the observational side, realistic galaxies show a quite different range of morphologies, typically not spherical and featuring disc-like shapes, oblate structures and even irregular traits. The complex non-spherical structure of galaxies allows for the existence of additional loss-cone refilling processes that are collisionless in nature.

Differently from collisional mechanisms, collisionless processes only depend on the global shape of the galactic structure and provide an efficient ``diffusion'' in angular momentum determined by large scale torques.\footnote{Those are directly generated by the global gravitational potential, which does not conserve angular momentum.} 
Such diffusion causes stellar angular momenta to change on a timescale which is generally longer than the radial period of stars, but much shorter than the relaxation time \citep{2013degn.book.....M}. 
The orbits typically found in non-spherical galaxies are rather diverse and appear in many different families. Specifically, in triaxial galaxies there exist a complete new class of centrophilic orbits, (box or pyramid orbits) that are able to attain arbitrarily low values of angular momentum without the need of invoking any relaxation. These orbits, by naturally crossing the galaxy centre, were identified as a promising way to easily repopulate the loss-cone in a collisionless fashion \citep{2002MNRAS.331..935Y, 2004ApJ...606..788M}, a possibility first observed through a suite of full N-body simulations by \citep{2006ApJ...642L..21B} and then confirmed and extended to systems with different MBH binary mass ratios, galaxy density profiles and orbits by various subsequent studies \citep{2012ApJ...749..147K, 2015ApJ...810...49V,2015MNRAS.454L..66S,2016ApJ...828...73K,2017MNRAS.464.2301G, 2018MNRAS.477.2310B}. As an example, as shown in Fig.~\ref{fig:refilling} and clearly demonstrated by \citet{2017MNRAS.464.2301G}, even a mild degree of triaxiality makes the loss-cone refilling much more effective when compared to the spherical case. 

In light of the above findings, the community has reached the broad consensus that MBH binaries forming in realistic galaxy mergers can reach the GW-driven inspiral within the Hubble time through the stellar hardening alone, \textit{i.e. there is no final parsec problem}. Still, the associated timescale heavily depends on characteristics of the host galaxy (e.g. the stellar density profile) and on the binary properties (e.g. the binary eccentricity). Those are generally difficult to pin down, therefore we are still left with a wide range of uncertainty. For instance, \citet{2019MNRAS.487.4985B} used scaling relations (calibrated with local Universe observations) to relate galactic properties to MBH binary masses in the range $10^5-10^8$ \citep[e.g.][]{2008MNRAS.386..864D,2013ARA&A..51..511K,2015ApJ...798...54G}. They found that depending on the steepness and normalisation of the stellar density profiles, the binary lifetime can range from 10 Myr up to more than 10 Gyr! Stellar hardening timescales are also expected to depend on redshift. Indeed, galaxies  in the early Universe are expected to be more concentrated in their dark matter halo and thus also on their stellar profile; such higher densities may increase the hardening efficiency at high $z$ \citep{2017JPhCS.840a2025M}.  

The MBH binary eccentricity is another important factor in determining the pace of the MBH binary path to coalescence. Indeed, a non-zero eccentricity can heavily boost the GW emission efficiency \citep{1964PhRv..136.1224P}, effectively increasing the scale at which GWs start to dominate the dynamics, with the timescales ratio for a circular binary versus an eccentric one with $e\sim 0.99$ that can be as high as $\sim 30$.
Stellar hardening is found to generally increase the binary eccentricity\footnote{The $K$ parameter found in scattering experiments is typically positive for high mass-ratio binaries, but the situation seems to reverse for very unequal binaries for which circularisation could happen \citep{2019ApJ...878...17R,2020MNRAS.493L.114B}.} if the latter at formation is not negligible (i.e. $\gtrsim 0.2-0.5$). Its subsequent growth as the binary shrinks will then anticipate the point where GWs can take over. On the contrary, if the eccentricity at formation is close to zero, then its growth is generally much milder. Despite the initial eccentricity of MBH binaries at their formation is very poorly constrained, there are indications towards a correlation with the eccentricity of the galaxy merger that brought the two MBHs in the same galaxy remnant \citep[see e.g.][]{2022MNRAS.511.4753G}. Another correlation seems to connect the initial eccentricity with the steepness of the stellar density profile, with steeper ones generally forming a lower eccentricity MBH binary \citep{2012ApJ...749..147K}. 

The above considerations apply to galaxy hosts that are supported by dispersion and have nearly isotropic velocity distributions. A game changer in the picture of the eccentricity evolution is the net rotation of the host galaxy (or at least of the nucleus). The net rotation sets a preferential direction in the galaxy, therefore co-rotating and counter-rotating orbits show qualitatively different behaviours. For instance, if a MBH co-rotates with the host, its large-scale dynamical friction inspiral is generally much faster due to lower relative velocities with respect to stars, which determines more energetic encounters able to efficiently extract the orbital energy. In addition, the orbit circularises fast, therefore in co-rotating nuclei we should expect nearly circular binaries at their formation, as already mentioned in Sec.~\ref{sec:discs_large_scale}. The opposite happens for counter-rotating orbits, which maintain their eccentricity and ultimately yield eccentric bound binaries. As they evolve further due to stellar hardening the eccentricity is expected to increase even more, but at the same time the binary orbital angular momentum is bound to flip and align with the angular momentum of the nucleus, again yielding co-rotating objects \citep{2012MNRAS.420L..38G,2017ApJ...837..135R,2020MNRAS.492..256K}. From the above findings, it seems that MBH binaries evolving in rotating environments should generally end up being almost co-rotating with their host and showing a quite low eccentricity, a fact quite relevant for the subsequent GW-driven evolution. Still, the associated timescale and the degree of binary alignment can vary depending on the host and binary properties, again leaving ample space to many different configurations.

A final aspect connected with the stellar hardening of MBH binaries is the possible influence of any kind of perturbers found in the galaxy nucleus. 
If the perturbers are normal stars, the main outcome we expect has been discussed in detail in the previous paragraphs.. Yet, though generally tiny, the scattering of stars is also responsible for a recoil of the MBH binary centre of mass, which should not be expected to perfectly remain at rest at the very centre of the galaxy potential well. Indeed, the MBH binary instantaneous centre of mass velocity is endlessly perturbed by the three-body encounters with the close-by stars, with this perturbation balanced by dynamical friction acting as a restoring force \citep{2001ApJ...556..245M}. The MBH binary centre of mass is therefore expected to wander around the galaxy center, being effectively subjected to a Brownian motion. The magnitude of the displacement actually depends on the ratio between the mass of the MBH binary and that of the perturber (typically a star) and it results to be effective only when $M/m_\star \leq 10^3$, thus pertaining only MBH binaries with a fairly small mass \citep{2016MNRAS.461.1023B}. For such binaries, the centre of mass wandering can enhance  the hardening rate. By moving around, the MBH binary can interact with a more numerous sample of stars, which   would not have found their way towards the MBH binary loss cone in spherical systems owing to inefficient relaxation \citep[see e.g.][]{1997NewA....2..533Q}. Nevertheless, for triaxial systems, for which the loss-cone is always full, the effect of the MBH binary Brownian wandering appears to be rather negligible. 
The situation can be different in rotating nuclei. Further to proceed at a faster rate, the stellar hardening in co-rotating nuclei seems also to displace MBH binaries from the centre in a coherent way (i.e. not a random Brownian motion). The physical interpretation of the phenomenon is the following: MBH binaries that co-rotate with their host encounter stars that generally give away angular momentum aligned with that of the binary. If a MBH binary is circular and it is shrinking due to hardening it cannot further increase its internal angular momentum. Such extra angular momentum is then acquired by the binary centre of mass, which inevitably settles on a rotating orbit around the centre. This situation does not happen for counter-rotating MBH binaries, which actually stay anchored to the very centre of the nucleus and shrink at a slower pace \citep{2021MNRAS.508.1533V}. 

When perturbers of much larger mass (compared to stars) are located in  the galaxy, their effect on MBH binaries can be more relevant  \citep{2007ApJ...656..709P}. ``Massive'' perturbers may come in different forms: e.g. as star clusters, giant molecular clouds or even additional MBHs (see next), all with masses close to or above that of the  MBHs forming the binary.
Given their large mass, such massive perturbers can influence the hardening phase of a MBH binary in different ways, some more direct than others.
For example,  they can efficiently scatter new stars into the loss-cone. Being much more massive than single stars and being there in fewer number, they can effectively shorten the two-body relaxation time and reduce the impact of loss-cone depletion; this results in more stars being sent toward the central region, so that the MBH binary hardening rate is enhanced \citep{1951ApJ...114..385S}. 
Furthermore, when massive perturbers come close enough to the binary, they may exert a strong perturbation on the MBH binary, possibly displacing it from the galaxy centre and again boosting the flux of stars reaching the loss-cone. Finally, if the massive perturber is an inspiralling stellar cluster, once it reaches the galactic centre and the MBH binary there hosted, it delivers new stars onto it, again effectively renewing the reservoir of stars and thus aiding the binary shrinking \citep{2018MNRAS.474.1054B,2019MNRAS.484..520A}.

\subsection{The small-scale inspiral: triplets and multiplets}

In this section we investigate deeper the situation when a MBH binary is perturbed by a third incoming MBH, a possibility that could generally arise when stellar hardening timescale are long with respect to the galaxy cosmic merger rate. Before analysing the specific case of MBHs, it is worth focusing on the three-body dynamics from a general point of view and highlighting some of its most interesting features.

\subsubsection{The three-body problem}

The three-body problem is one of the oldest open question in astrophysics, and has resisted a general closed-form analytic solution for centuries. 
The statement of the problem concerns the time evolution of a closed system (i.e. no external forces present) with three gravitational sources (three planets, three stars, three black holes or a combination thereof), each of them subjected to the gravitational forces generated by the other two. The aim is to find a solution that would give the value of position and momentum of the three bodies at any instant of time. Unfortunately, no such solution exists. Deterministic chaos \citep{1892mnmc.book.....P}, i.e. the fact that tiny perturbations in the initial conditions lead to exponentially divergent outcomes, prevents the existence of any closed-form solution that can be expressed in terms of a finite number of standard mathematical functions. 
Analytical solutions exists only in portions of the full parameters space and generally require particular initial configurations or hierarchies in terms of either masses or separations. Notable examples involve the Euler and Lagrange solutions of the collinear and equilateral triangular configurations, respectively \citep[see e.g.][]{1999thor.book.....B}; the restricted three-body problem, where one of the bodies is taken as massless and therefore unable to perturb the motion of the other two \citep{1967torp.book.....S}; the hierarchical three-body problem, where two of the bodies are bound in a tighter binary (the inner binary), while the third one orbits far apart, effectively forming an outer binary with the centre of mass of the inner one \citep{1962AJ.....67..591K,1962P&SS....9..719L}.

Another and orthogonal way to analytically tackle the problem involves the search for a statistical solution. By exploiting the chaotic nature of the interaction, which quickly looses any memory of the initial conditions, one can make the assumption of ergodicity and note that the distribution of outcomes is uniform across the accessible phase volume. A complete solution, but in the statistical sense, of the three-body problem can be obtained, providing closed-form distributions of outcomes (e.g. binary orbital elements) given the conserved integrals of motion \citep{2019Natur.576..406S}.

Still, if the time evolution of a specific system is the goal, the only viable approach has to rely on the numerical integration of the equations of motion, a technique that was not applicable when the original statement of the problem was conceived (owing to the need of computers able to perform many computations in a short timescale), while nowadays is largely employed. Once the numerical noise is kept small enough through various algorithms and software implementations, the numerical integration of the three-body problem successfully provides us a complete description of the time evolution of positions and momenta of the three involved bodies \citep[see e.g.][]{1983ApJ...268..319H,2003MSAIS...1...54S}.

Despite during the years many efforts have been directed towards a more formal characterisation of the three-body problem, the interest in triple systems is not a purely mathematical exercise. Those system are indeed common in the Universe and they are found in many different astrophysical settings covering a large range of masses and physical scales, such as: planetary dynamics \citep{2013MNRAS.431.2155N,1997Natur.386..254H,2011PhRvL.107r1101K,2012ApJ...754L..36N}
triple stars \citep{1997AstL...23..727T,2007IAUS..240..347E,2014AJ....147...86T,2014AJ....147...87T,2020A&A...640A..16T}, accreting compact binaries with a companion \citep[e.g.][]{1988ApJ...334L..25G}, interactions of stellar size objects in globular clusters \citep{2016ApJ...816...65A,2016MNRAS.456.4219A} and around MBHs \citep[see e.g.][]{2012ApJ...757...27A}, and even triple MBH systems, that we are now going to describe in detail.

\subsubsection{The formation of MBH triplets}

In the context of the hierarchical formation of  cosmic structures, especially at high-redshift, the environment in which MBHs live is highly dynamical, as halo interactions and mergers are quite more frequent than in the  $z=0$ Universe \citep[see e.g.][where a system at $z\sim 2$ containing several AGNs in the same 400 kpc-wide Ly-$\alpha$ Jackpot nebula was discovered]{2015Sci...348..779H}. The outcome of these encounters could lead to either the formation of an MBH binary or, at least temporarily, a wandering MBH. This translates into a multiple MBH population in the grown galaxy halo, each inherited from a different merger \citep{2019MNRAS.486..101P}.  
The specific environmental conditions of each galaxy determine the evolutionary pace of each  MBH. Failures in the binary formation/shrinking process are likely when the galaxy morphology is rather messy or the stellar densities are low or gaseous coupling is rather inefficient. Such bottlenecks are expected to affect the specific merger rate as a function of redshift, MBH mass, and mass ratio \citep{2016PhRvD..93b4003K,2019MNRAS.486.4044B,2020ApJ...904...16B}. 
When a bound MBH binary happens to shrink inefficiently and it is thus  long-lived, the formation of MBH triplets (or even multiplets) in the same galaxy is a likely outcome. Within a triple system we witness a richer and more complex phenomenology with respect to binaries, possibly driving MBHs towards a prompt coalescence \citep{1990ApJ...348..412M,2001A&A...371..795H,2002ApJ...578..775B,2007MNRAS.377..957H,2010MNRAS.402.2308A,2012MNRAS.422.1306K,2017ApJ...840...53R,2018MNRAS.473.3410R,2018MNRAS.477.3910B,2021ApJ...912L..20M}.

\subsubsection{The evolution of MBH triplets}

\begin{figure}
    \centering
    \includegraphics[scale=0.4]{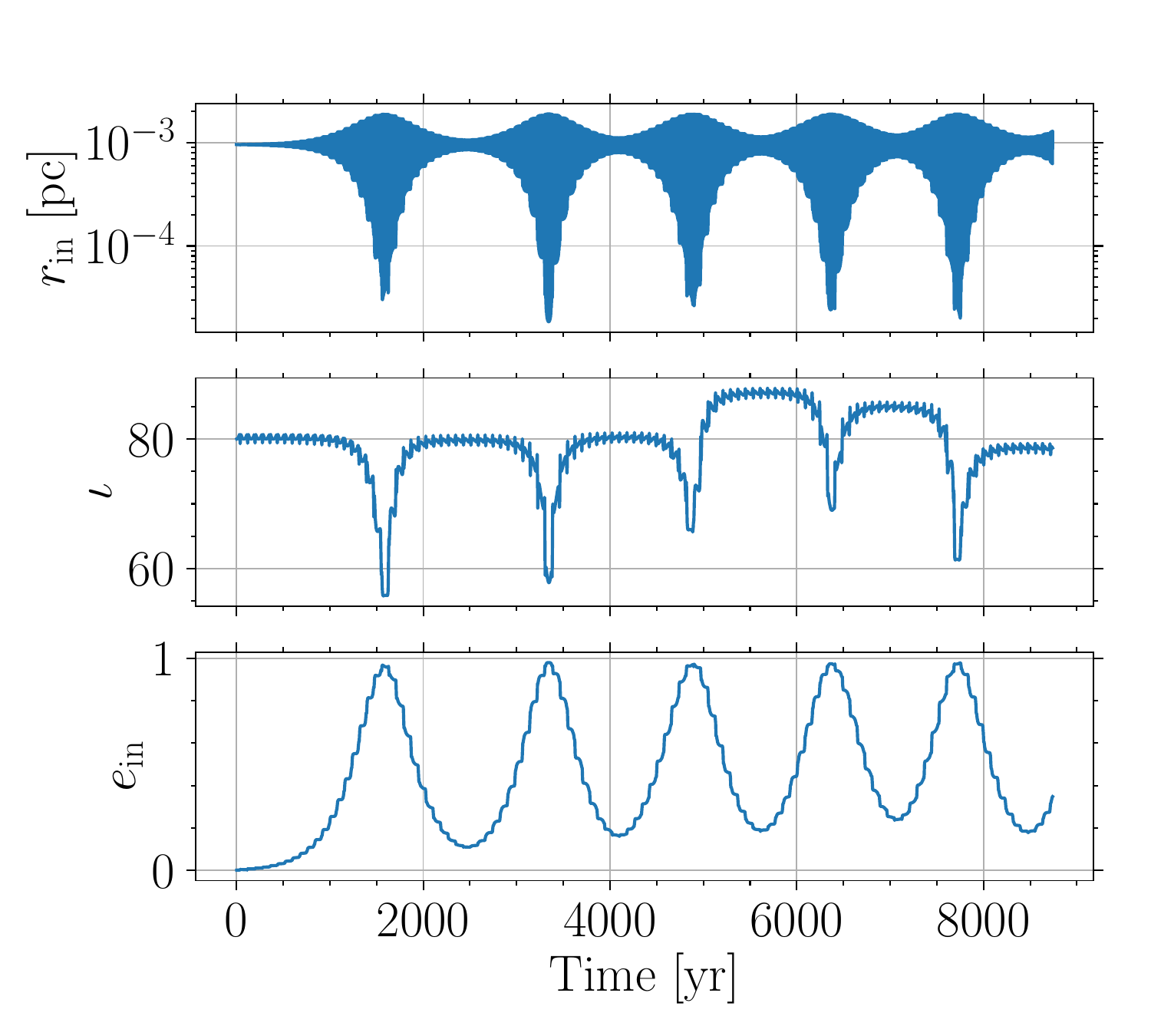}
    \caption{Example of triplet evolution under von Zeipel-Kozai-Lidov oscillations. The system features an inner MBH binary with $m_1 = 10^6, m2 = 10^5 \rm M_\odot$, perturbed by a third object with $m_3 = 10^7$; the initial relative inclination is $\iota =80^\circ$, while the inner and outer semi-major axes are $\sim 10^{-3}$ pc and $\sim10^{-2}$ pc, respectively. From top to bottom we show the relative separartion of the inner binary, the relative inclination and the inner eccentricity, all as a function of time. As clearly visible, during a von Zeipel-Kozai-Lidov cycle the inner eccentricity grows significantly (from $\sim 0$ up to $\sim 0.98$) while the relative inclination decreases.}
    \label{fig:ZKL}
\end{figure}

Triplets of MBHs generally start their evolution as hierarchical systems, where a third body binds to an already formed MBH binary. Given the hierarchy in scales two well separated binaries can be identified: the inner binary characterised by masses $m_1,m_2$, semi-major axis $a_{\rm in}$ and eccentricity $e_{\rm in}$, and the outer binary with elements $a_{\rm out}, e_{\rm out}$ and formed by the intruder MBH ($m_3$) and the centre of mass of the inner binary, effectively appearing as a point mass for the third MBH. An additional important parameter is the relative inclination of the orbital planes of the two binaries ($\iota$), which depending on its value determines a quite interesting phenomenology. 
In the simplified case of   three-body systems in vacuum (with no additional forces further to those among the three bodies are present), the bodies torque each other and exchange angular momentum on a timescale much longer than the orbital periods of both binaries; as a result, oscillations of the inner binary orbital eccentricity are periodically excited at the expense of the relative inclination (see Fig.~\ref{fig:ZKL}). Those are the so-called von Zeipel-Kozai-Lidov   oscillations \citep{1910AN....183..345V,1962AJ.....67..591K,1962P&SS....9..719L,2016ARA&A..54..441N}.
To gain physical insight into the phenomenon, this can be described in an analytical fashion. Given the spatial separation of the inner and outer binary, the full set of equation of motion can be simplified by expanding them in terms of the small parameter $a_{\rm in}/a_{\rm out}$ \citep[see e.g.][]{2016ARA&A..54..441N}. Moreover, since the von Zeipel-Kozai-Lidov oscillations happen on a long timescale, one generally averages out over the orbital timescale of the two binaries, effectively revealing only the long-term or secular variations of the orbital elements. The first non-vanishing order in the above expansion is proportional to $(a_{\rm in}/a_{\rm out})^2$ and is known as the quadrupole term. At this level of approximation, von Zeipel-Kozai-Lidov oscillations happen only when the initial inclination $\iota$ lies in the range
\begin{equation}
    \cos \iota = \pm \sqrt{\frac{3}{5}},
\end{equation}
or more explicitly when $39.2^\circ \leq \iota \leq 140.77^\circ$. The growth of $e_{\rm in}$ also only depends on the initial $\iota$, i.e.
\begin{equation}
    e_{\rm in, max} = \sqrt{1-\frac{5}{3}\cos \iota},
\end{equation}
meaning that (i) von Zeipel-Kozai-Lidov oscillations are  typical of inclined orbits and (ii)  polar orbits generally determine a much higher eccentricity growth. 
This is a crucial results (and one of reasons for which the von Zeipel-Kozai-Lidov mechanisms has raised much attention) in view of the small scale GW-driven dynamics. A highly eccentric binary is known to efficiently dissipate its orbital energy due to GW emission on a timescale that is a very steep function of the eccentricity \citep{1964PhRv..136.1224P}, therefore any mechanism that can produce very eccentric binaries is quite important for determining the rate of expected GW events. 
Finally, at the quadrupole level of approximation, the regular oscillations of the eccentricity and inclination yield a well defined associated timescale given by
\begin{equation}
    T_{\rm quad} \sim \dfrac{15}{16} \frac{a_{\rm out}^3 (1-e_{\rm out}^2)^{3/2} \sqrt{m_1+m_2}}{a_{\rm in}^{3/2} m_3 \sqrt{G}},
\end{equation}
with $G$ the gravitational constant.
When the next order in the $a_{\rm in}/a_{\rm out}$ expansion is taken into account (the octupole order), the richness of the phenomenology of hierarchical triplets further increases at the expense of  analytical treatments. To evolve the dynamics of the system we generally need to numerically integrate the secular equations of motion, revealing a series a non-linear phenomena like chaotic variations of the relative inclination. Specifically, orbital flips between prograde and retrograde orbits are now allowed, possibly causing an even more severe enhancement of the eccentricity growth \citep[see][for a comprehensive review]{2016ARA&A..54..441N}.

Despite the effectiveness of the von Zeipel-Kozai-Lidov mechanism in growing the eccentricity of the inner binary, we need to keep in mind that the period of the von Zeipel-Kozai-Lidov oscillations has a strong dependence on the outer semi-major axis ($\propto a_{\rm out}^3$), therefore wider outer binaries can take a significant time to actually excite the oscillations. This is a crucial point since any other physical phenomenon perturbing the inner binary on a shorter timescale can actually interfere with a successful von Zeipel-Kozai-Lidov cycle. One of such phenomena is any form of apsidal precession. In practice, if the pericentre precession is faster than a full von Zeipel-Kozai-Lidov cycle, then the apsidal motion destroys the coherent accumulation of secular torques, hindering therefore any relevant eccentricity growth \citep[see e.g.][]{2000ApJ...535..385F,2016ARA&A..54..441N,2020PhRvD.102f4033L}.
Given the relativistic nature of MBHs, the relativistic precession of the inner binary is in many situation non-negligible. It is therefore crucial to consider it in order not to overestimate the eccentricity growth of MBH binaries. Still, we need to remember that the same processes that have formed (and shrunk) the inner binary also operate on the outer binary. The interaction with the host galaxy environment determines the orbital decay of the intruder MBH and the consequent decrease of $a_{\rm out}$. This can  effectively re-enhance the von Zeipel-Kozai-Lidov mechanism as the triplet becomes less hierarchical \citep[see e.g.][]{2018MNRAS.477.3910B}. This translates into a shortening of the von Zeipel-Kozai-Lidov oscillation timescale and a gradual strengthening of the perturbation acting on the inner binary, again promoting the eccentricity oscillations with subsequent strong GW emission and possible coalescence at the peak of a von Zeipel-Kozai-Lidov cycle \citep{2018MNRAS.477.3910B}.

If on the one side the evolution of the outer binary can help the effectiveness of the von Zeipel-Kozai-Lidov mechanisms, on the other side there are many initial conditions under which no merger can occur during the secular evolution phase of MBH triplets. For example, the mutual inclination may not be high enough to excite the oscillations, the perturber can be too light, or the binary might remain too wide for efficient emission of GWs. In these situations, the final fate of many MBH triplets is thus dynamical instability. As the orbit of the outer MBH  gradually shrinks, the third body gets closer and closer to the inner binary. At some point, the three bodies will be too close for the triplet to remain stable, resulting in a  chaotic dynamics characterized by strong encounters, exchanges, and ejections (i.e. the same processes described in Sec.~\ref{sec:dyn_proc_MBHB}). 
Again, this may not represent the end of the story. An ejected MBH may be kicked on a wide but bound trajectory, in which case it may return back and perturb the leftover binary, this time through a series of energetic encounters, the specific strength and occurrence depending on the galactic potential (spherical, axisymmetric or triaxial), the outgoing trajectory and also on the dynamical friction efficiency.
The repeated chaotic perturbations can again increase the orbital eccentricity of the leftover binary, therefore promoting coalescence in a non-negligible fraction of cases, but also leaving a considerable number of ejected MBHs that may keep wandering inside galaxies \citep[see, e.g.][]{2018MNRAS.477.3910B}. 

As a final consideration, we can consider the situation when the lifetime of hierarchical triplets is long enough for new galaxy mergers to provide additional MBHs, leading to hierarchical quadruplets (and even higher-order multiplets).
In the specific case of quadruplets, a quite natural configuration can be produced when two merging galaxies both already host MBH binaries.
In this particular case, the system can be effectively described as a hierarchical triplet until the four-body nature of the system becomes manifest, leading again to chaotic dynamics characterised by highly stochastic and largely non-predictable encounters, requiring therefore numerical investigations.

Despite the large uncertainties and range of possible outcomes, a likely signature of a MBH binary coalescence triggered by dynamical interactions should be always present: the very high acquired eccentricity. Such a high eccentricity could be retained (at least in residual form) well inside the GW-dominated phase (e.g. at a  separation of a few tens of gravitational radii), therefore providing a decisive smoking gun of this exciting  MBH binary path to coalescence \citep[][]{2018MNRAS.473.3410R,2019MNRAS.486.4044B}.
 
\subsection{The small-scale inspiral: Hardening in gaseous environment}

In the previous Sections we have considered the evolution of MBH binaries in the presence of a purely stellar background and accounting for the formation of possible MBH multiplets. Here instead we focus on another set of phenomena that can deeply affect the evolution of MBH binaries, the hydrodynamical interactions between the  bound, small-scale MBH binary and the reservoir of gas that may surround it. Discs of gas are often present in the surroundings of single MBHs (consider for instance AGNs and quasars) thus it is likely that at least a fraction of MBH binaries is also surrounded by gas, which may be brought down to small scales by the galaxy merger itself. In order to detail the evolution of MBH binaries embedded in gaseous discs, we first need to introduce the concept of \textit{accretion disc}. In the beginning, we do not specialise our description to the binary case; instead, we introduce in a very broad manner the idea of an astrophysical accretion disc.

\subsubsection{Accretion discs} 

Accretion discs are ubiquitous in astrophysics, from large scales, e. g. surrounding MBHs in AGN, down to small scales, e. g. protostellar discs and discs around planets. In particular, the existence of discs in AGN has been found through observations of water masers emission \citep{Miyoshi1995}.

The formation of disc-like geometries is related to the angular momentum content of the molecular cloud that surrounds a central object. In the absence of rotation, the gas in the cloud could in principle freely collapse towards the central object. Even a small amount of angular momentum prevents such radial collapse and allows the gas to sink down to a minimum distance from the centre. 

The dissipative processes that occur in these discs between fluid elements provide an efficient source of angular momentum and energy transport throughout the disc itself. 
This allows the fluid elements to loose angular momentum, spiral inwards and eventually accrete onto the central gravitating object. \\

\paragraph{Accretion disc structure and dynamics }

Accretion discs are usually assumed to be geometrically thin \citep{Pringle1981}. The typical scale for the disc thickness $H$ is set by the disc temperature $T$, which varies considerably, even within individual discs. 
In general, accretion discs have a large surface and thus they can cool down fast through radiation.
This implies that the temperature and the pressure of the disc can reach low values rapidly and thus are not able to support the disc against gravity unless the disc is flat, thus if $H/R\ll1$ where $R$ is the radial distance.

Requiring that $H/R \ll 1$ is equivalent to requiring the sound speed $c_{\rm s}$ to be much smaller than the rotational velocity $v_{\phi}$. Moreover, assuming that  accretion takes place on a long timescale, the radial velocity $v_{\rm R}$ turns out to be smaller than both the sound speed and the rotational speed. Thus in thin accretion discs we have the ordering: $v_{\rm R}\ll c_{\rm s}\ll v_{\phi}$.
We now treat the thin disc approximation according to which most of the equations we are going to use can be integrated in the vertical direction. Thus rather than dealing with quantities per unit volume (such as the density $\rho$), we deal with quantities per unit surface (such as the surface density $\Sigma$) instead. When volume quantities are needed (for example the viscosity $\nu$), these will generally be understood as vertically averaged. 

The evolution of accretion discs can be described by the basic equations of viscous fluid dynamics: the continuity and the Navier-Stokes equations. The continuity equation reads:

\begin{equation}
\frac{\partial\rho}{\partial t}+\mathbf{\nabla}\cdot(\mathbf{\rho v})=0
\label{eq:continuity-eq}
\end{equation}
while the Navier-Stokes equation is:

\begin{equation}
\frac{\partial\mathbf{v}}{\partial t}+(\mathbf{v}\cdot\mathbf{\nabla})\mathbf{v}=-\frac{1}{\rho}(\mathbf{\nabla}P-\mathbf{\nabla}\cdot\mathbf{T})-\mathbf{\nabla}\Phi
\label{eq:navier-stokes}
\end{equation}
where $\mathbf{T}$ is the stress tensor that describes the effect of viscous forces, $P$ is the pressure, $\rho$ is the density and $\Phi$ is the gravitational potential.
The left hand side is the acceleration with the second term describing the momentum convected into the fluid by velocity gradients. On the right hand side we find the various forces acting on the fluid: pressure, viscous forces and gravity. 

The condition for centrifugal balance in the radial direction can be derived by writing the vertically integrated form of Eq. (\ref{eq:navier-stokes}) in cylindrical coordinates (see for details section 4.3 in \cite{Lodato2008}). This condition is essentially a restatement of Kepler's third law, meaning that accretion discs in centrifugal equilibrium are Keplerian discs. Note that a Keplerian disc is strongly shearing in the radial direction, since the angular velocity is $\Omega=(GM/R^3)^(1/2)$ and therefore decreases relatively fast with $R$. 

The condition for hydrostatic equilibrium in the vertical direction can be instead obtained from the Navier-Stokes equation where the left hand side becomes negligible provided that the velocity in the vertical direction is very small. 
The viscous force vanishes as well, since the only non-zero component of the stress is in the $R\phi$ direction and we are therefore left with only two terms to balance: gravitational force and pressure force in the vertical direction.
Solving the equation with these remaining terms gives the disc vertical density profile, which turns out to be a Gaussian 
\begin{equation}
\rho(z)=\rho_{0}e^{-z^{2}/2H^{2}}
\end{equation}
where $\rho_{0}$ is the mid-plane density and $H=c_{\rm s}/\Omega=R c_{\rm s}/v_{\phi}$ is the disc scale height.
The disc aspect ratio is then 

\begin{equation}
\frac{H}{R}=\frac{c_{\rm s}}{v_{\phi}}\,,
\label{eq:vertical_hoverr}
\end{equation}
which then demonstrates that requiring that the disc is geometrically thin is equivalent to requiring that the disc rotation is highly supersonic. 
The speed of sound is given by $c_{\rm s}^{2}=P/\rho$, where in general the pressure $P$ is the sum of gas and radiation pressures.\\

\paragraph{Disc viscosity}

The nature of viscosity in accretion discs has been debated for decades in the scientific community. It is now commonly known, by means of hydrodynamical simulations, that accretion discs are turbulent and the transport occurs because of fluctuations associated with turbulence. While in non turbulent flows the exchange of angular momentum occurs through collisions between individual gas particles, in turbulent flows the mixture of fluid elements enhances the transport of angular momentum. Turbulence might arise from magneto-hydrodynamical instabilities in the accretion disc, in particular the magneto-rotational instability (MRI) \citep{Balbus1991}.
Let us consider a vertical weak magnetic field in a well-ionized plasma. In this configuration, the magnetic field lines are frozen within the disc. If now we perturb two fluid elements with a small radial displacement the magnetic field would tend to bring them back to the original configuration. Due to the shear in the disc, the inner fluid element coupled to the field moves azimuthally faster than an outer one. Magnetic tension along the field line then acts to remove angular momentum from the inner element, and add angular momentum to the outer one. This causes further radial displacement, leading to an instability.

\citet{SS1973} provided an estimate of the magnitude of viscosity using the simplest assumption to take the viscous stress tensor to just be proportional to the vertically integrated pressure $\Sigma c_{\rm s}^{2}$ : 

\begin{equation}
T_{\rm R\phi}=\frac{d\ln\Omega}{d\ln R}\alpha\Sigma c_{\rm s}^{2}\,,
\label{eq:stress_tensor}
\end{equation}
where $d\ln\Omega/d\ln R=-3/2$ for a Keplerian disc, and where $\alpha$ is the proportionality factor between the stress tensor and the pressure.
Assuming that the turbulence in the disc is isotropic and subsonic,  \cite{SS1973} introduced a prescription for the kinematic viscosity $\nu$ given by

\begin{equation}
\nu=\alpha c_{\rm s}H\,.
\label{eq:alpha_presc}
\end{equation}
The magnitude of the turbulent viscosity is given roughly by $\nu\sim v_{\rm t}l$, where $l$ is the typical size of the largest eddies in the turbulent pattern and $v_{\rm t}$ is the typical turbulent velocity. It is unlikely that the turbulence is highly supersonic since otherwise it would easily be dissipated through shocks. We thus have $v_{\rm t}\lesssim c_{\rm s}$. An upper limit to the size of the largest eddies $l$ is  given by the disc thickness $H$ (if we consider an isotropic viscosity). These two upper limits, taken together, clearly imply that $\alpha<1$. 
It is important to point out that this is not a theory of viscosity in accretion discs but rather a simple parameterization based on dimensional analysis and on the fact that it is natural to assume that to a first approach the local stresses are proportional to the local rate of strain. This parameterization allows to put all the uncertainties on the mechanism that provides the source of viscosity in the disc in this parameter $\alpha$.

Non-local mechanisms of angular momentum transport may exist and for these the simple linear proportionality in Eq. (\ref{eq:stress_tensor}) is no longer valid. 
For instance, disc instabilities may play a crucial role in sustaining turbulence and providing a source of energy and angular momentum transport in accretion discs. 
The two main instabilities are the MRI and the gravitational instability (GI) \citep{Kratter2016}. The condition for the MRI to take place, in the regime of ideal magnetohydrodynamics (MHD), is that $\partial\Omega^2/\partial R <0$, which is generally satisfied in accretion discs, provided that the gas is perfectly coupled with the magnetic field. The gravitational instability might take place when $M_{\rm disc}/M \gtrsim H/R$.  Especially in the colder outer parts, the disc self-gravity might affect its behaviour due to the propagation of density waves which lead to the formation of spirals. These waves are expected to provide a non negligible contribution to the angular momentum transport. In some cases the instability is strong enough that the disc might also fragment producing bound clumps. The investigation of the dynamics of self-gravitating discs is crutial to understand the processes that lead to the feeding of young stars and supermassive black holes in AGN.\\   

\paragraph{Angular momentum conservation}

The azimuthal component of the Navier-Stokes equation combined with the continuity equation and integrated in the vertical direction yields the equation for the radial velocity profile within the accretion discs. 
This can be then inserted back into the continuity equation to give:

\begin{equation}
\frac{\partial \Sigma}{\partial t} = - \frac{1}{R} \frac{\partial}{\partial R} \left[ \frac{1}{\left(R^2\Omega\right) '} \frac{\partial}{\partial R} \left(\nu \Sigma R^3 \Omega ' \right) \right]\,.
\label{eq: discs_sigma_evol}
\end{equation}
Equation (\ref{eq: discs_sigma_evol}) is one of the key ingredients in accretion discs theory. It is a diffusion equation for the surface density of the disc, whose temporal evolution is determined only by the kinematic viscosity $\nu$.

In stationary conditions \footnote{i.e. conditions under which the relevant quantities do not change with time and we can therefore assume their time derivative to be zero.}, $v_{\rm R}<0$  and thus the matter is moving inward. We can then define the accretion rate as $\dot{M}=-2\pi\Sigma v_{\rm R}R={\rm const}$ and use the angular momentum conservation equation to obtain 

\begin{equation}
\dot{M}\Omega R^{2}+2\pi T_{R\phi}R^{2}\equiv\dot{J}
\label{eq:ang-mom-flux}
\end{equation}
where the first term represents the amount of angular momentum transported by the accreting matter. The forces that are leading to the angular momentum transport in a rotating system are also inducing an energy flow equal to $G(R)=-2\pi R^{2}T_{R\phi}$. In Keplerian discs $\Omega$ increases inward and $T_{R\phi}<0$, thus the energy flow is directed outwards. 

The flux of angular momentum becomes, considering the definition in Eq. (\ref{eq:stress_tensor}),

\begin{equation}
\dot{J}=\dot{M}\Omega R^{2}-3\pi\nu\Sigma R^{2}\Omega=R^{2}\Omega(\dot{M}-3\pi\nu\Sigma)
\label{eq:dotmSigma}
\end{equation}
The value of the flux of angular momentum can be obtained from the so called `no torque boundary condition' assuming that the disc extends all the way down to the surface $R=R_{\rm in}$ of the central object. In a realistic situation, the central object must rotate more slowly than the break-up speed at its equator, i.e. with angular velocity $\Omega_{\rm{in}}<\Omega(R_{\rm in})$. In this case the angular velocity of the disc material remains Keplerian and thus increases inwards, until it begins to decrease to the value $\Omega_{\rm{in}}$ in a `boundary layer' of radial extent $b$.
Thus there exists a radius $R=R_{\rm in}+b$ at which $d\Omega/dR=0$. At this radius, the viscous torque vanishes due to the presence of a boundary layer where the disc connects to the central object.
For the thin disc approximation described above, we have that $b\ll R_{\rm in}$. 

At large radii, $R\gg R_{\rm in}$ , the surface density and the viscosity satisfy the following simple relation: 

\begin{equation}
\dot{M}=3\pi\nu\Sigma
\end{equation}
that is, surface density and viscosity are inversely proportional.

We note that, since a black hole does not have a `hard surface' at $R_{\rm in}$, the `no torque boundary condition' might not be necessary. However, numerical simulations performed by \cite{Shafee2008}  have shown that, in the steady state, the specific angular momentum profile of the accreting magnetized gas (around a non-spinning black hole) at the Innermost Stable Circular Orbit (ISCO) deviates by only 2\% from the standard thin disc model.

\begin{figure}
    \centering
    \includegraphics[width=0.7\textwidth]{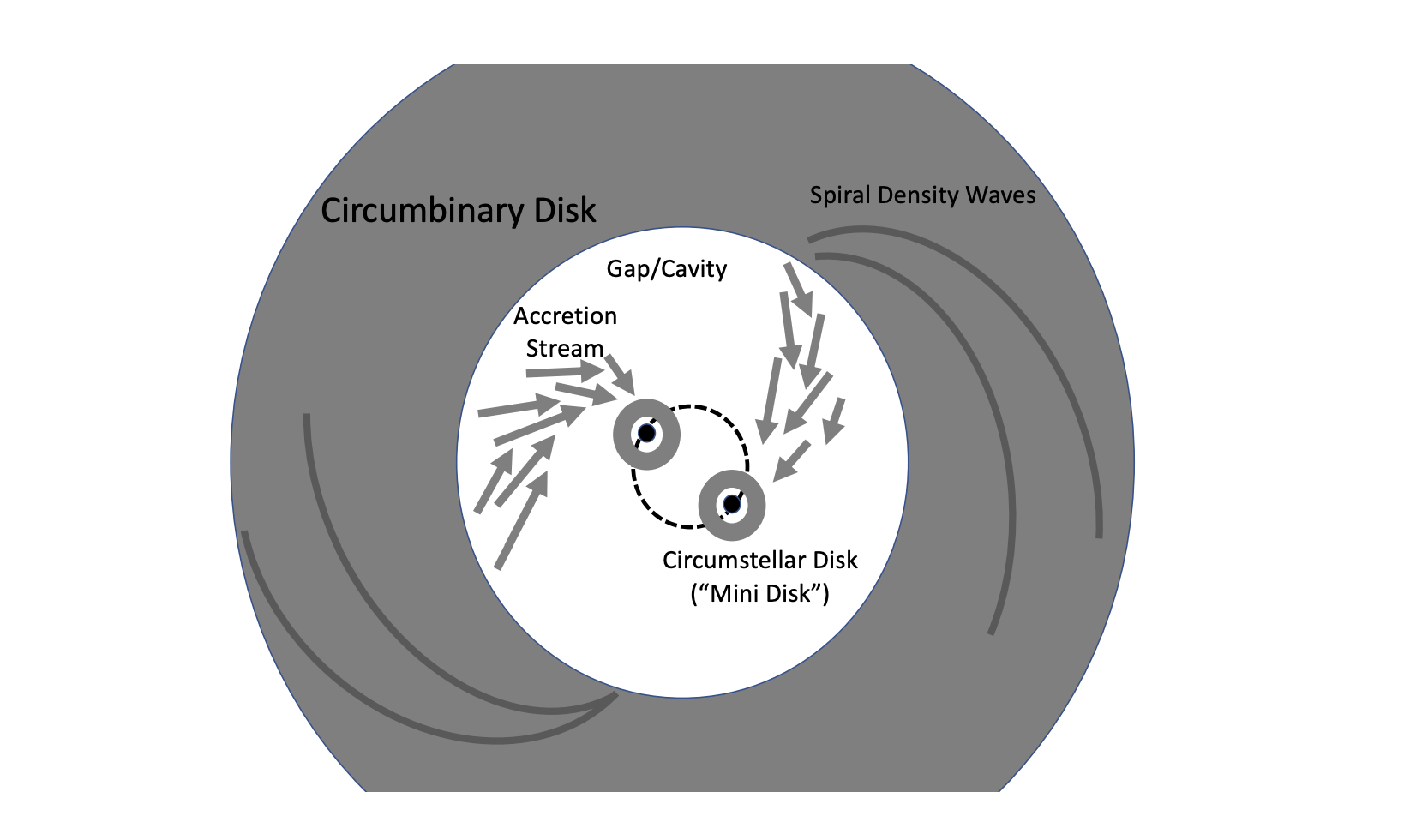}
    \caption{Schematic representation of a circumbinary disc. Figure adapted from Figure 1 in \cite{Lai2022}.}
    \label{fig:cbd}
\end{figure}

\subsubsection{Accretion discs around massive black hole binaries}

During a galaxy merger, each of the two galaxies brings with it a significant amount of gas that eventually sinks to the center of the remnant forming a circumbinary accretion disc, i.e. a disc surrounding two objects that orbit each other, around the bound MBH binary \citep{Mayer2007}. This gaseous disc might facilitate the MBH binary merger and give rise to electromagnetic counterparts of the gravitational wave (GW) emission \citep{1980Natur.287..307B,2002ApJ...567L...9A,2009MNRAS.398.1392L}. 

A coplanar circumbinary disc can extend down to a few times the binary separation where it is truncated by the varying gravitational force from the binary. 
Very early numerical simulations investigated the interaction of a binary with its gaseous circumbinary disc finding that only a small amount of material is able to enter the cavity carved by the binary \citep{1994ApJ...421..651A,1996ApJ...467L..77A}. This material then forms circum-single discs (also referred to as mini-discs) and eventually accretes onto the binary components (see Fig. \ref{fig:cbd}). \\

\paragraph{Theory of circumbinary disc accretion}

The most significant additional piece of physics of circumbinary discs compared to standard accretion discs is the presence of resonances inside the disc.
In the linear regime, i.e. when the perturbation of the disc by the binary is weak, the gravitational potential produced by a binary of mass $M=M_1+M_2$, eccentricity $e$ and semi-major axis $a$ can be written as \citep{Goldreich1979}

\begin{equation}
    \Phi({\bf r},t)=\sum_{m=0}^{\infty} \sum_{n=-\infty}^{\infty}\Phi_{\rm m,n}(r)\cos\left[m\phi - (m\Omega_{\rm b}+n\kappa_{\rm b})t\right]
\end{equation}
where $\Omega_{\rm b}=(GM/a^3)^{1/2}$ is the mean angular frequency of the binary and $\kappa_{\rm b}$ is the radial epicyclic frequency (for Keplerian binaries $\Omega_{\rm b}=\kappa_{\rm b}$).
The potential component $\Phi_{\rm m,n}$ depends on the binary eccentricity and semi-major axis and for $m>0$ it rotates with the pattern frequency $\omega_{\rm p}=(m\Omega_{\rm b}+n\kappa_{\rm b})/m$ exciting spiral density waves at the Lindblad resonances (LRs) where
\begin{equation}
    \omega_{\rm p}-\Omega(r) = \pm \frac{\kappa(r)}{m}
    \label{eq:res}
\end{equation}
where the upper (lower) sign corresponds to the outer (inner) LR.
Since the circumbinary disc is approximately Keplerian, Eq. (\ref{eq:res}) corresponds to $r_{\rm LR}\simeq((m \pm 1)/(m+n))^{2/3}a$.

The disc orbits resonate with the binary orbit at these discrete locations leading to the exchange of angular momentum between the disc and the binary \citep{1972MNRAS.157....1L,1986ApJ...309..846L}. 
The torque on the disc at the outer Lindblad resonance is positive, meaning that the disc particles gain angular momentum through resonant gravitational torques.
The magnitude of the resonant torques depends on the binary potential, i.e. its mass ratio and eccentricity, and is proportional to the disc surface density at the resonance locations \citep{1980ApJ...241..425G}. Therefore, the amount of angular momentum transferred from the binary to the disc at the resonances depends on the disc properties as well. 

The particles in the disc also lose angular momentum owing to the presence of viscosity, therefore the cavity can be opened at the LR locations if the resonant torque is stronger than the viscous torque.
Numerical simulations found the cavity size to be $r_{\rm cav}\simeq 2-3 a$  for coplanar discs \citep{1994ApJ...421..651A} while the cavity can be smaller for misaligned discs \citep{Franchini2019c}.\\

\paragraph{The impact of gravitational and accretion torques on the binary orbital evolution}

The gravitational torque that the distribution of gas particles exerts on the binary would vanish in a perfectly symmetric system. However, its contribution to the binary angular momentum change arises because of the asymmetries in the flow around the binary. Furthermore, the circumbinary disc exerts a negative gravitational torque owing to the presence of resonances that extract angular momentum from the binary.

Since the gas particles within the disc also lose angular momentum through viscous processes, they will eventually accrete onto the binary.
Angular momentum conservation implies that the sum of the gravitational torque with the accretion torque regulates the change of the binary angular momentum $L_{\rm z}=\mu\sqrt{GMa(1-e^2)}=L_{\rm z,grav}+L_{\rm z,acc}$.
We can therefore write
\begin{equation}
    \frac{\dot{a}}{2a} = \frac{\dot{L}_{\rm z,acc}}{L_{\rm z}} + \frac{\dot{L}_{\rm z,grav}}{L_{\rm z}} - \frac{\dot{\mu}}{\mu} - \frac{\dot{M}}{2M} + \frac{e\dot{e}}{(1-e^2)}
    \label{eq:adot}
\end{equation}
where $\mu=M_1M_2/M$ is the reduced mass and $e$ is the binary eccentricity. 
The first term is always positive and represents the accretion of angular momentum onto the binary. The second term is given by the sum of the positive contribution of the discs around the two components and the negative contribution of the circumbinary disc. The terms due to the accretion of mass (i.e. third and fourth terms) are additional negative contributions to the semi-major axis evolution. The last term comes from the binary eccentricity evolution.
Therefore we essentially have two positive terms whose effect is to drive the binary apart and two negative contributions that remove angular momentum from the binary driving it towards merger.
In more physical terms, the angular momentum change due to the gravitational and accretion torques translates into a change of the different elements of the system, i.e. the masses and orbital elements. 
Since the eccentricity contribution is negligible for initially circular binaries, the evolution of the binary separation depends on the fraction of the angular momentum exchanged with the disc that goes into the mass accretion terms.\\

\paragraph{Numerical simulations}

Numerical simulations are needed in order to fully capture the complexity of binary disc interactions as eccentric Lindblad resonances and secular interaction also play a role exciting the disc eccentricity \citep{Lubow1991, Lubow2022} which may have a non-negligible impact on the torques acting on the binary.

Simulations of circumbinary accretion discs are challenging because of the wide spatial range involved and the different timescales on which variability takes place. Furthermore, since the flow in the disk and near the binary is highly dynamical, to determine the long-term effect of the disc on the binary evolution, the simulations need to be run for a sufficiently large number of binary orbits.
    
The main finding of early numerical works is that the binary semi-major axis decreases with time owing to the interaction with the disc. This means that the disc is able to extract angular momentum and energy from the binary orbit during the interaction.
This picture has been recently challenged by a few works \citep{2019ApJ...871...84M,2020ApJ...901...25D} employing 2D (and one 3D simulation, see \citealt{2019ApJ...875...66M}) static or moving-mesh grid numerical simulations with fixed binary orbits. In particular, these studies found that the torque exerted by the circum-single discs is positive and may overcome the negative torque from the circumbinary disc effectively causing the binary to gain angular momentum and expand its orbit.  More recently \cite{2020ApJ...900...43T} found, using the same numerical techniques, that the overall sign of the torque exerted by the disc onto the binary depends on the disc temperature, i.e. on its aspect ratio $H/R$, for locally isothermal discs. In particular, they found discs with $H/R \leq 0.04$ to shrink the binary. 
Using 3D smoothed particle hydrodynamics (SPH) simulations of locally isothermal discs, \cite{2020A&A...641A..64H} found instead the threshold value for binary expansion to be $H/R \simeq 0.2$.
Other works that employed SPH simulations in the regime where GIs are the source of angular momentum transport and the disc temperature changes with time, found binary shrinking as a result of the interaction with massive discs regardless of the initial disc temperature \citep{2009MNRAS.393.1423C,2012A&A...545A.127R,2021MNRAS.507.1458F}.
The discrepancy in the results inferred using different numerical techniques has been argued to originate from the lack of numerical resolution in SPH simulations which, in constrast with grid-based ones, were unable to properly resolve the dynamics of the gas streams entering the cavity, artificially suppressing the positive torque associated to such gas that forces the binary to expand.

Recently, \cite{2022ApJ...929L..13F} employed the code {\sc gizmo} \citep{2015MNRAS.450...53H} in its mesh-less finite mass (MFM) mode, coupled with adaptive particle-splitting for numerical refinement of the gas dynamics inside the disc cavity; the aim of this work has been to accurately measuring the gravitational and accretion torques that originate from the circum-single discs onto the binary itself.
They found the binary orbit to decay with time for very cold (i.e. $H/R=0.03$) and very warm (i.e. $H/R=0.2$) discs and that the result of the interaction in the intermediate regime where $H/R=0.1$ to be strongly influenced by the disc viscosity as this essentially regulates the fraction of mass contained in the discs around the binary components as well as the fraction that is accreted by the binary. The balance between these two quantities ultimately determines whether the binary inspirals.
A detailed and comprehensive review of all the numerical simulations existent so far in the literature has been written by \cite{Lai2022}.

Although whether there exist a threshold for binary expansion in terms of the disc aspect ratio still remains to be determined, such large disc thicknesses are well beyond the typical aspect ratio we may expect for circumbinary discs surrounding MBH binaries, if their structure resembles that of AGN discs \citep{Collin1990}.

Finally, we note that even in the worst case scenario where the interaction with the disc is able to increase the binary angular momentum content, the presence of a stellar background coupled with GW emission can ultimately bring the two MBHs to merge within a Hubble time \citep{Bortolas2021}.

\section*{Acknowledgements}
JR acknowledges support from the Irish Research Council Laureate programme under grant number IRCLA/2022/1165. JR also acknowledges support from the Royal Society and Science Foundation Ireland under grant number URF\textbackslash R1\textbackslash 191132.
AF acknowledges financial support provided under the European Union’s H2020 ERC Consolidator Grant ``Binary Massive Black Hole Astrophysics" (B Massive, Grant Agreement: 818691). MB acknowledges support provided by MUR under grant ``PNRR - Missione 4 Istruzione e Ricerca - Componente 2 Dalla Ricerca all'Impresa - Investimento 1.2 Finanziamento di progetti presentati da giovani ricercatori ID:SOE\_0163'' and by University of Milano-Bicocca under grant ``2022-NAZ-0482/B''.

\interlinepenalty=10000 
\section{Bibliography}
\setlength{\bibsep}{6pt}
\bibliography{Biblio,mbh,mbh2}

\end{document}